\pdfoutput=1
\documentclass[twocolumn,twocolappendix]{aastex63}
\usepackage{afterpage}
\usepackage{multirow}
\usepackage{listings}
\usepackage[tbtags]{amsmath}
\usepackage{hyperref}
\usepackage[normalem]{ulem} % allows strikeout (\sout)
\usepackage{tabularx}

\newcommand{\dg}{$^\circ$}
\newcommand{\red}[1]{#1}%{\textcolor{red}{#1}}
\newcommand{\eg}{{e.g.}}

\newcommand{\ignore}[1]{{}}

\graphicspath{{./}{figures/}}

\received{\today}
\submitjournal{ApJ}

\shorttitle{MaDCoWS cluster observations using MUSTANG2}
\shortauthors{Dicker et al.}

\begin{document}
\title{The Massive and Distant Clusters of \textit{WISE} Survey X: \red{Initial Results from a Sunyaev-Zeldovich Effect Study of Massive Galaxy Clusters at $z > 1$ using MUSTANG2 on the GBT}}

\correspondingauthor{Simon R.\ Dicker}
\email{sdicker@hep.upenn.edu}
%Author list TBD - please add info to authors.txt
%\input{authors.tex}
\author[0000-0002-1940-4289]{Simon R.\ Dicker} 
\affiliation{Department of Physics and Astronomy, University of Pennsylvania, 209 South 33rd Street, Philadelphia, PA, 19104, USA}
\author[0000-0001-5725-0359]{Charles E. Romero}
\affiliation{Department of Physics and Astronomy, University of Pennsylvania, 209 South 33rd Street, Philadelphia, PA, 19104, USA}
\affiliation{Green Bank Observatory, Green Bank, WV, 24944, USA}
\author[0000-0003-3586-4485]{Luca Di Mascolo}
\affiliation{Max-Planck-Institut f\"{u}r Astrophysik (MPA), Karl-Schwarzschild-Strasse 1, Garching 85741, Germany}
\author[0000-0003-3816-5372]{Tony Mroczkowski}
\affiliation{ESO - European Southern Observatory, Karl-Schwarzschild-Str.\ 2, D-85748 Garching b.\ M\"unchen, Germany}
\author[0000-0001-6903-5074]{Jonathan Sievers}
\affiliation{Department of Physics, McGill University, 3600 University Street Montreal, QC H3A 2T8, Canada}
\author[0000-0001-9793-5416]{Emily Moravec}
\affiliation{Astronomical Institute of the Czech Academy of Sciences, Bo\v cn\'i II 1401/1A, 14000 Praha 4, Czech Republic}
%%
%% 
%%%%%%%%%%%%%%
\author{Tanay Bhandarkar}
\affiliation{Department of Physics and Astronomy, University of Pennsylvania, 209 South 33rd Street, Philadelphia, PA, 19104, USA}
\author[0000-0002-4208-798X]{Mark Brodwin}
\affiliation{Department of Physics and Astronomy, University of Missouri, 5110 Rockhill Road, Kansas City, MO 64110, USA}
\author[0000-0002-7898-7664]{Thomas Connor}
\affiliation{Jet Propulsion Laboratory, California Institute of Technology, 4800 Oak Grove Dr., Pasadena, CA 91109, USA}
%%%
\author{Bandon Decker}
\affiliation{Department of Physics and Astronomy, University of Missouri, 5110 Rockhill Road, Kansas City, MO 64110, USA}
\author[0000-0002-3169-9761]{Mark Devlin}
\affiliation{Department of Physics and Astronomy, University of Pennsylvania, 209 South 33rd Street, Philadelphia, PA, 19104, USA}
%%
%\author{Peter Eisenhardt}% 	Peter.R.Eisenhardt@jpl.nasa.gov 	
%\affiliation{Jet Propulsion Laboratory, California Institute of Technology, 4800 Oak Grove Dr., Pasadena, CA 91109, USA}
%%
\author[0000-0002-0933-8601]{Anthony H. Gonzalez}
\affiliation{Department of Astronomy, University of Florida, 211 Bryant Space Center, Gainesville, FL 32611, USA}
%%%
\author{Ian Lowe}
\affiliation{Department of Physics and Astronomy, University of Pennsylvania, 209 South 33rd Street, Philadelphia, PA, 19104, USA}
\author[0000-0002-8472-836X]{Brian S.\ Mason}
\affiliation{National Radio Astronomy Observatory, 520 Edgemont Rd., Charlottesville VA 22903, USA}
\author[0000-0003-0167-0981]{Craig Sarazin} 
\affiliation{Department of Astronomy, University of Virginia, P.O. Box 400325, Charlottesville, VA 22901, USA}
\author{Spencer A.\ Stanford}
\affiliation{Department of Physics, University of California, One Shields Avenue, Davis, CA, 95616, USA}
\author[0000-0003-2686-9241]{Daniel Stern}
\affiliation{Jet Propulsion Laboratory, California Institute of Technology, 4800 Oak Grove Dr., Pasadena, CA 91109, USA}
\author{Khunanon Thongkham}
\affiliation{Department of Astronomy, University of Florida, 211 Bryant Space Center, Gainesville, FL 32611, USA}
\author[0000-0003-2212-6045]{Dominika Wylezalek}
\affiliation{ESO - European Southern Observatory, Karl-Schwarzschild-Str.\ 2, D-85748 Garching b.\ M\"unchen, Germany}
%%%%
\author{Fernando Zago}
\affiliation{Department of Physics, McGill University, 3600 University Street Montreal, QC H3A 2T8, Canada}
\begin{abstract}
The properties of galaxy clusters as a function of redshift can be utilized as an important cosmological tool. We present initial results from a program of follow-up observations of the Sunyaev-Zeldovich effect (SZE) in high redshift galaxy clusters detected at infrared wavelengths in the Massive and Distant Clusters of \textit{WISE} Survey (MaDCoWS).  Using typical on-source integration times of 3--4 hours per cluster, MUSTANG2 on the Green Bank Telescope was able to measure strong detections of SZE decrements  and statistically significant masses on 14 out of 16 targets. On the remaining two, weaker (3.7$\sigma$) detections of the SZE signal and strong upper limits on the masses were obtained. In this paper we present masses and pressure profiles of each target and outline the data analysis used to recover these quantities.  Of the clusters with strong detections, three show significantly flatter pressure profiles while, from the MUSTANG2 data, five others show signs of disruption at their cores. 
However, outside of the cores of the clusters, we were unable to detect significant amounts of asymmetry.  Finally, there are indications that the relationship between optical richness used by MaDCoWS and SZE-inferred mass may be significantly flatter than indicated in previous studies.
\end{abstract}

\section{Introduction} \label{sec:intro}
The study of clusters of galaxies is important to our understanding of the Universe.  For example, the mass of clusters as a function of redshift helps us understand the formation of structure and constrains cosmological parameters such as $\sigma_8$ \citep{Allen2011}.  By looking at the internal dynamics of clusters one can obtain insight to processes such as the interaction between active galaxies and the intra-cluster medium (ICM) where most of the baryons of a cluster reside. With merging clusters one is able to make inferences as to the nature of dark matter. To utilize clusters as a tool in this way, extensive catalogs of galaxy clusters are needed.  Of particular value are catalogs with well known selection functions that include massive objects at high redshifts.  Early catalogs were made by measuring the over-density of galaxies observed at optical wavelengths \citep{abell1958}. At low redshift, massive clusters are also easily identified by the strong X-ray emission from the hot gas in the ICM\@. All-sky surveys from instruments such as {\it ROSAT} have made extensive catalogs \citep[e.g.,][]{Cruddace_2002}. At higher redshifts cosmic dimming becomes significant and even the most massive clusters have low X-ray surface brightness.  Above $z\gtrsim 1$ clusters are expected to be hotter and denser and \red{ surface brightness should not decrease further} \citep[e.g.,][]{Churazov2015}, but these clusters will still have low X-ray surface brightness.
Thus, long integration times are required and X-ray surveys are limited to small areas ($<100\mbox{ deg}^2$) \citep[e.g.,][]{Fassbender2011}.  As a result current X-ray surveys do not probe the volume required to meaningfully sample the high-mass end ($M>5\times 10^{14}M_\odot$) of the high-redshift cluster mass  function.  There are alternative methods of searching for clusters such as using the Sunyaev-Zel’dovich effect (SZE) which do not suffer from cosmological dimming \citep[for a review, see][]{review2002}.
However current surveys from the SPT \citep{Williamson_2011, Bleem2015, Bleem_2020, Huang_2020} and ACT \citep{Marriage2011, Hasselfield2013, Hilton2018} still only cover a few thousand square degrees.  All-sky SZE surveys from experiments such as {\it Planck} \citep{Planck2015} are  limited to low redshift clusters due to large beam sizes.

Clusters at redshifts $z>1$  provide far greater leverage on cosmological parameters -- for example the discovery of even a few massive clusters at high enough redshifts could rule out some cosmological models \citep{Holz_2012}.  Also the current relationships between cluster mass and cluster properties are based on extrapolations from low to moderate redshifts.  In order to find the high redshift counterparts of the lower redshift clusters that dominate current cluster surveys, the Massive and Distant Cluster of \textit{WISE} Survey (MaDCoWS) project was conceived. IR galaxies are selected from the {\it WISE} all-sky IR survey \citep{Wright2010}. Using color cuts and additional data from the Sloan Digital Sky survey, likely clusters were located by looking for peaks in the number density of high redshift galaxies. A full description of the methods used and the clusters found in the 10\,000~deg$^2$ searched can be found in \cite{Gonzalez2019}. Follow-up observations have included SZE measurements using the Combined Array for mm-wave Astronomy (CARMA) \citep{Brodwin2015} and the Atacama Compact Array (ACA) \citep[also known as the Morita Array; ][]{DiMascolo2020} in order to more easily compare masses obtained from richness to the mass scales used by SZE experiments.  In this paper we present initial results of a follow-up program in the SZE using MUSTANG2 on the 100~m Green Bank Telescope (GBT), operated by the Green Bank Observatory.  \red{The cluster sample was chosen based on visibility from the GBT and so will contain the same selection biases (e.g. Malmquist bias at low richness) as MaDCoWS. The 16 clusters presented here were the first observed from this sample with a priority partly chosen due to scheduling constraints and the availability (or lack of) auxiliary data sets.} 
%\footnote{The Green Bank Observatory is a major facility supported by the National Science Foundation and operated under cooperative agreement by Associated Universities, Inc.} - I think this goes at the end....
As well as obtaining masses from the total integrated SZE signal ($Y$) these high resolution ($10''$) measurements of the SZE allow one to measure the cluster profiles, identify and remove point sources, and classify cluster dynamical states \citep[see][for a review of the application of high resolution studies of the SZE]{Mroczkowski2019}.

The paper is organized as follows. In Section~\ref{sec:obs} we outline our observations and in Sections~\ref{sec:dataReduction} and \ref{sec:sims} we overview the data reduction pipelines used and tests used to confirm our ability to recover masses and pressure profiles. Our results, including recovered pressure profiles, masses, comparisons of our SZE masses with optical richness, and notes on the symmetry of the clusters, are presented in Sections~\ref{sec:results}, \ref{sec:mass-richness}, \& \ref{sec:segments}. Our conclusions are in Section~\ref{sec:concl}.   
Throughout this paper we assume a $\Lambda$CDM concordance cosmology with: $H_0=70$~km~s$^{-1}$~Mpc$^{-1}$, $\Omega_M=0.3$, $\Omega_{\Lambda}=0.7$.% We define $h_{70} \equiv H_0$~(70 km~s$^{-1}$~Mpc$^{-1}$)$^{-1}$ and $h(z) \equiv H(z) H_0^{-1}$. 
%Statistical error bars assume Gaussian distribution when presented with $\pm$ format, or when values are expressed as ${M}^{+U}_{-L}$, $M$ is the median, and $U$ and $L$ express the difference from the median to reach the 16$^\textsc{th}$ and 84$^\textsc{th}$ percentiles. We report literature values in the latter format, even if the original work provides results in the former format.

\section{Observations}\label{sec:obs}
\begin{figure}
    \centering
    \includegraphics[width=\columnwidth]{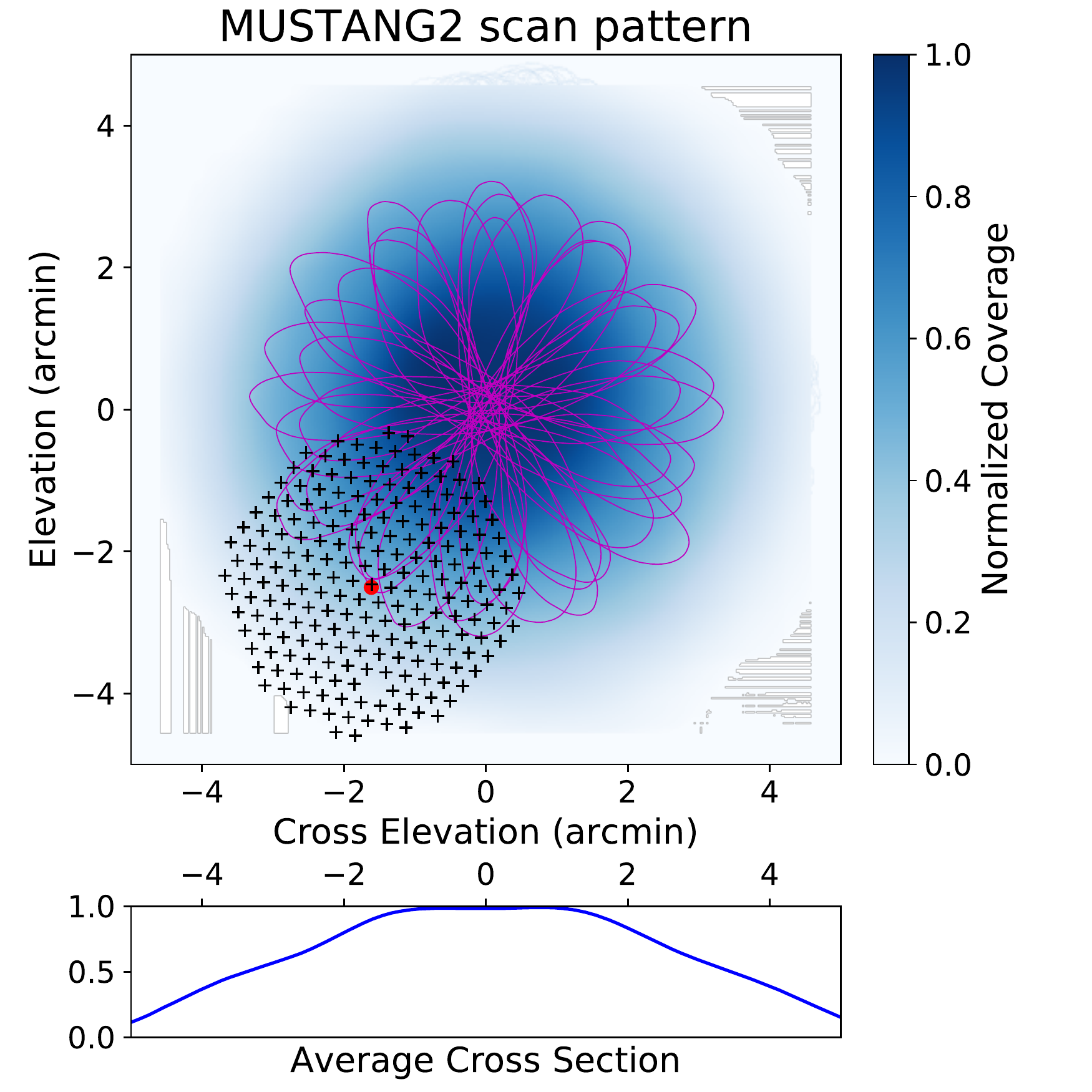}
    \caption{The scan pattern used on these clusters superimposed on the normalized depth of the coverage. The magenta lines represent the path of the central detector of the array.  Shown in the bottom left is the footprint of the array with the central detector marked as the red dot. The scan pattern is designed to provide cross-linking on many different timescales,  between all parts of the array. The scan period (10 seconds) is set as short as possible within the limits of the GBT's servo system so as to reduce the effects of $1/f$ noise from the atmosphere and receiver. In later observation, four different pointing centers at $\pm 1.5'$ were used for each cluster.  This enabled more robust removal of residual atmospheric noise (Section~\ref{sec:obs}). \red{The bottom plot shows a cross-section through the center of the normalized depth.}}
    \label{fig:scan}
\end{figure}

From October 2018 to February 2020, a total of 95 hours (69 hours excluding setup \& calibration) were spent observing clusters from the MaDCoWS sample using the MUSTANG2 bolometer camera \citep{dicker2014} on the GBT\@.  MUSTANG2 has a bandpass of 75 to 105~GHz and a resolution of $\sim 10''$. A summary of the on-source integration time, the noise level, signal to noise ratio (SNR), and detection significance for each cluster observed is given in Table~\ref{tab:obs}. Observations were carried out using a daisy scanning pattern (shown in Figure~\ref{fig:scan}) designed to cross the cluster on timescales faster than expected atmospheric noise (10 seconds), to provide many redundant observations, and to ensure all detectors get off-source.  Scan radii of $2.5'$ and $3'$ were used which, when the detector array's field-of-view (FoV) is taken into account, provided good coverage over the $\sim7'$ diameter of the maps.  Each daisy scanning pattern, referred to as a scan, takes 500 seconds to complete. During early data analysis (Section~\ref{sec:dataReduction}), it was found that, under some circumstances, incomplete removal of common mode atmospheric emission from the time ordered data of each scan (timestreams) could leave structure in the elevation direction. When combined with sky rotation, this resulted in cluster-sized noise features in the maps. To reduce these features, instead of a single pointing center at the location of the MaDCoWS cluster location, later data were collected with four centers, offset from the MaDCoWS center by $\pm 1.5'$ in RA or Dec.   

Approximately every 20 minutes a nearby bright point source was observed. This was used to check the focus in real time and later on, during data reduction, to calibrate the raw detector timestreams.  Most of these secondary calibrators were quasars with unknown, possibly variable, flux at 90~GHz so at least once a night their flux was tied to an absolute calibrator. When available, we made use of the planet Uranus while at other times we used flux calibrators commonly used by the Atacama Large Millimeter/submillimeter Array (ALMA).  These flux calibrators are regularly monitored by ALMA at 100 and 91~GHz\footnote{\url{https://almascience.eso.org/alma-data/calibrator-catalogue}} and their flux was extrapolated in both time between ALMA observations and in frequency (to account for the difference in ALMA's and MUSTANG2's bandpasses).  The flux of the secondary calibrators (in Jy) was assumed to be constant over each night. 
%calibrators used include 1224+2122 ; 1229+0203 ; 1159+2914 ; 1256-0547 ; 0854+2006 ; 0319+4130 ; 1058+0133 ; 0750+1231 ; 1751+0939

%mustang scan pattern (fig:scan)

%time on source and noise levels (tab:obs)
\begin{table*}[htb]%TBD - look into changing to deluxe table and separating out RA/DEC column
    \begin{center}
    \caption{A summary of our MUSTANG2 observations.} 
    \label{tab:obs}
    \begin{tabular}{ccclccccc}
    \hline\hline Cluster ID\tablenotemark{a} & RA & Dec\tablenotemark{b} & Redshift\tablenotemark{c} & Time On Src &
    Map Noise\tablenotemark{e} & Peak SNR & Overall & Jy to K \\
       & \multicolumn{2}{c}{(J2000)} & & (hr)\tablenotemark{d}  & ($\mu \rm K_{\mbox{\tiny{Main beam}}}$) & per 10$''$ beam & detection &(K/Jy)\tablenotemark{f}\\ \hline 
            MOO 0105 & 01:05:30.45  &$+$13:24:01.9 & ~~1.143 s  & 2.0 & 42~ &    ~-7.3      & ~8.0$\sigma$  & 1.43\\
            MOO 0135 & 01:35:04.31  &$+$32:07:27.2 & ~~1.46     & 4.9 & 29~ &    ~-4.6      & ~5.9$\sigma$  & 1.46\\
            MOO 1014 & 10:14:07.49  &$+$00:38:30.2 & ~~1.230 s  & 2.7 & 23~ &    -11.1      & 29.3$\sigma$  & 1.49\\
            MOO 1031 & 10:31:48.23  &$+$62:55:30.5 & ~~1.33     & 3.4 & 45~ &    ~-3.8      & ~3.7$\sigma$  & 1.50\\
            MOO 1046 & 10:46:52.82  &$+$27:58:02.9 & ~~1.16     & 2.3 & 36$^*$ & ~-9.5      & ~9.5$\sigma$  & 1.20\\
            MOO 1052 & 10:52:15.30  &$+$08:23:53.0 & ~~1.41     & 3.6 & 23$^*$ & ~-6.1      & ~7.2$\sigma$  & 1.48\\
            MOO 1054 & 10:54:56.00  &$+$05:05:39.0 & ~~1.44      & 5.4 & 19$^*$ & ~-3.3     & ~5.3$\sigma$  & 1.47\\
            MOO 1059 & 10:59:50.83  &$+$54:54:58.4 & ~~1.14      & 7.1 & 11~ &    ~-7.5     & 53.1$\sigma$  & 1.33\\
            MOO 1108 & 11:08:48.00  &$+$32:43:35.8 & ~~1.12     & 6.7 & 14~ &    ~-5.5      & 19.2$\sigma$  & 1.37\\
            MOO 1110 & 11:10:57.15  &$+$68:38:30.7 & ~~0.93     & 7.3 & 12~ &    ~-8.0      & 15.5$\sigma$  & 1.44\\
            MOO 1142 & 11:42:45.51  &$+$15:27:15.4 & ~~1.189 s  & 5.2 & 13$^*$ & -14.5      & 20.9$\sigma$  & 1.40\\
            MOO 1203 & 12:03:07.00  &$-$09:09:13.0 & ~~1.24     & 4.6 & 23$^*$ & ~-4.5      & ~3.7$\sigma$  & 1.43\\
            MOO 1322 & 13:22:56.30  &$-$02:28:15.0 & ~~0.82     & 2.7 & 28$^*$ & ~-8.1      & ~9.5$\sigma$  & 1.39\\
            MOO 1329 & 13:29:48.00  &$+$56:47:39.0 & ~~1.43     & 1.5 & 46$^*$ & ~-7.8      & 19.2$\sigma$  & 1.59\\
            MOO 1354 & 13:54:51.70  &$+$13:29:36.0 & ~~1.48     & 3.8 & 13$^*$ & ~-9.2      & ~9.4$\sigma$  & 1.58\\
            MOO 1506 & 15:06:20.35  &$+$51:36:53.6 & ~~1.09     & 5.7 & 36$^*$ & ~-6.5      & 11.3$\sigma$  & 1.38 \\ \hline
    \end{tabular}\end{center}
    \tablenotetext{a}{Full MaDCoWS cluster names are given in table~\ref{tab:mass_richness}}
    \tablenotetext{b}{Coordinates are from the MaDCoWS survey \citep{Gonzalez2019} and represent the center of the galaxy over density.}
    \tablenotetext{c}{Redshifts from \citet{Gonzalez2019}. Redshifts are photometric, except when marked with an ``s,'' in which case they are spectroscopic.}
    \tablenotetext{d}{Data were taken over more than 20 separate nights. As MUSTANG2's sensitivity is limited by the atmosphere, some nights have far greater weight than others.  Totals only include data used in each map.}
    \tablenotetext{e}{Map noise refers to the standard deviation at the center of the maps while the overall significance of the detections comes from the fitting of pressure profiles to the calibrated  timestreams in Minkasi.  A $^*$ indicates  a 4 pointing centers observational strategy was used.}
    \tablenotetext{f}{\red{At 90~GHz, thermal effects cause the beam of the GBT to degrade over an observing session and the quality of the focus achievable each night varies at a 25\% level. These values are weighted averages of the scans used for each source.}}
\end{table*}

\section{Data Reduction}\label{sec:dataReduction}
This paper makes use of two different data reduction pipelines, MIDAS and Minkasi.\@  MIDAS is based on MUSTANG-1's IDL pipeline \citep{mason2010,korngut2011,young2015,romero2015}.   Minkasi is a maximum likelihood pipeline based on the one used by the ACT collaboration \citep{Dunner2012}.  Although both pipelines can produce maps, fit point sources, and find cluster surface brightness profiles, in this paper we only present maps made by MIDAS and profiles and point source fits using Minkasi.\@  Calibration of raw MUSTANG2 data for both pipelines is carried out using the following steps in MIDAS:
\begin{enumerate}
    \item An initial flat-fielding of the array is made using a skydip taken at the beginning of each night's observations.  Unresponsive detectors are flagged and their data discarded.  %Detector sky offsets are obtained using the tuning files from the microwave multiplexer (umux) readout \citep{dober_2017}.
    \item At this stage most of the signal in the timestreams is atmospheric emission and should be the same in all detectors.  To account for relative gain drifts between detectors, the timestreams are renormalized such that the first principle component has the same amplitude in each detector.
    \item Maps of all calibrator sources are made and fits to the peak height and beam volume carried out.  A calibration factor to Jy is obtained by taking the ratio of the expected peak to the measured one. Additionally, a calibration to (main beam) brightness temperature is calculated by using the fitted beam volume. The atmospheric opacity is obtained using archival weather data via the GBT's observing tools - this is then used to adjust for any differences in elevation between calibrators and our clusters.
    \item The calibration factors from step 3 are extrapolated between observations of the secondary calibrators and are applied to the cluster scans to produce calibrated timestreams. 
    \item A number of heuristics are used to detect glitches (such as jumps caused by readout errors) in individual detector timestreams.  For small glitches, only part of a timestream is masked out while timestreams showing many glitches or excess noise are dropped completely.
\end{enumerate}
The resulting timestreams from this process are passed on to the map making stage of MIDAS (described next) or saved to disk for later use by Minkasi.\@  Overall, we estimate there is a 10\% error in our absolute calibration of the observations presented in this paper.  Sources of this error include the fits to the primary and secondary calibrators, the assumed flux of the absolute calibrator in our band, and our knowledge of the atmospheric opacity.

\subsection{MIDAS map making}
In addition to calibrating the timestreams, MIDAS was used to make the maps presented in this paper. In the calibrated detector timestreams the cluster's signal can easily be a factor of $10^5$ below atmospheric emission and $1/f$ noise from the detectors.  However, our scanning pattern means that point sources pass through the beam at $\sim$10~Hz, while the entire map is crossed once every 10 seconds, so frequencies $f \gg 10\mbox{~Hz}$\ and $f \ll 0.1\mbox{~Hz}$ contain very little astronomical signal and can be filtered out.  This still leaves a significant noise due to the change in optical depth as the GBT scans in elevation. However, this noise, along with contributions to the noise from the readout electronics, are all highly common mode.  Using a principle component analysis it is possible to subtract these contributions before binning the data into a map. In the 16 cluster maps presented in this paper, filter bandpasses of 0.065--41~Hz were used and the first 3 principle components were subtracted before map making.

Because of the Fourier filter, MIDAS maps are not unbiased.  Features on the largest angular scales can have Fourier components below 0.065~Hz and will be slightly attenuated. Features of the order of the map size or greater are mostly DC and are thus not detected at all. To quantify this we create fake MUSTANG2 data using real timestreams reversed in time to smear out any astronomical signals. A fake sky containing random structure on all angular scales is sampled then added to the reversed timestreams and the results passed through the MIDAS pipeline to obtain a recovered sky map.  A transfer function can be defined by the ratio between the FFTs of the fake and recovered sky maps averaged over many versions of the noise and sky.  Tests with the data reduction parameters used in this paper show that we recover all angular scales up to $5'$ (diameter) although at angular scales larger than $R_{500}$ ($\sim3'$ diameter for typical MaDCoWS clusters) small corrections are needed.  To circumvent this problem, in this paper we choose to present cluster profiles made using the Minkasi pipeline discussed  next. 

\subsection{Minkasi and brightness profiles}\label{sec:profiles}
As described in \citet{Romero2020}, surface brightness profiles can be calculated by Minkasi.  Minkasi operates on the calibrated but unfiltered timestreams output by MIDAS.  It \red{simultaneously fits multiple} parameters directly to timestreams -- for example the location, width and amplitudes of point sources, without going through map space.  A full covariance matrix is recovered and the results are unbiased by Fourier filtering -- to the extent that the signal is present in the raw data, no transfer function needs to be considered.  

Fitting the brightness profiles to clusters is a two step process.  First, the centers of the cluster and any point sources detected at greater than $4\sigma$ in the MIDAS maps are found using a weighted least squares fit to all data on each cluster.  Symmetrical Gaussian shapes are assumed.  Weights for each detector in each scan are derived using a singular value decomposition (SVD) technique. The timestreams are rotated into SVD space, a power spectrum taken, and the results smoothed in order to obtain a better estimate of the true {\it underlying} power spectrum. 
Alternatives to smoothing would be the averaging of power spectra from different scans but the noise in MUSTANG2 data can vary significantly and this was found to give incorrect weights. Once smoothed power spectra in SVD space are obtained, they are rotated back into timestream space and used in a weighted least squares fit for the cluster's (and any point sources') amplitude, width, and location.   This is done iteratively so that the astronomical signal does not bias the noise estimate. In each iteration, the results from the previous iteration are subtracted from the timestreams before recalculation of the noise and then added back in.  Tests showed convergence in as few as five iterations; however, we used a conservative 15 iterations.  

With the centers fixed, this process is repeated with parameters of the amplitude of each point source (if any) and the surface brightness in fixed annuli around the cluster center found in the last step.  Our initial results (Section~\ref{sec:results}) assumed circular symmetry  but later azimuthally segmented annuli were used (Section~\ref{sec:segments}).

\subsection{Pressure profiles}
Of more intrinsic interest than surface brightness profiles in understanding the physics of clusters are cluster masses and shapes of the pressure profiles.  Using the method described in greater detail in \citet{Romero2020}, it is possible to deproject the brightness profiles and obtain pressure profiles.  A fit \red{to the brightness profile} assuming a non-parametric model of the cluster with six shells spaced logarithmically in radius between 10$''$  and 200$''$ is carried out.  If either of the outer bins had a fitted pressure less than $2\sigma$ from zero, then the fit was repeated excluding these bins.   Within each bin the pressure is assumed to follow a power law with the slope in the last bin constrained to go to zero at infinity.  Assuming the redshifts from \citet{Gonzalez2019} (reproduced in Table \ref{tab:obs}), adopting a temperature of 4-5~keV  \citep[appropriate for clusters in this mass range e.g.][]{Bulbul2019}, the relativistic corrections to the thermal SZE given in \citet{itoh1998}, and using the known MUSTANG2 beamshape (found in the initial calibration), these bins can be integrated analytically to obtain a brightness profile. Iterations using the Markov Chain Monte Carlo (MCMC) package \texttt{emcee} \citep{ForemanMackey2013} make use of the covariance matrix from Minkasi to find the most likely pressure profile.
%In parallel with this process, %a generalized NFW profile\cite{Nagia2007} is fit with the  
%an \citet{arnaud2010} (hereafter A10) profile is fit and from this it is possible to recover self-consistent values for $\mbox{R}_{500}$ and $\mbox{M}_{500}$.  
%In doing so we assume the relationship between integrated brightness in the SZE out to R$_{500}$ (Y$_{500}$) and mass in \citet{arnaud2010}. 
\red{From  the  non-parametric  profile,  a spherically integrated $Y(r)$ is calculated.  Using the approach in \citet{Romero2020}  a $Y_{500}$ is self-consistently calculated assuming a Y-M relation  derived from the universal pressure profile  \citep[equation 22 in ][hereafter A10]{arnaud2010}. Other Y-M relationships could have been used but, as \citet{Romero2020} shows, in our mass range differences between scaling relations are less than 20\%. All these relationships have potential biases. In this work there is no strong astrophysical reason to prefer one Y-M relation over another so we chose the relation from A10 as it is well established, making comparisons with past work \citep[e.g.,][]{ 2019MNRAS.484.1946G,2020A&A...639A..73C} easier.   The A10 pressure profiles shown in Figure~\ref{fig:results} is thus the A10 pressure profile for the $M_{500}$ we derived from the Y-M relation.
As stated in \citet{Romero2020}, this process has been shown to be robust against initial assumptions on a cluster's shape, mass, and electron temperature. }

One concern when using SZE measurements to measure masses is that point sources cancel out the SZE decrement, biasing masses low. Interferometers such as the ACA, CARMA, or ALMA can constrain the flux contributions from such sources using their long baselines.  Likewise, MUSTANG2's high resolution allows the easy removal of any source significantly above the noise floor in the maps (20--50~$\mu$K see Table~\ref{tab:pntSrc}).  No attempt was made to fit sources detected at $4\sigma$ or less but these will have fluxes below 200~$\mu$K.\@  In the central few bins of a cluster's pressure profile  such sources can slightly lower the pressure (as sources will always be positive and the SZE signal is negative) but when averaged over the whole cluster (several arcminutes squared) their effect on our mass determinations was expected to be negligible.  To confirm this assumption, fake sources that were not modeled when fitting the rings were added to \red{our real timestreams and the data reanalyzed using identical steps as those used to recover the masses in Table~\ref{tab:mass_richness}}. Source amplitudes up to 400~$\mu$K ($>8\sigma$ depending on the cluster) were used and these changed recovered masses by less than 3\%. The largest errors on our recovered fluxes are 260~$\mu$K so errors in the recovered cluster masses due to poor fits to point sources should also be negligible.  

%main results figure (fig:results)
\begin{figure*}[h!]
\begin{center}
\begin{tabularx}{\textwidth}{ccc}
\includegraphics[width=0.32\textwidth]{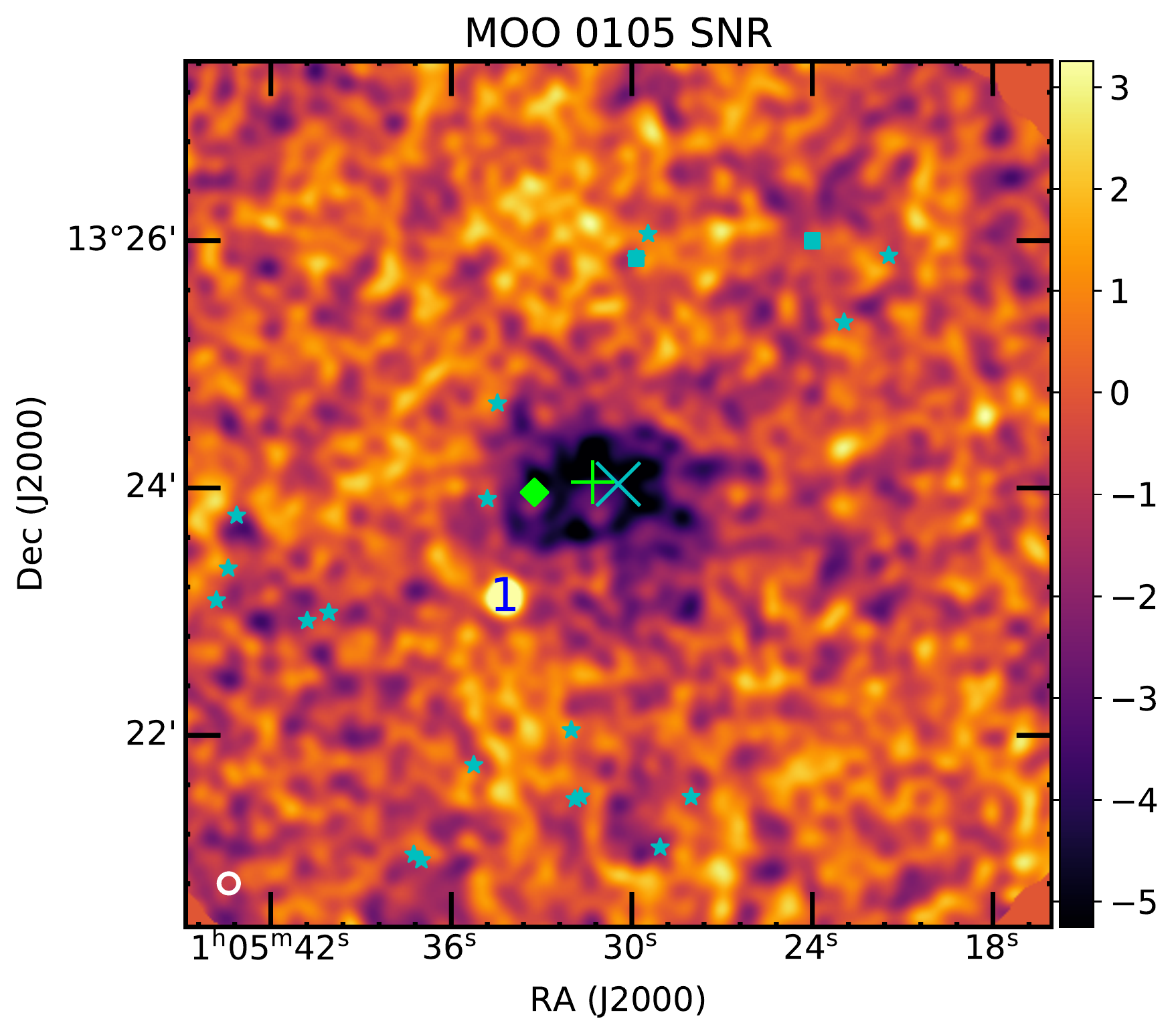} & 
\includegraphics[width=0.3\textwidth]{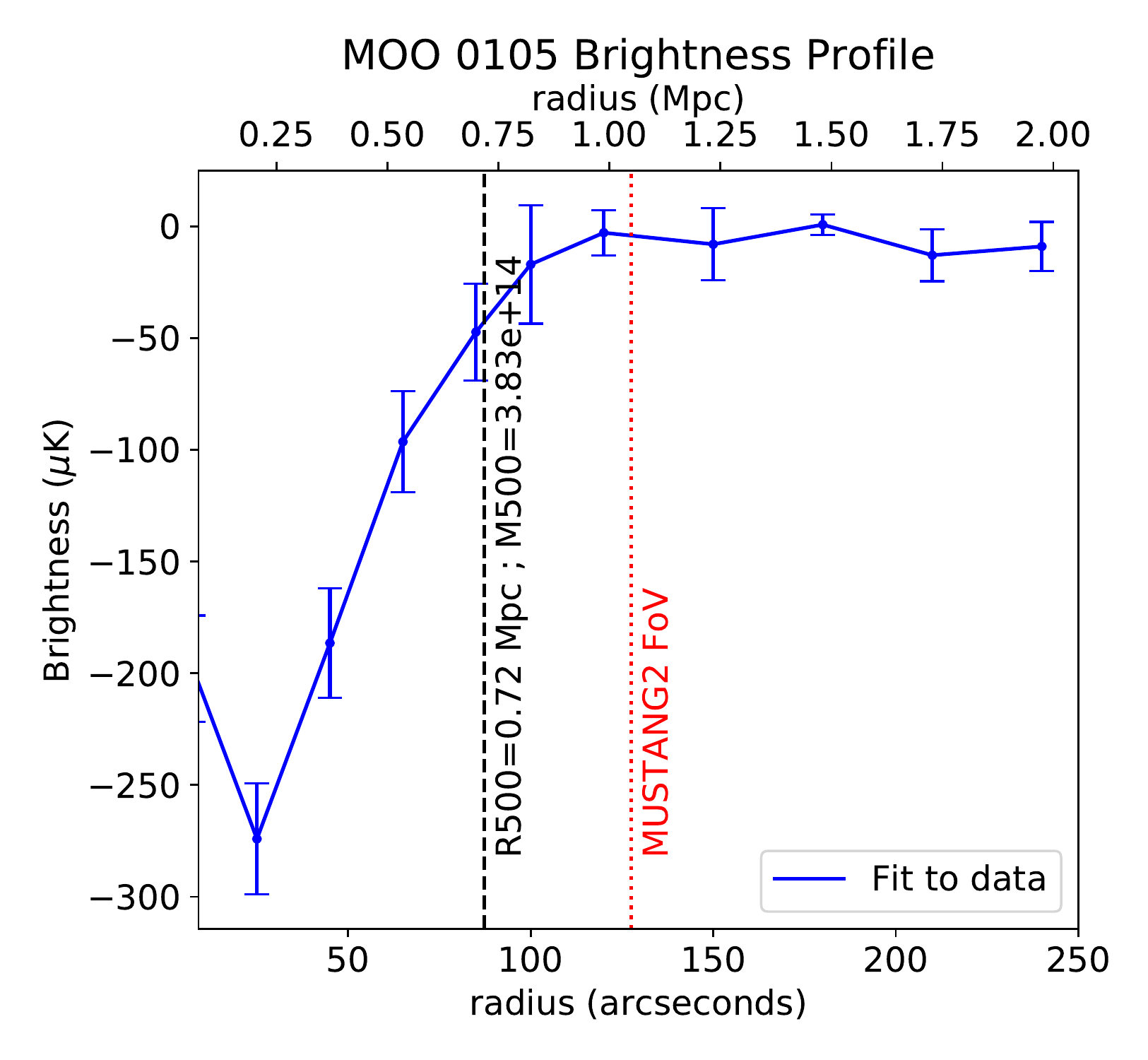} &
\includegraphics[width=0.3\textwidth]{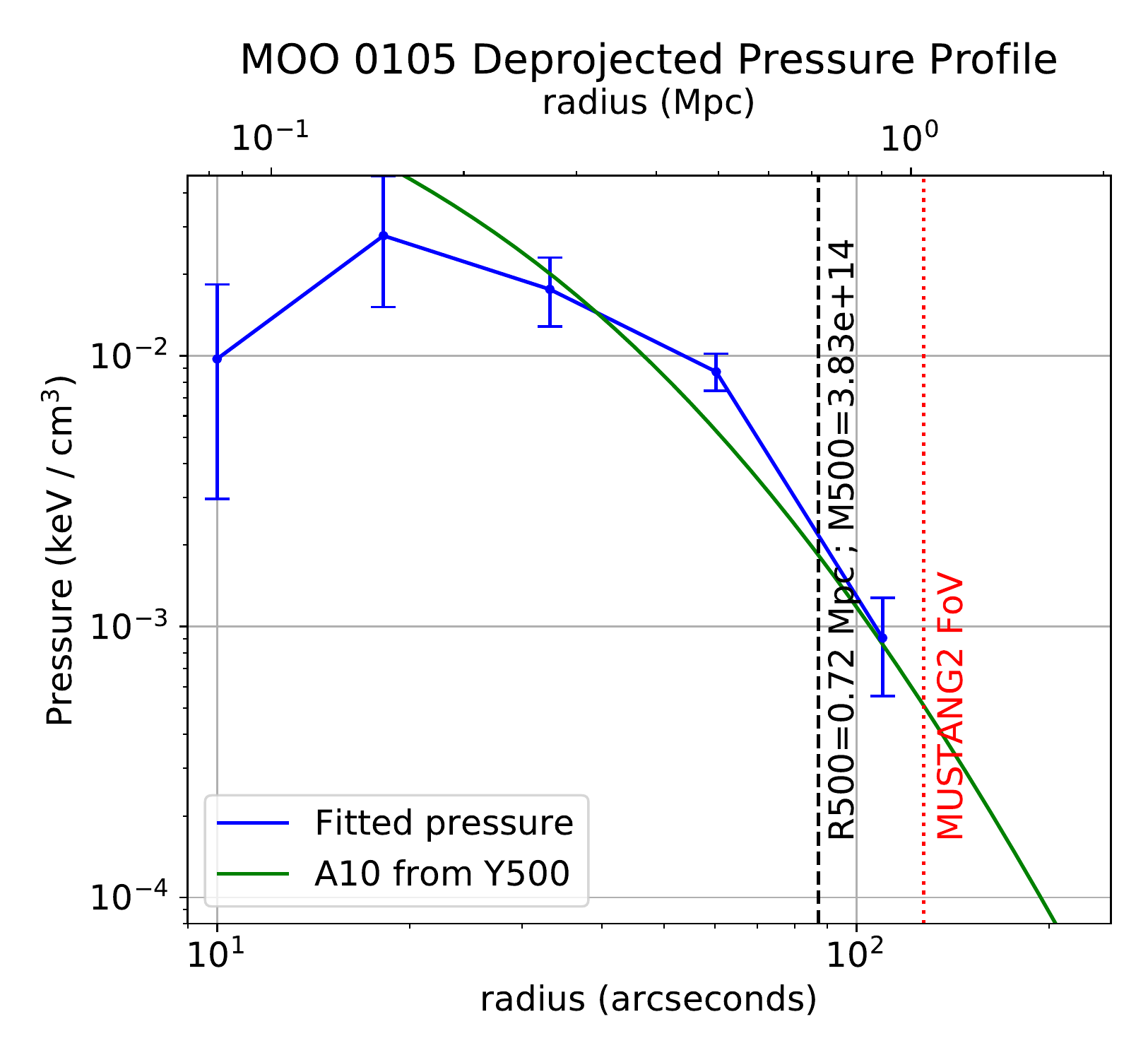}
\end{tabularx}
%{{\bf MOO 0105}}
\begin{tabularx}{\textwidth}{ccc}
\includegraphics[width=0.32\textwidth]{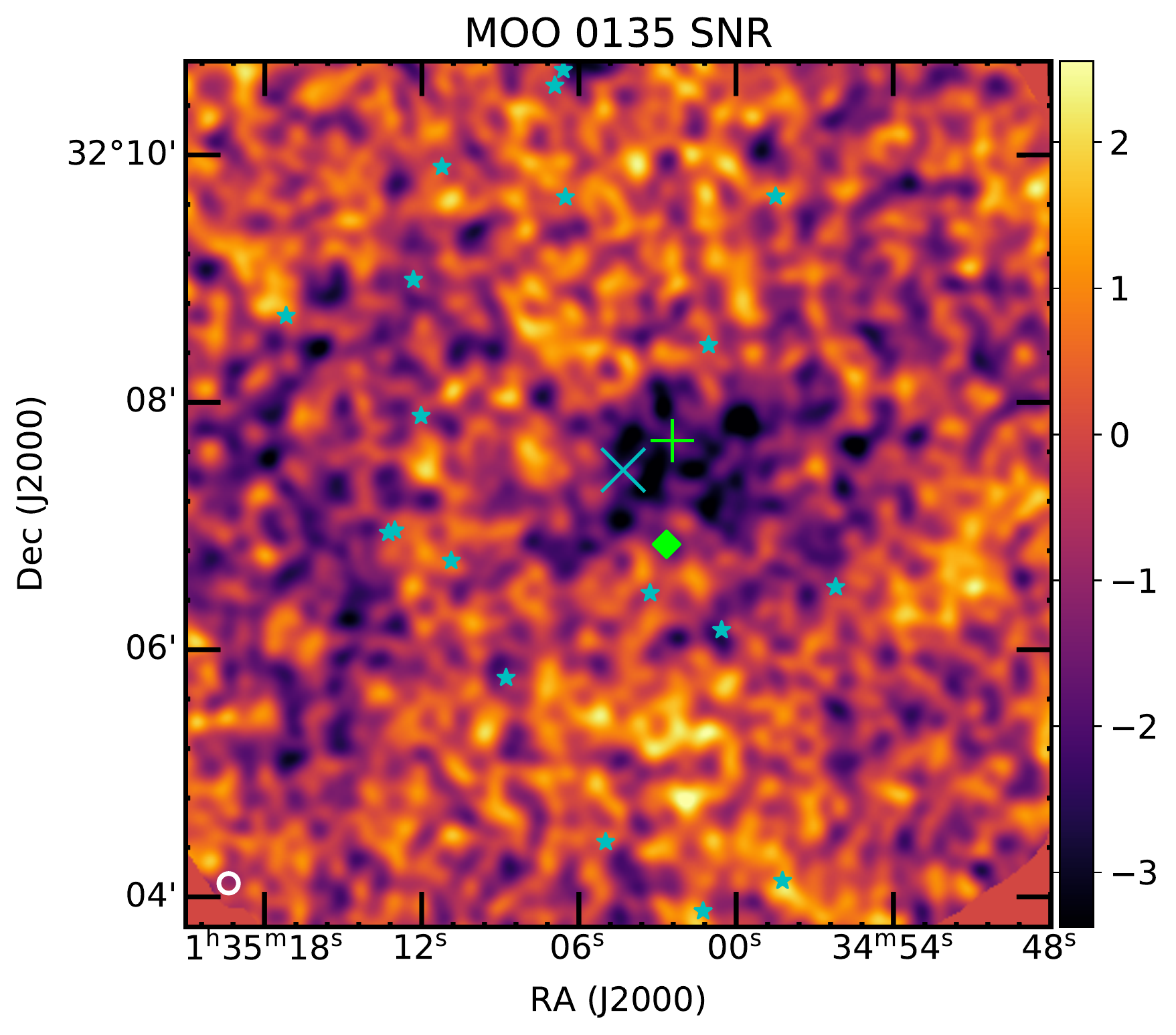} & 
\includegraphics[width=0.3\textwidth]{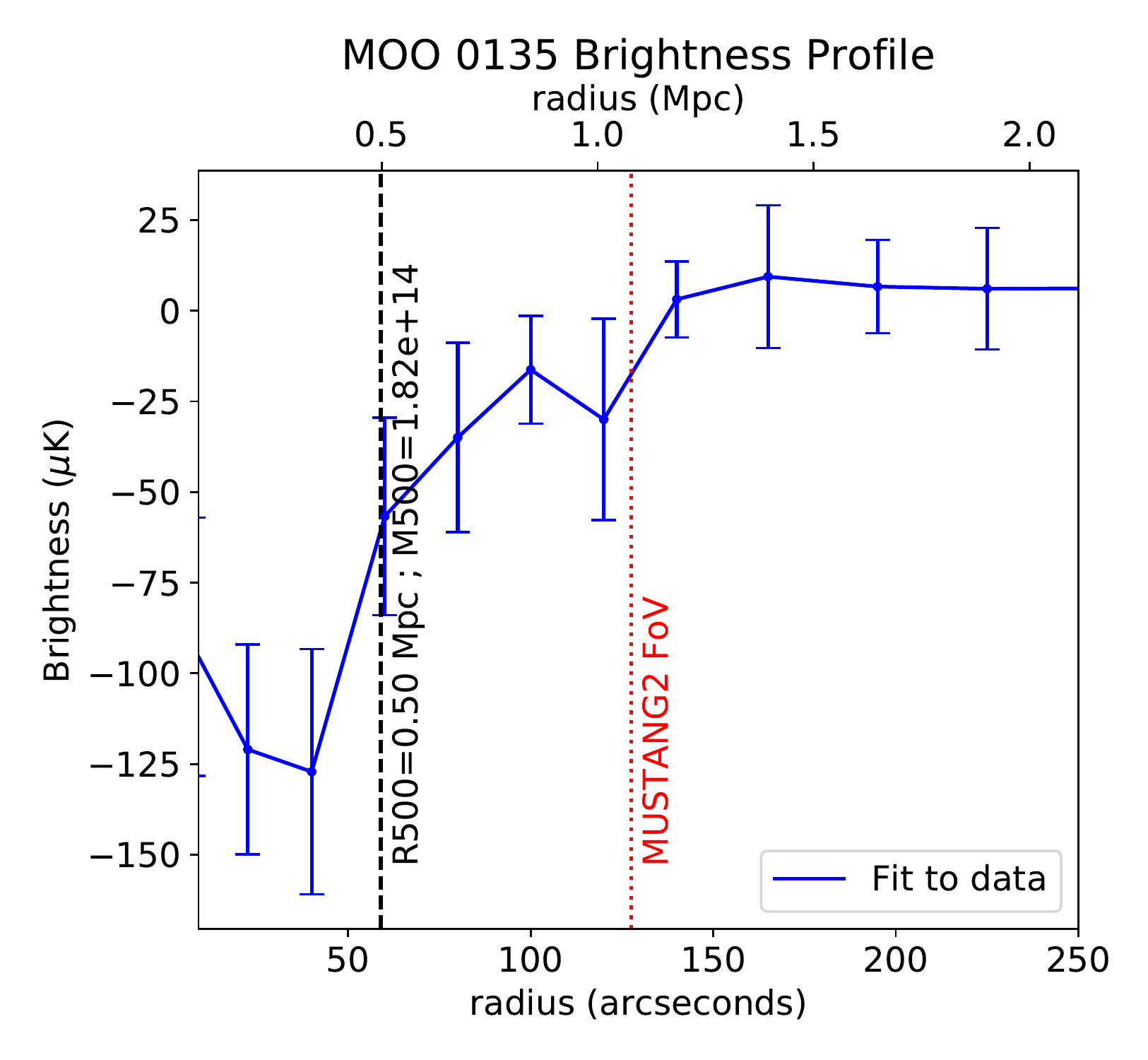} &
\includegraphics[width=0.3\textwidth]{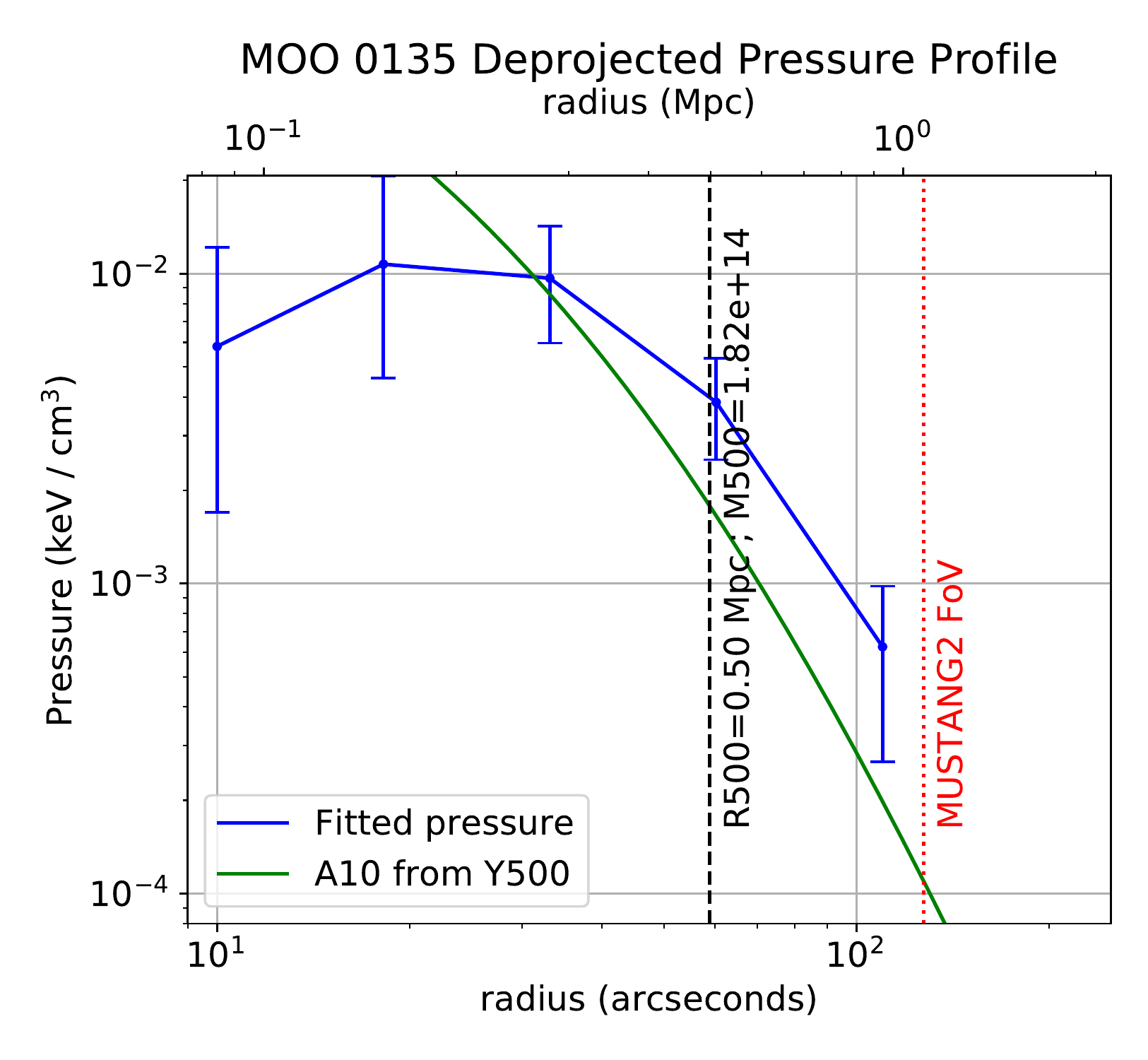}
\end{tabularx}
%{{\bf MOO 0135}}
\begin{tabularx}{\textwidth}{ccc}
\includegraphics[width=0.32\textwidth]{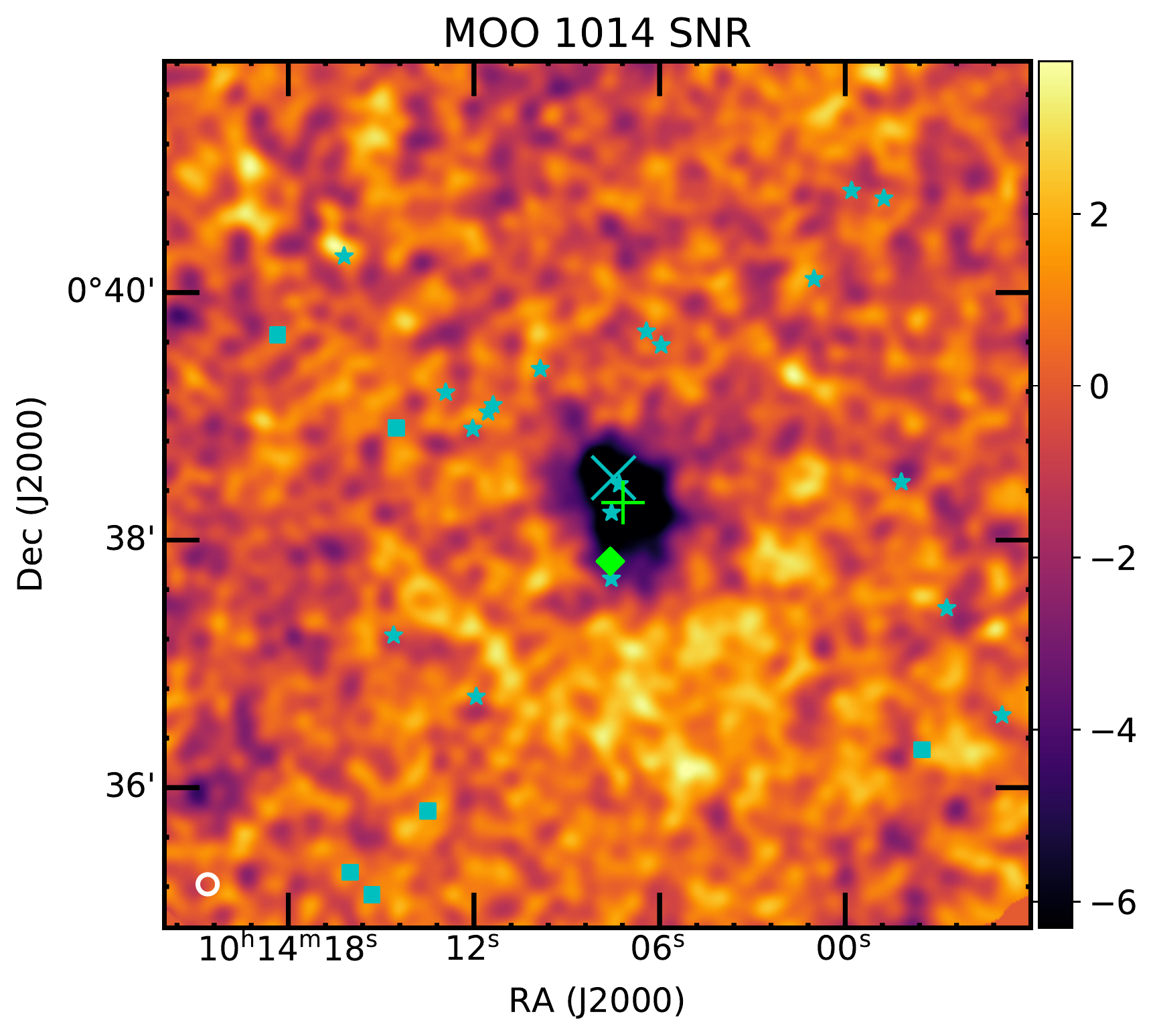} & 
\includegraphics[width=0.3\textwidth]{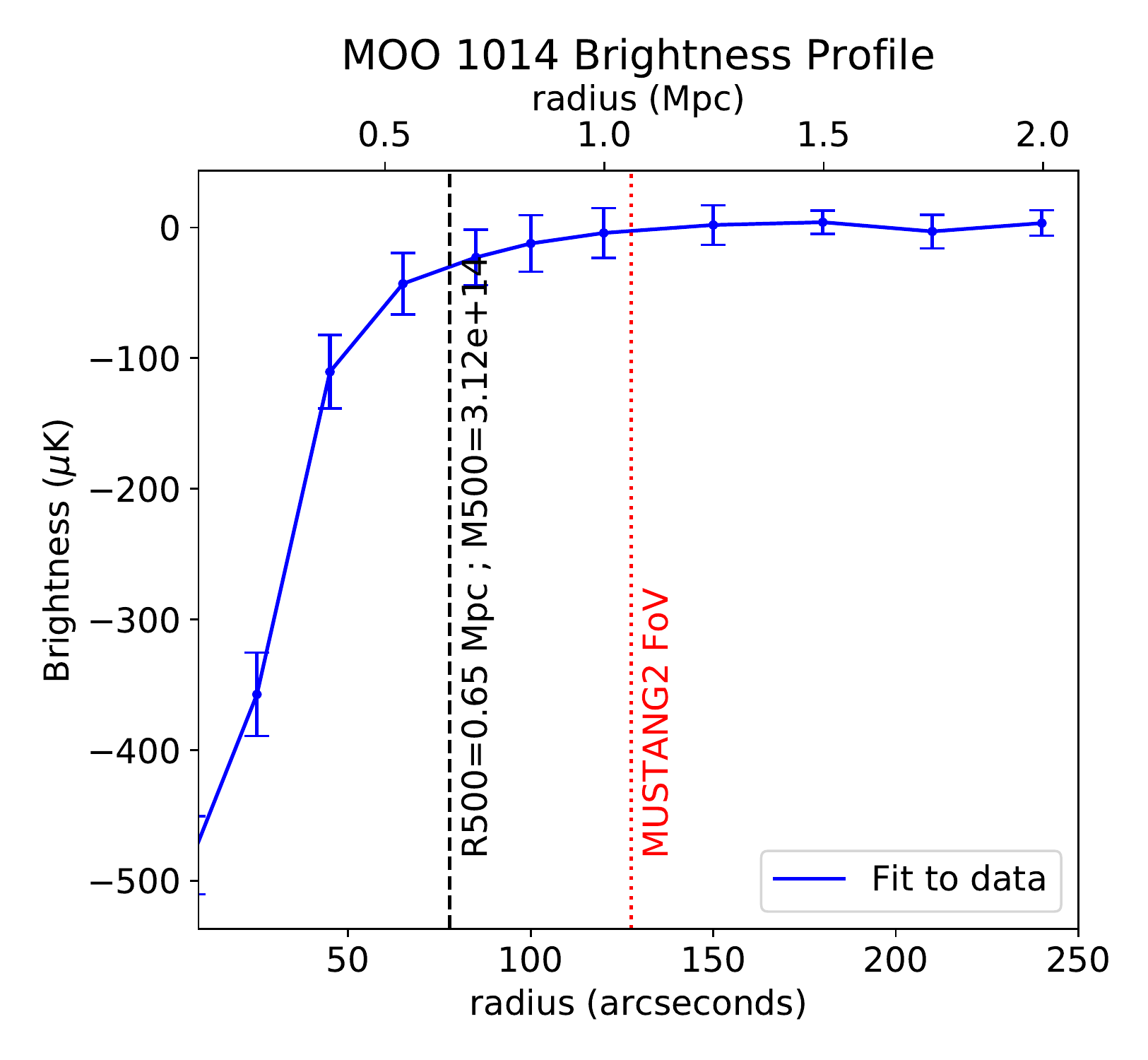} &
\includegraphics[width=0.3\textwidth]{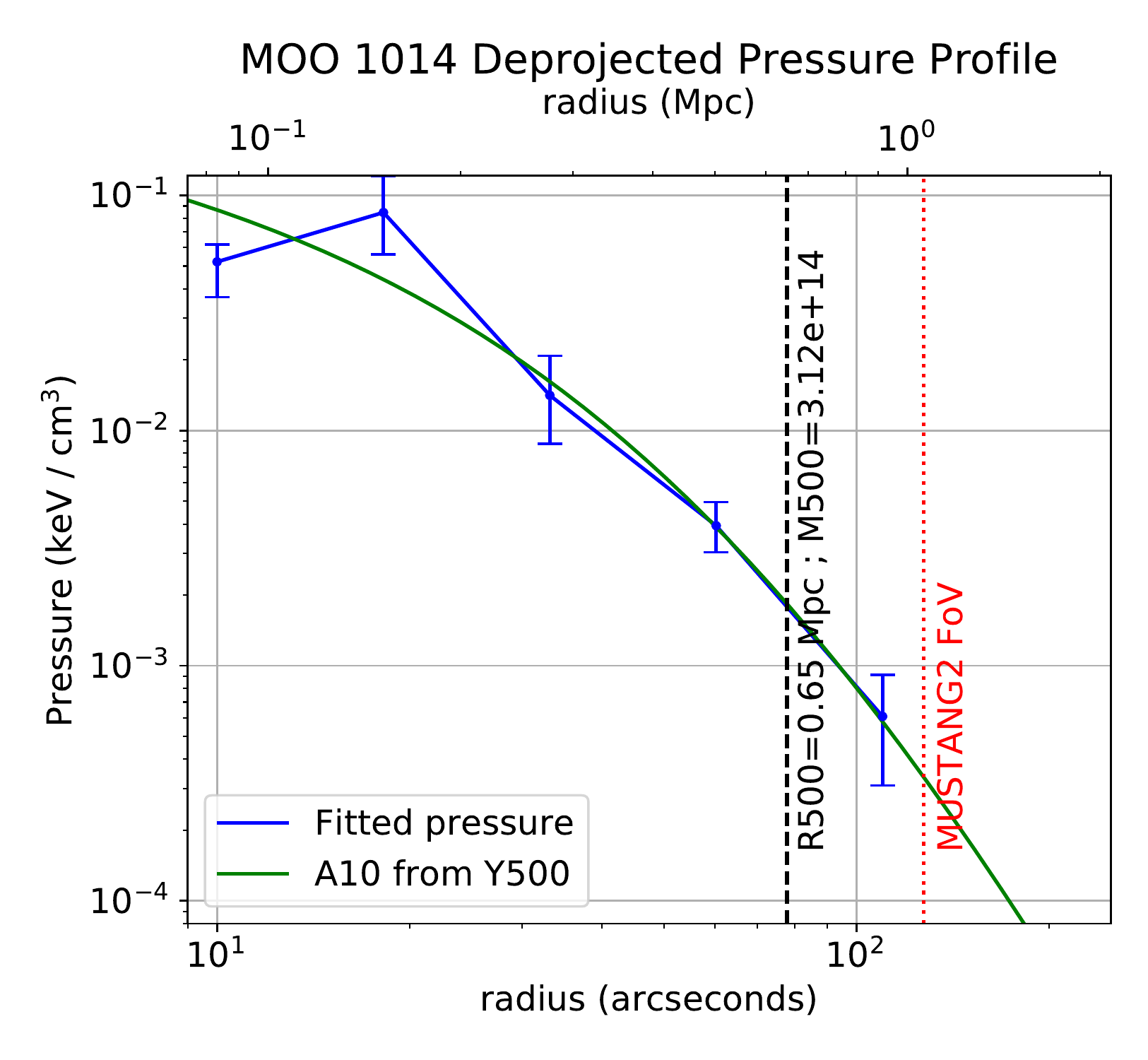}
\end{tabularx}
%{{\bf MOO 0105}}
\begin{tabularx}{\textwidth}{ccc}
\includegraphics[width=0.32\textwidth]{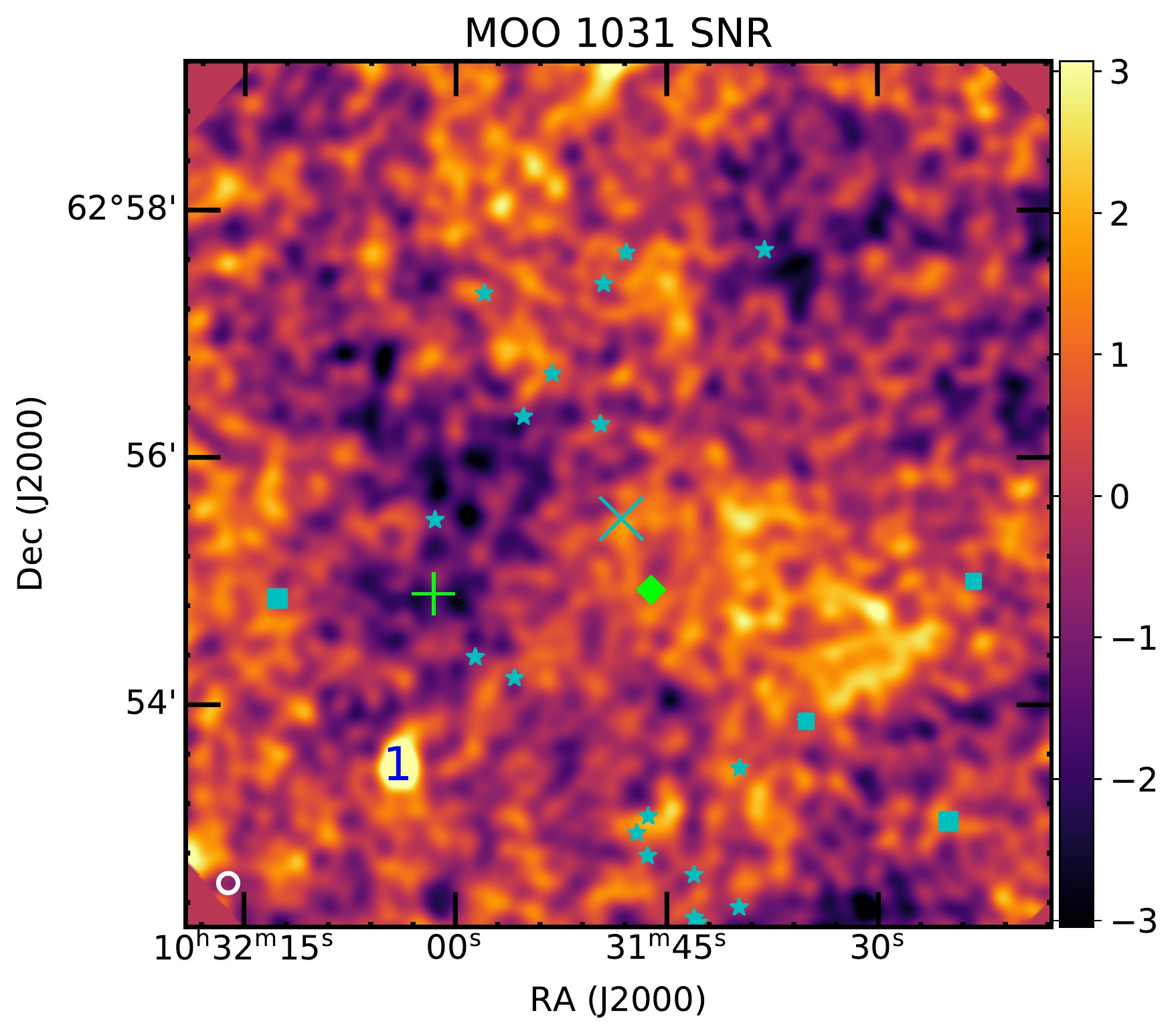} & 
\includegraphics[width=0.3\textwidth]{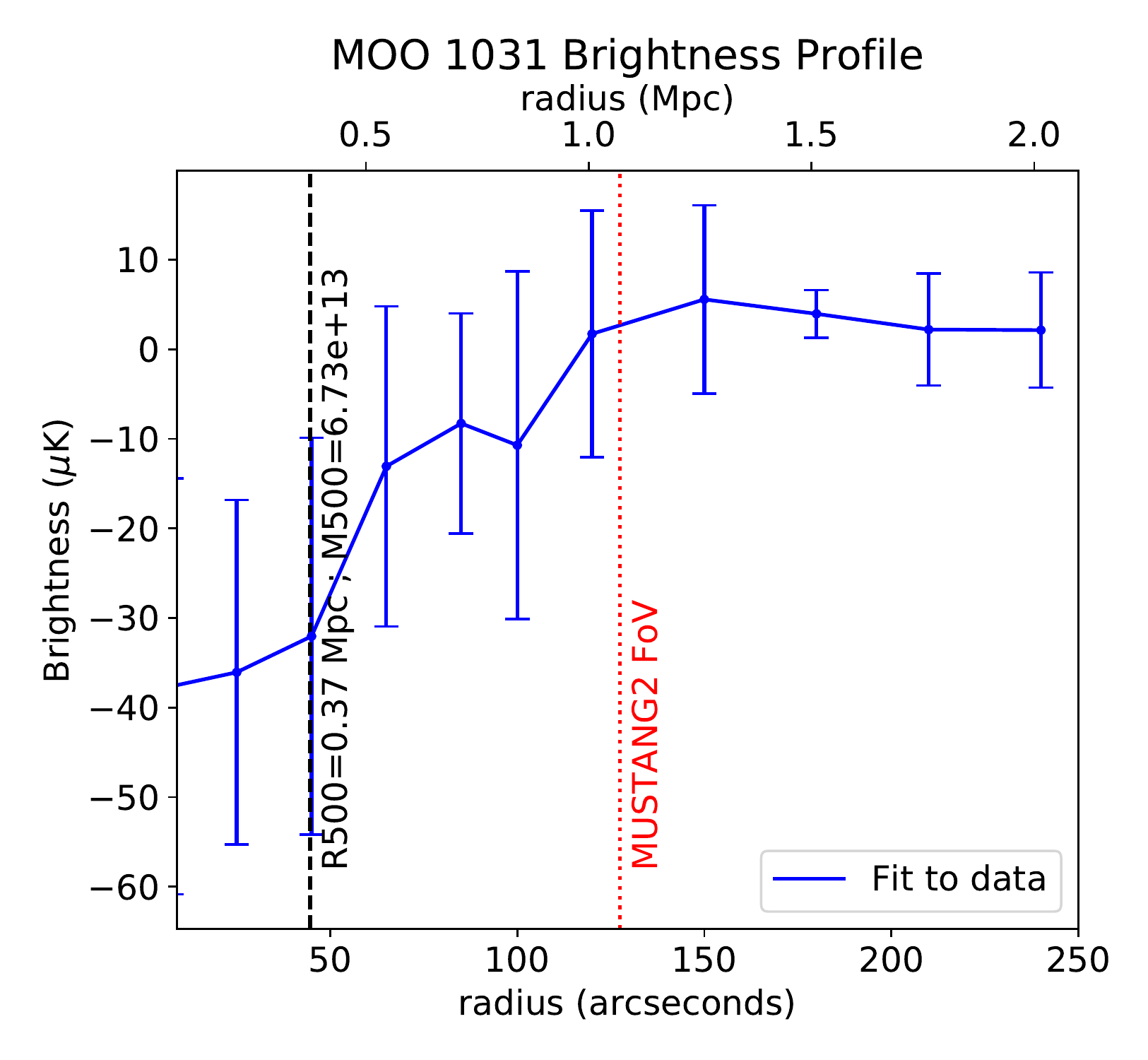} &
\includegraphics[width=0.3\textwidth]{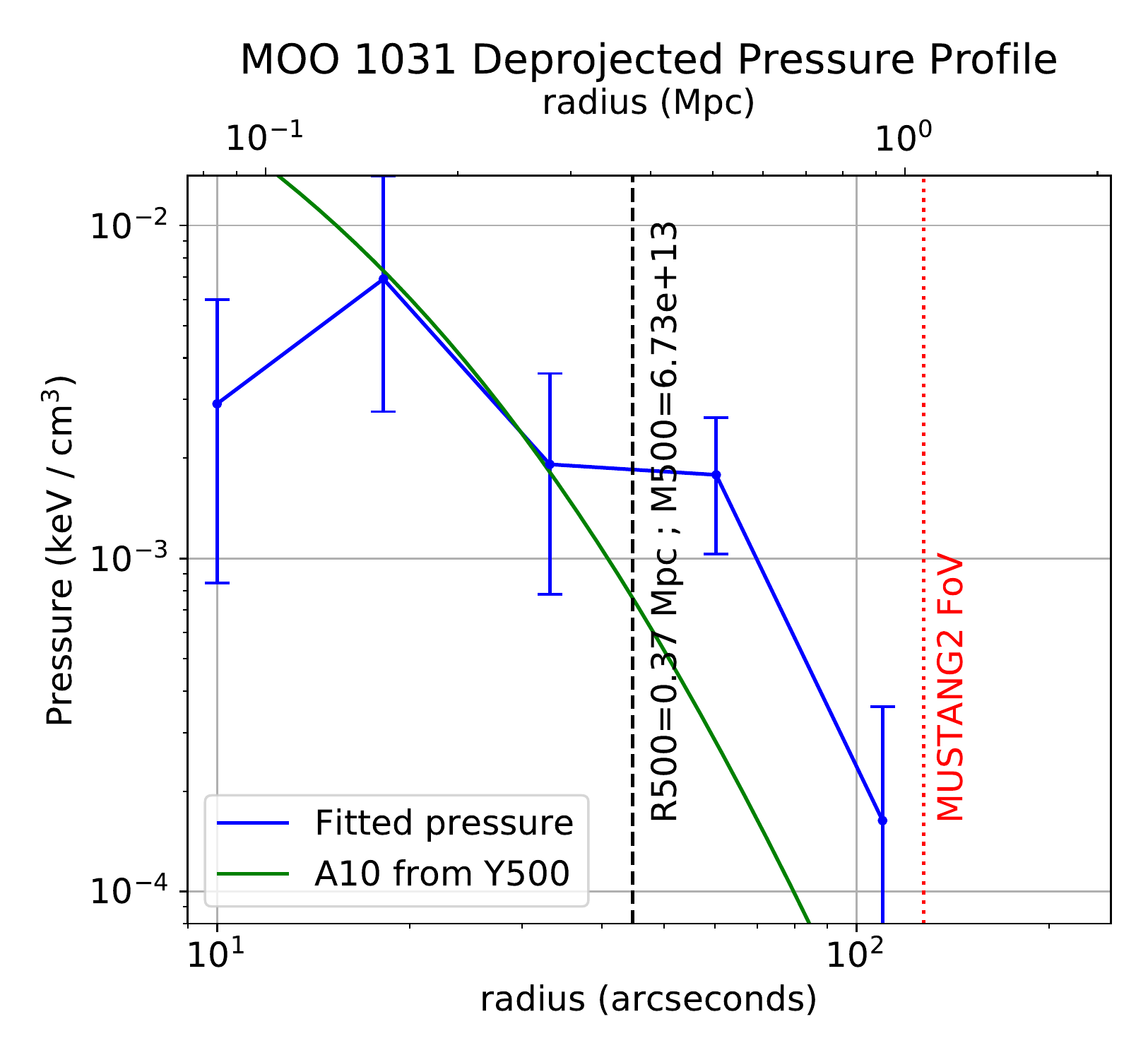}
\end{tabularx}
%{{\bf MOO 0105}}
\caption{\label{fig:results}  {\bf Left:} MUSTANG-2 images made using the MIDAS pipeline.  A cyan X marks the original center found by MaDCoWS, while the green cross marks the best fit SZE centroids.  Cyan stars and squares mark the locations of bright galaxies detected by \textit{Spitzer} and SDSS, respectively.  The BCG is marked as a green diamond and the MUSTANG2 beam is shown as a white circle on the lower left. The bright sources are clipped at $+4\sigma$ and are labeled with numbers to match Table~\ref{tab:pntSrc}. {\bf Center:} Brightness profiles of our clusters from Minkasi.\@ {\bf Right:} Pressure profiles derived from each data set.  The MUSTANG2 FoV is marked as a red line, while the black dashed line represents the $R_{500}$ for our recovered mass. \red{The A10 profile that corresponds to this mass is shown in green}.}
\end{center}
\end{figure*}

\begin{figure*}
\addtocounter{figure}{-1} %\ContinuedFloat
\begin{center}
\begin{tabularx}{\textwidth}{ccc}
\includegraphics[width=0.3\textwidth]{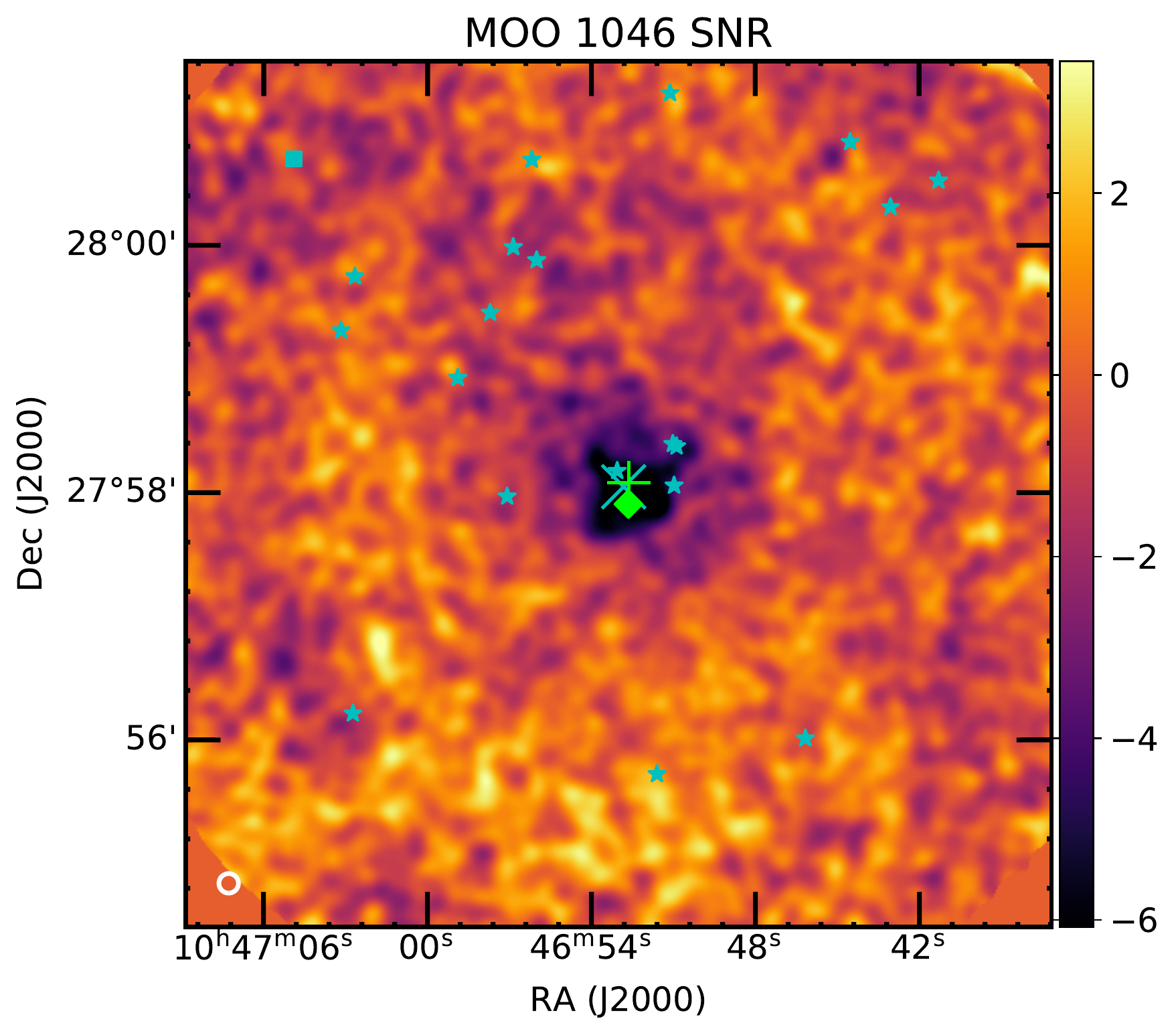} & 
\includegraphics[width=0.3\textwidth]{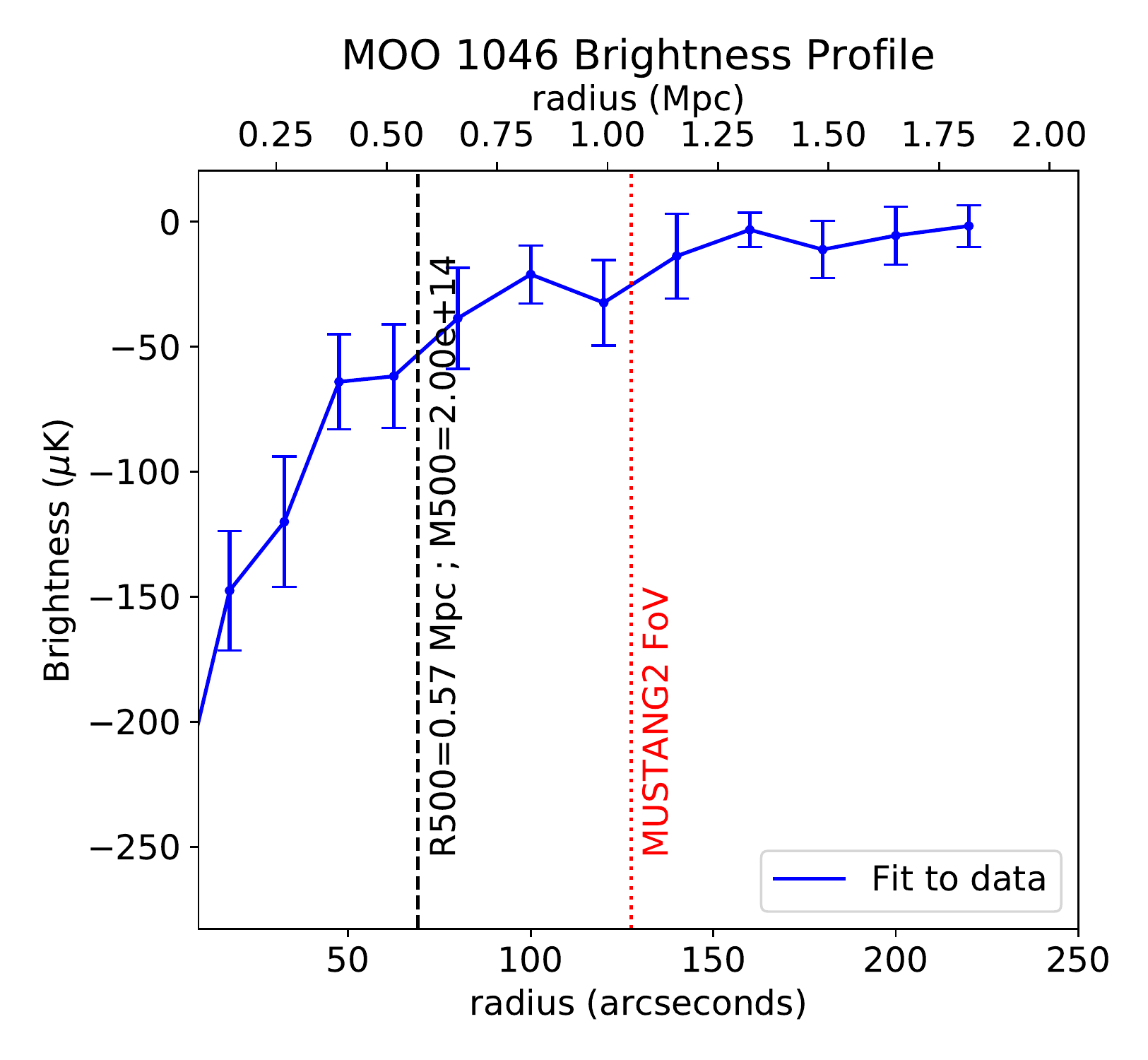} &
\includegraphics[width=0.3\textwidth]{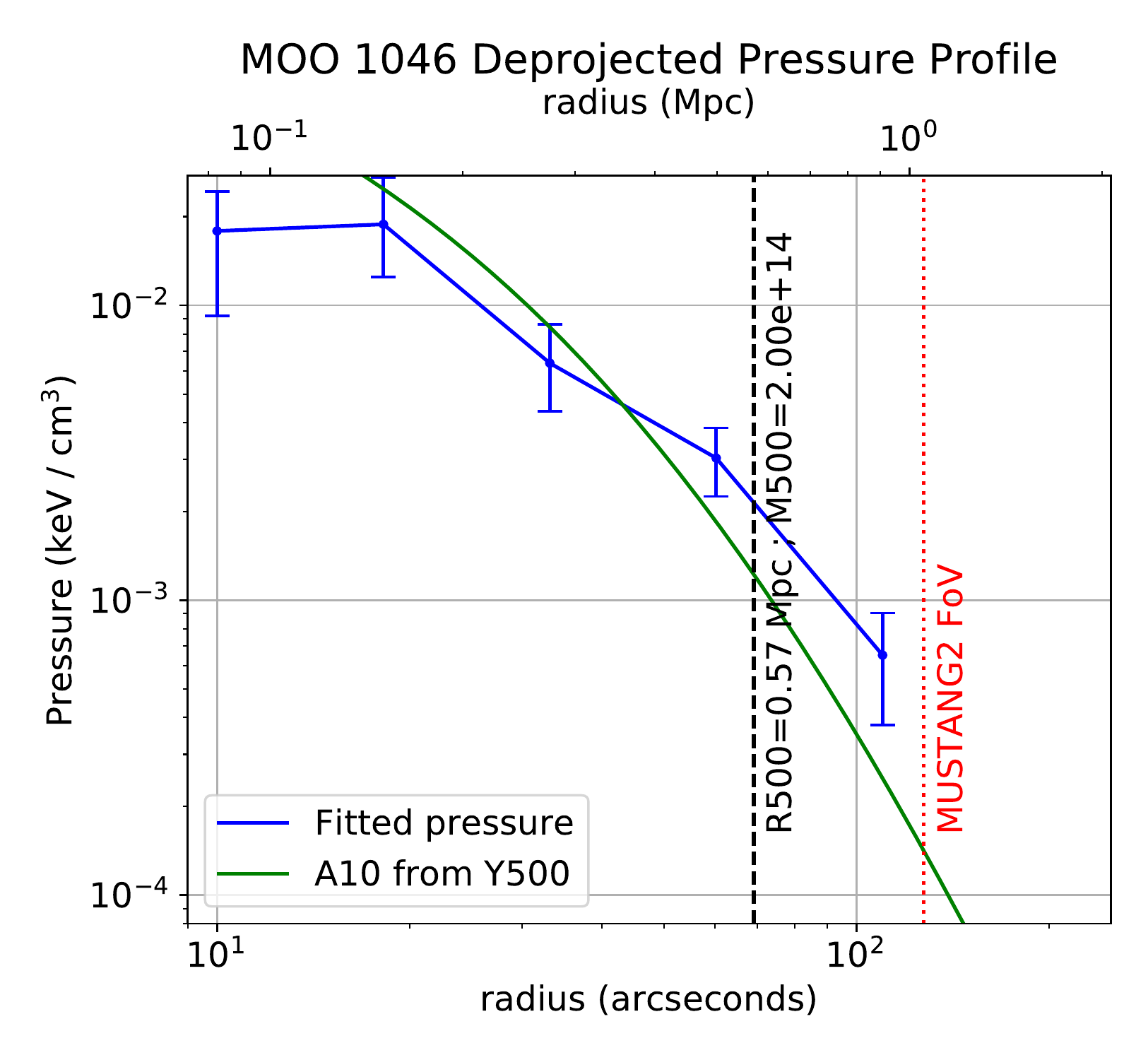}
\end{tabularx}
\begin{tabularx}{\textwidth}{ccc}
\includegraphics[width=0.3\textwidth]{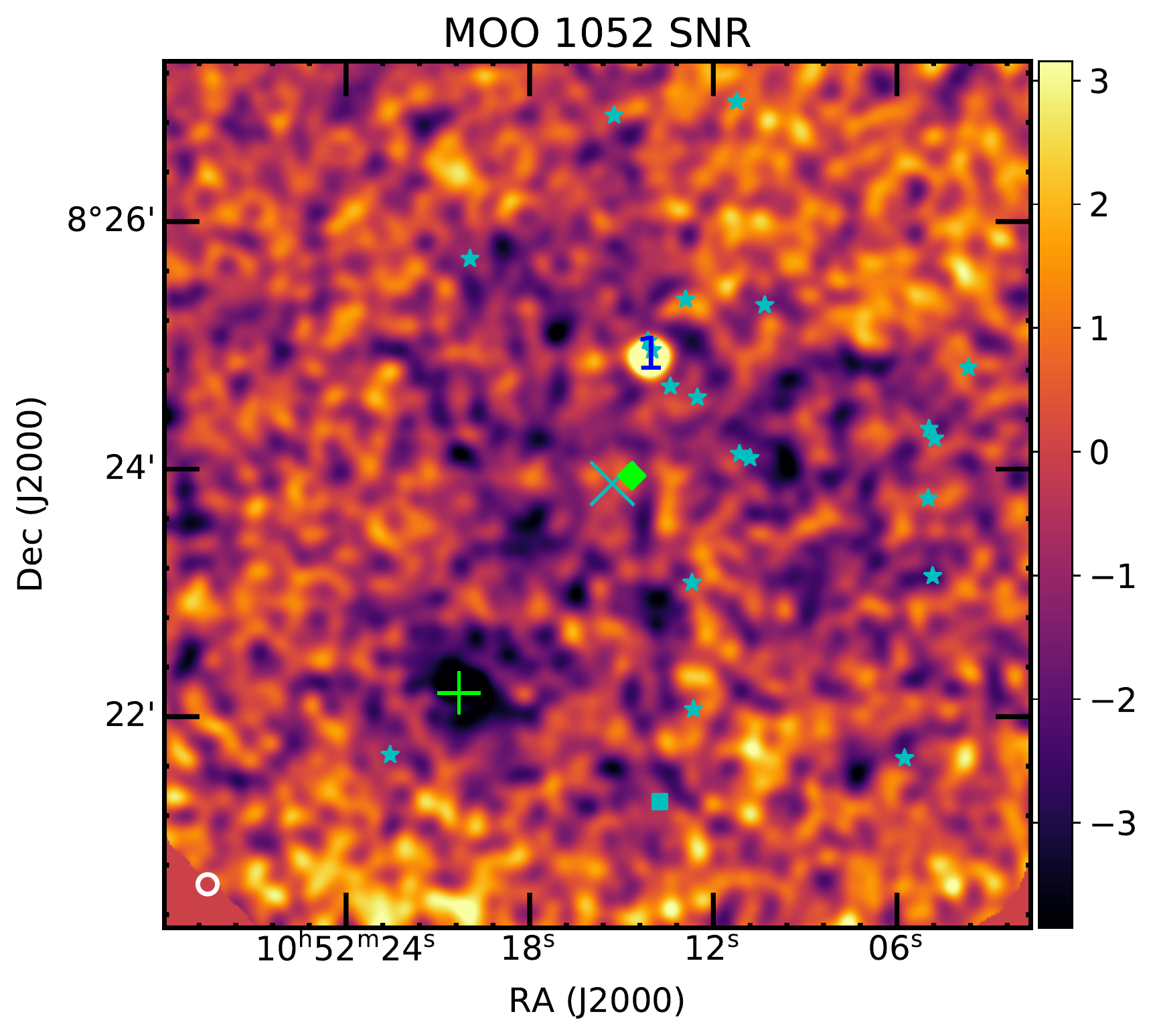} & 
\includegraphics[width=0.3\textwidth]{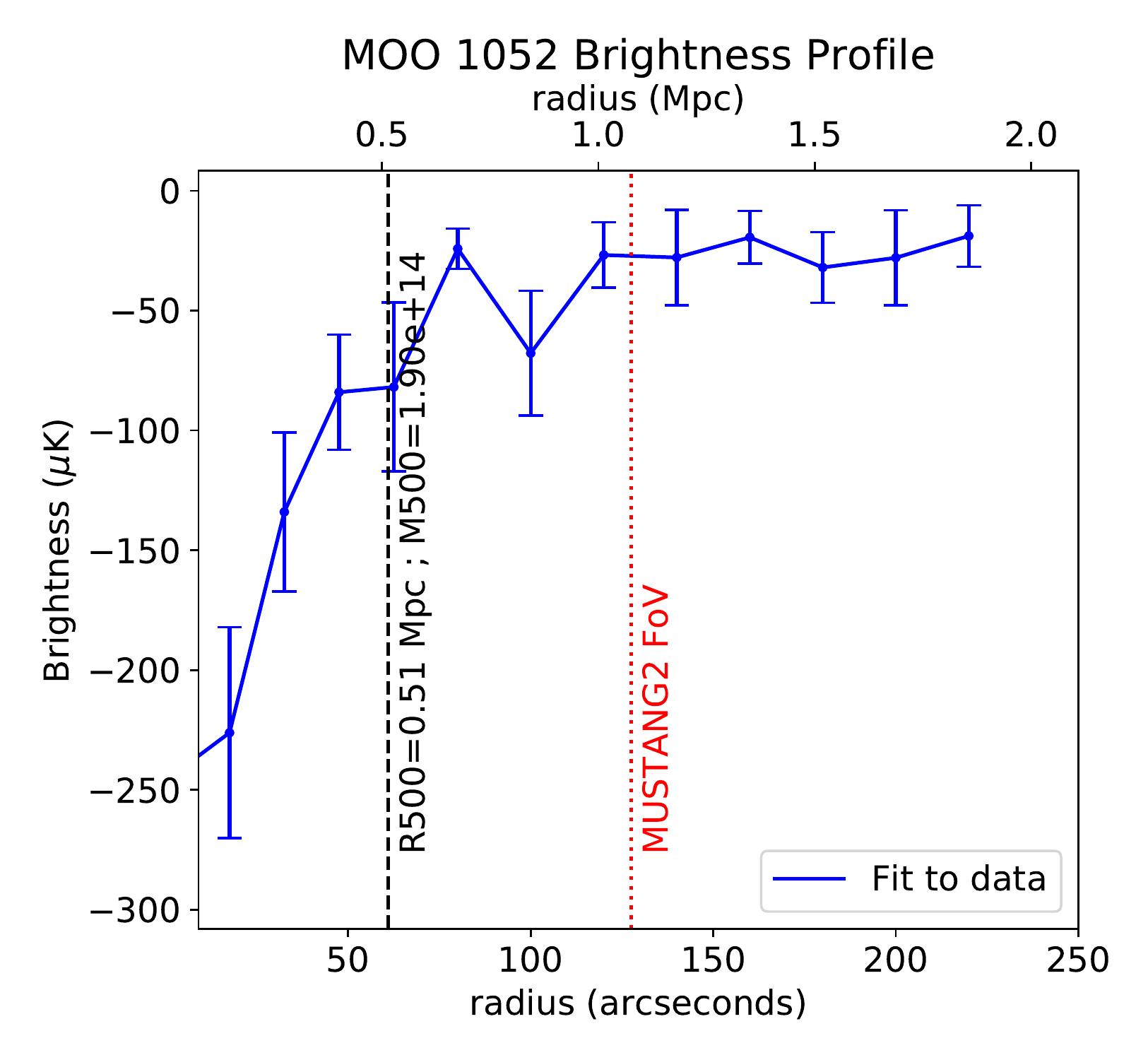} &
\includegraphics[width=0.3\textwidth]{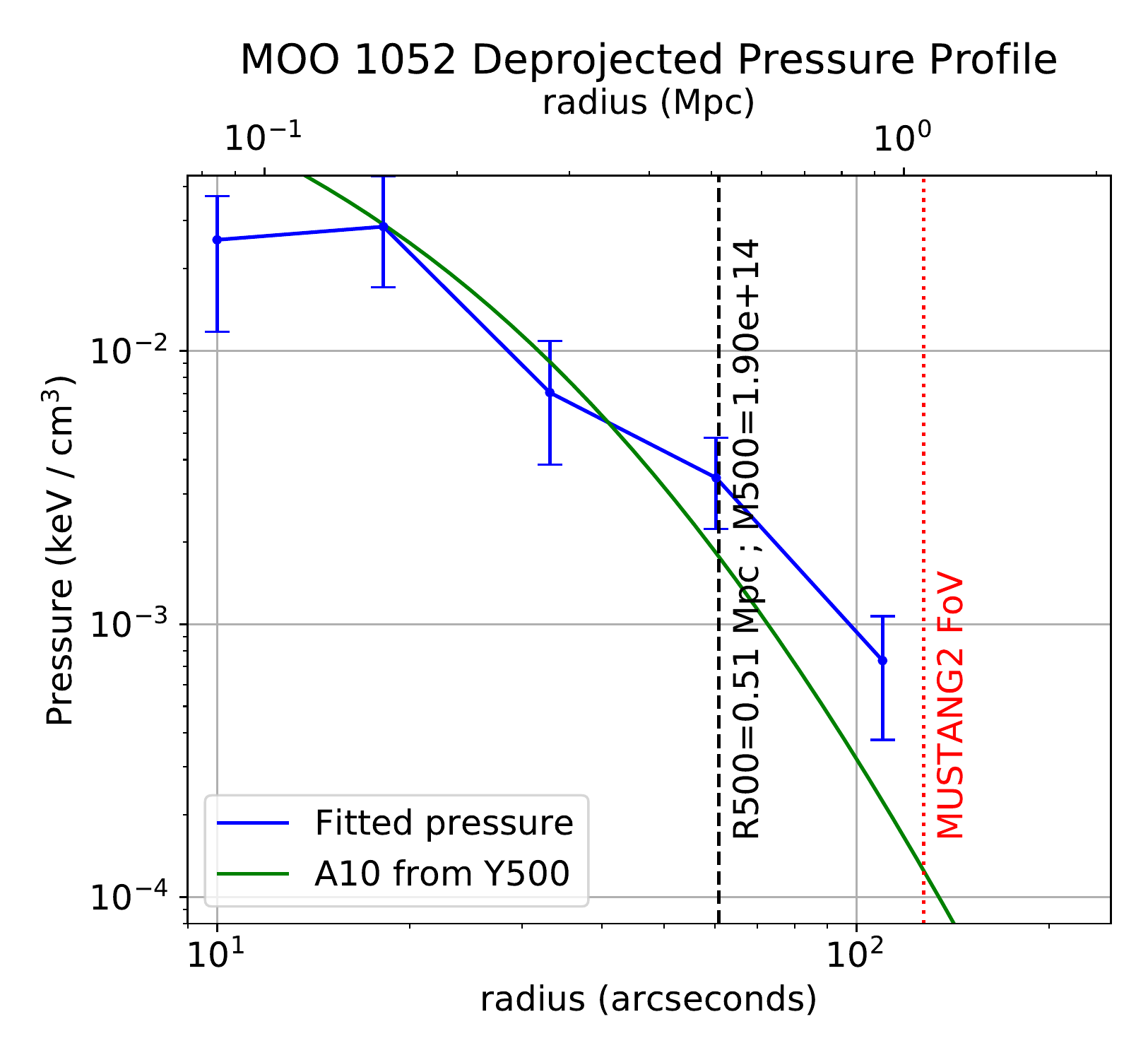}
\end{tabularx}
%{{\bf MOO 0135}}
\begin{tabularx}{\textwidth}{ccc}
\includegraphics[width=0.3\textwidth]{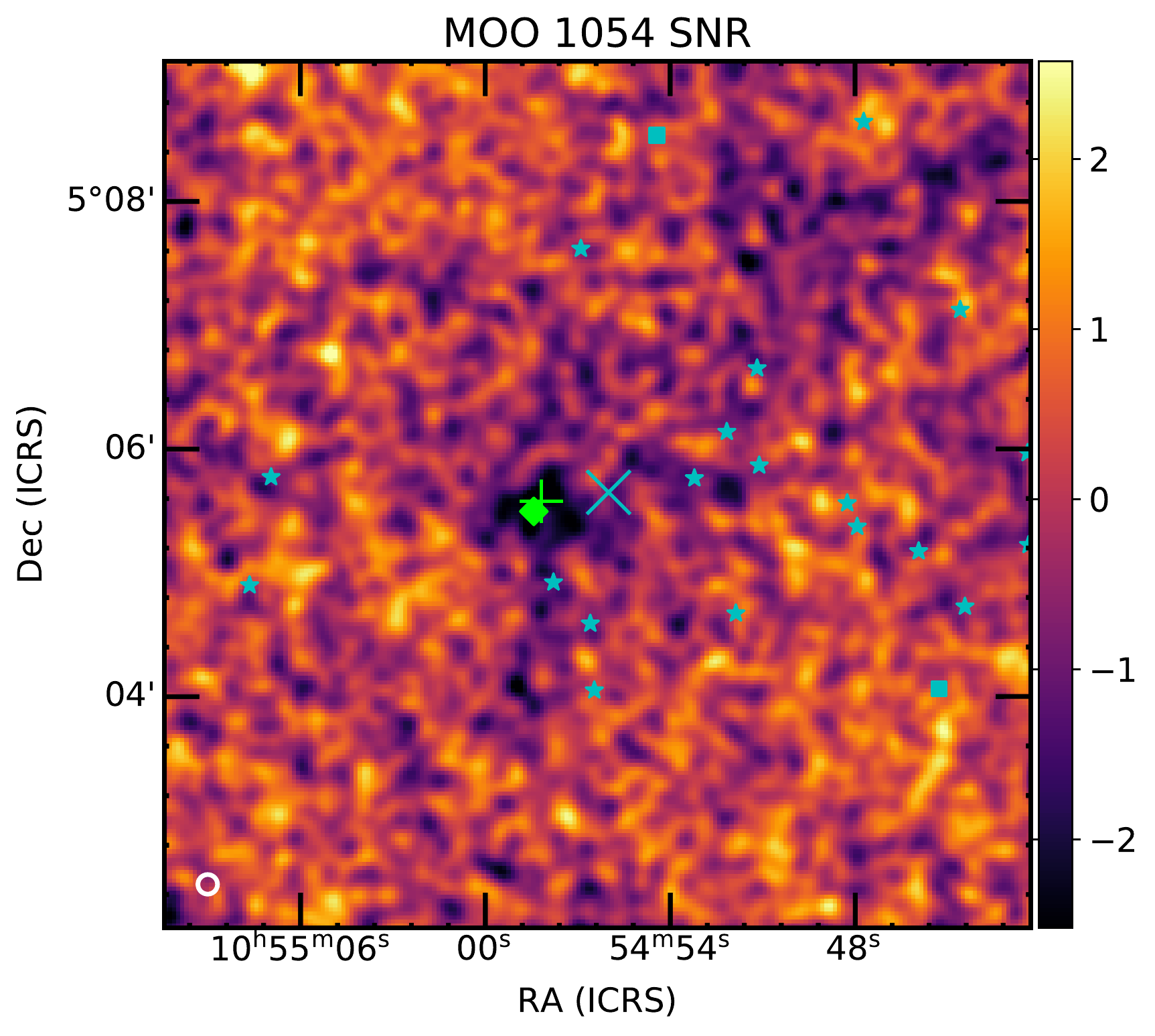} & 
\includegraphics[width=0.3\textwidth]{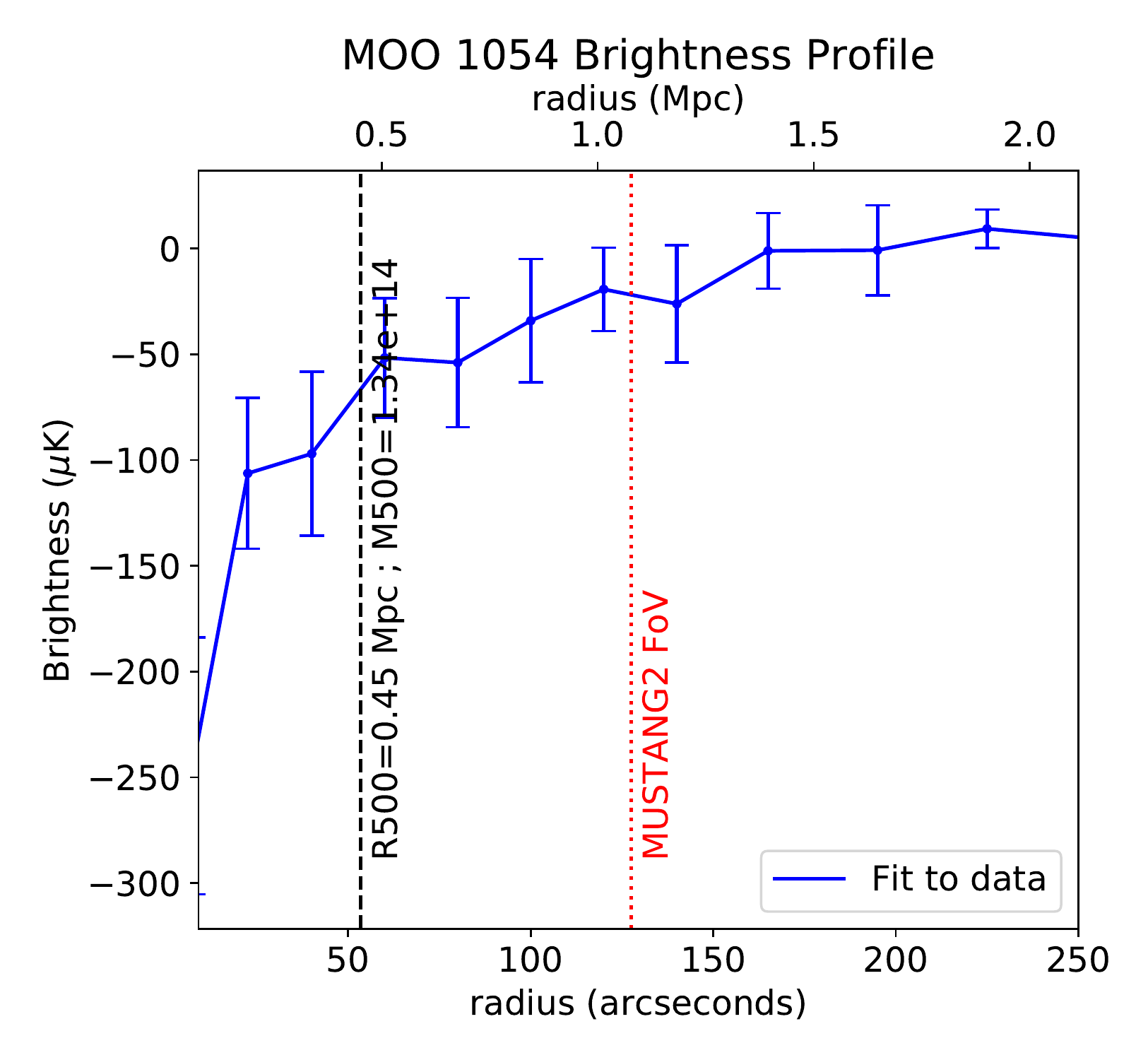} &
\includegraphics[width=0.3\textwidth]{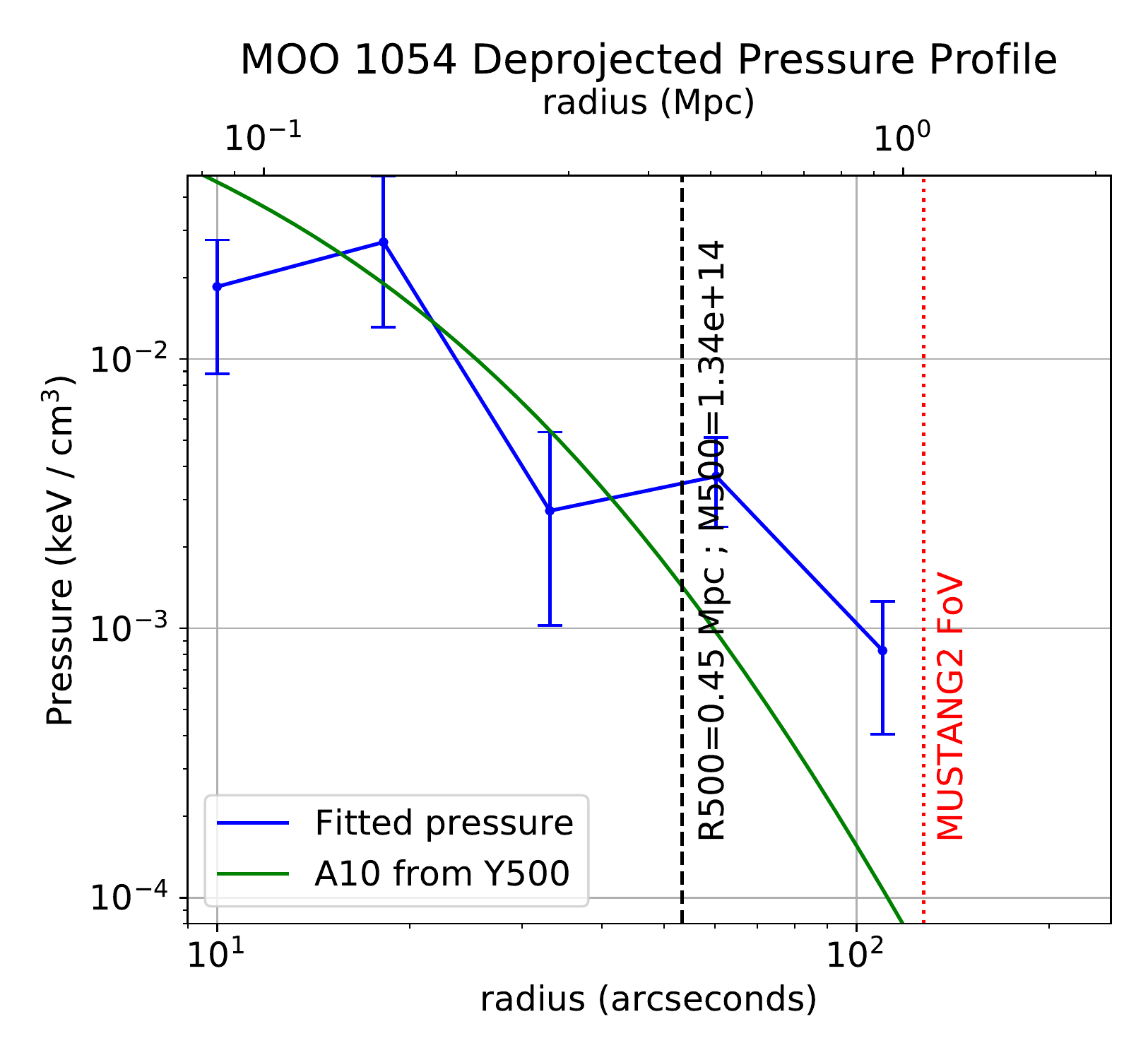}
\end{tabularx}
%{{\bf MOO 0105}}
\begin{tabularx}{\textwidth}{ccc}
\includegraphics[width=0.3\textwidth]{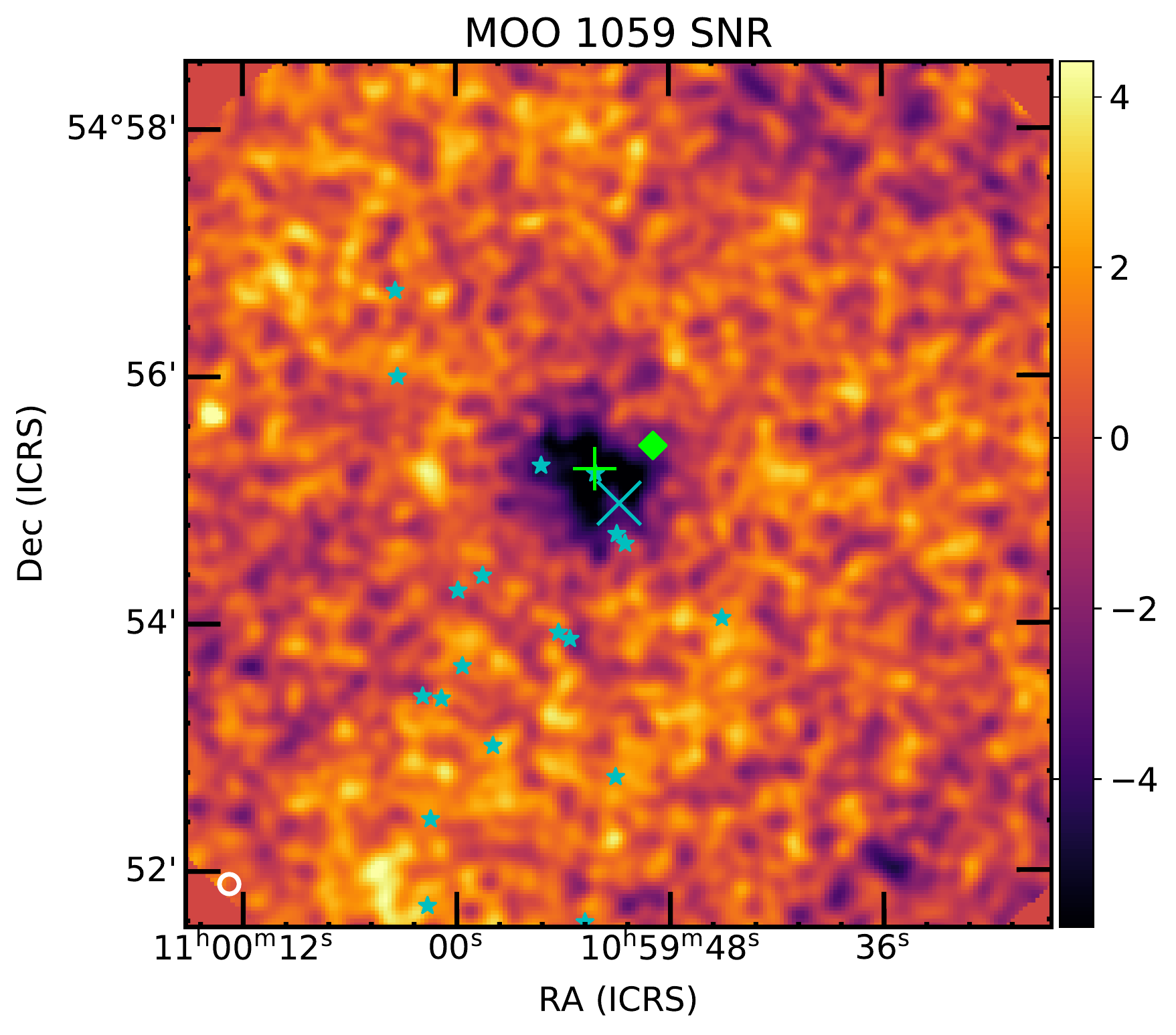} & 
\includegraphics[width=0.3\textwidth]{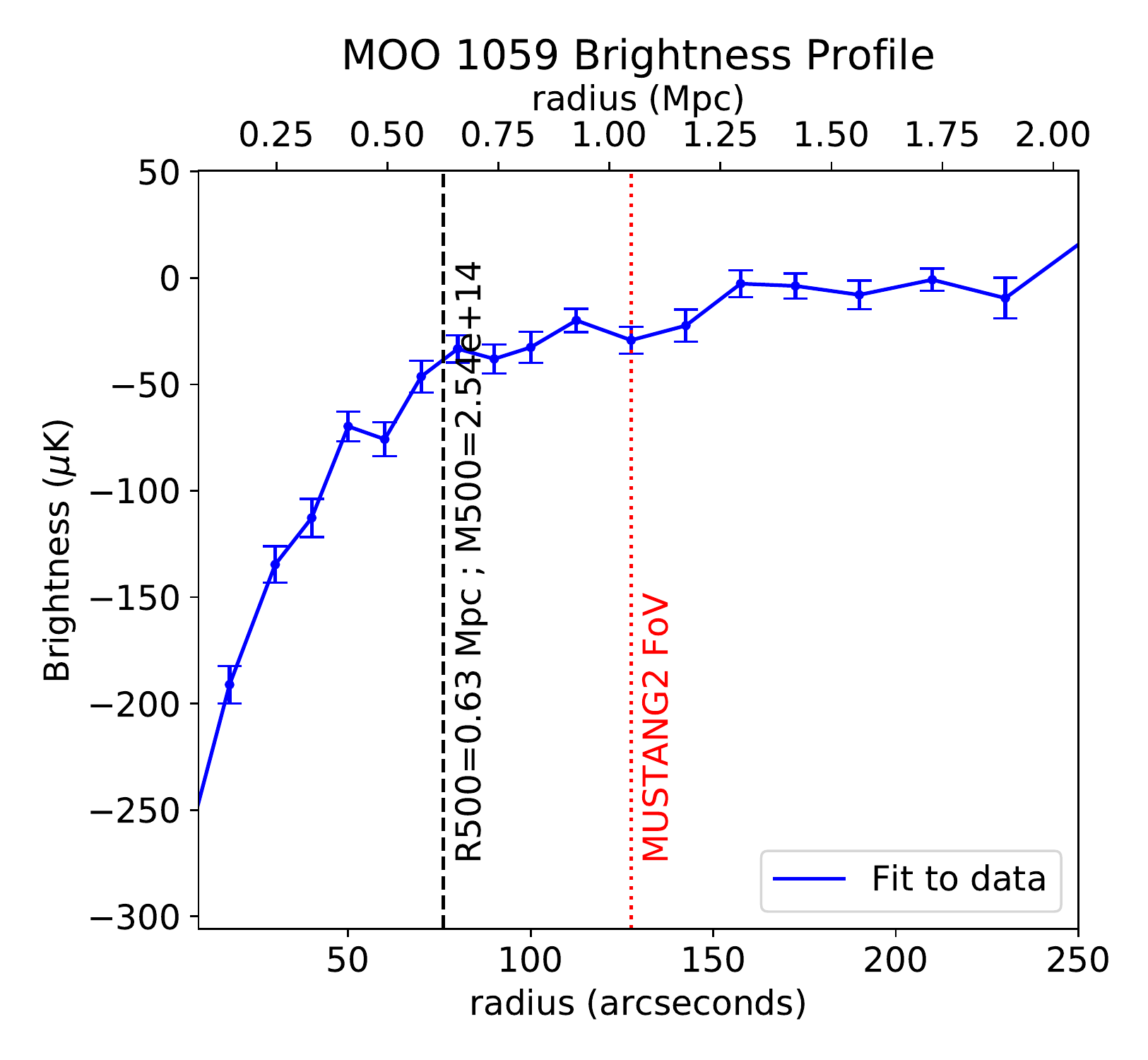} &
\includegraphics[width=0.3\textwidth]{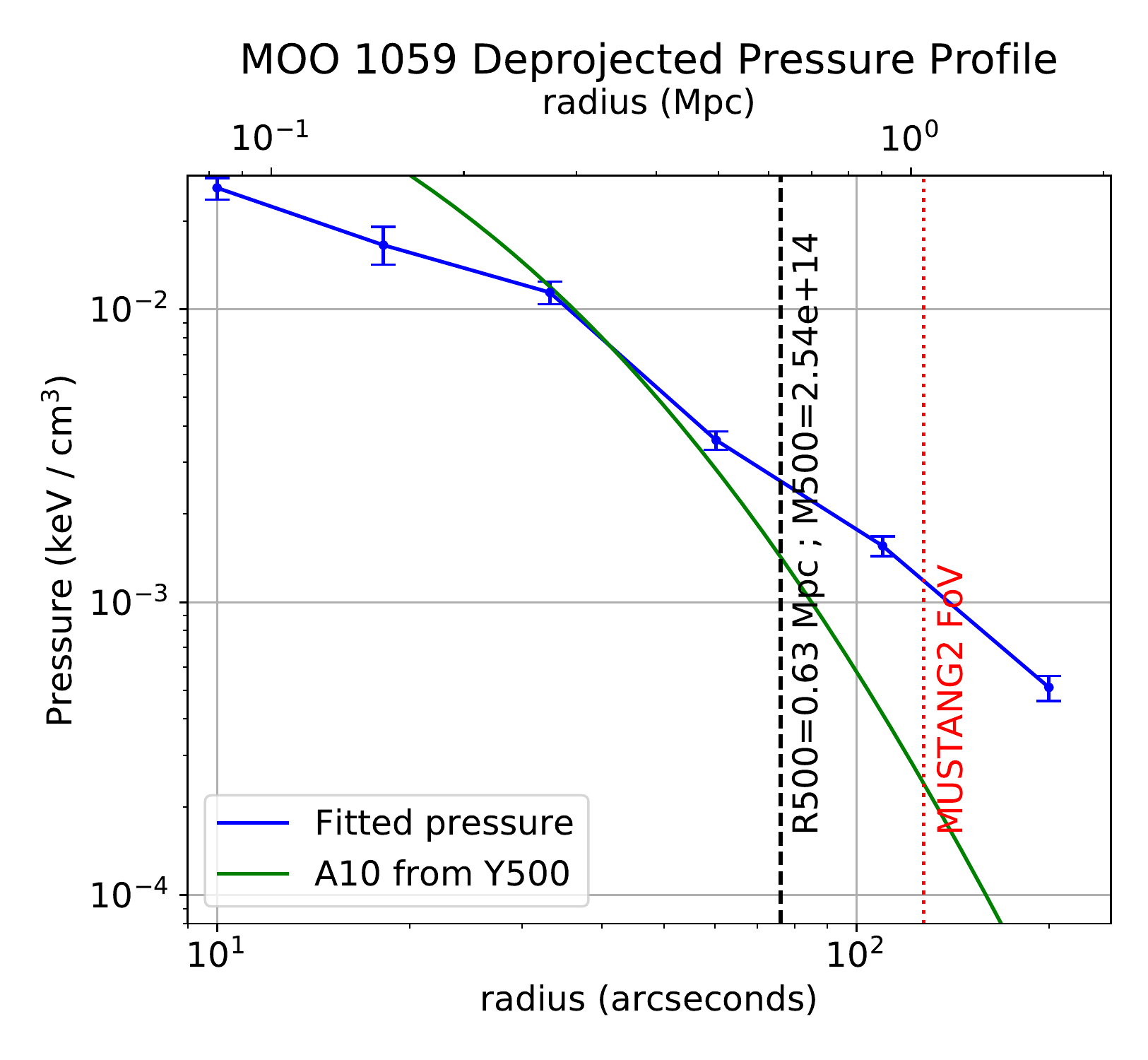}
\end{tabularx}
%{{\bf MOO 0105}}
\caption{(continued) {\bf Left:} MUSTANG-2 images made using the MIDAS pipeline.  A cyan X marks the original center found by MaDCoWS, while the green cross marks the best fit SZE centroids.  Cyan stars and squares mark the locations of bright galaxies detected by \textit{Spitzer} and SDSS, respectively.  The BCG is marked as a green diamond and the MUSTANG2 beam is shown as a white circle on the lower left. The bright sources are clipped at $+4\sigma$ and are labeled with numbers to match Table~\ref{tab:pntSrc}. {\bf Center:} Brightness profiles of our clusters from Minkasi.\@ {\bf Right:} Pressure profiles derived from each data set.  The MUSTANG2 FoV is marked as a red line, while the black dashed line represents the $R_{500}$ for our recovered mass. \red{The A10 profile that corresponds to this mass is shown in green}.}
%Continued from Figure \ref{fig:results}.}
\end{center}
\end{figure*}

\begin{figure*}
\addtocounter{figure}{-1} %\ContinuedFloat
\begin{center}
\begin{tabularx}{\textwidth}{ccc}
\includegraphics[width=0.3\textwidth]{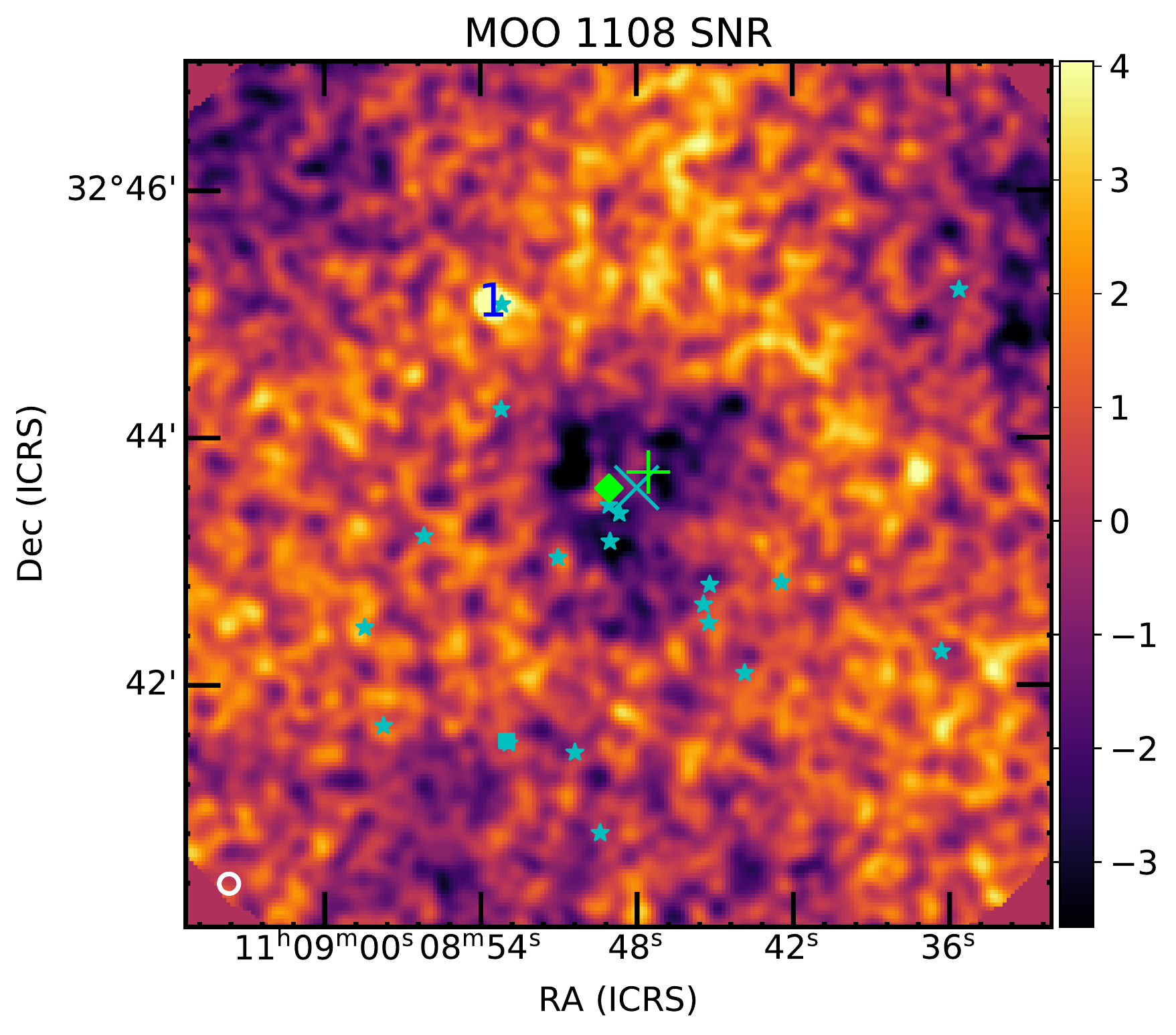} & 
\includegraphics[width=0.3\textwidth]{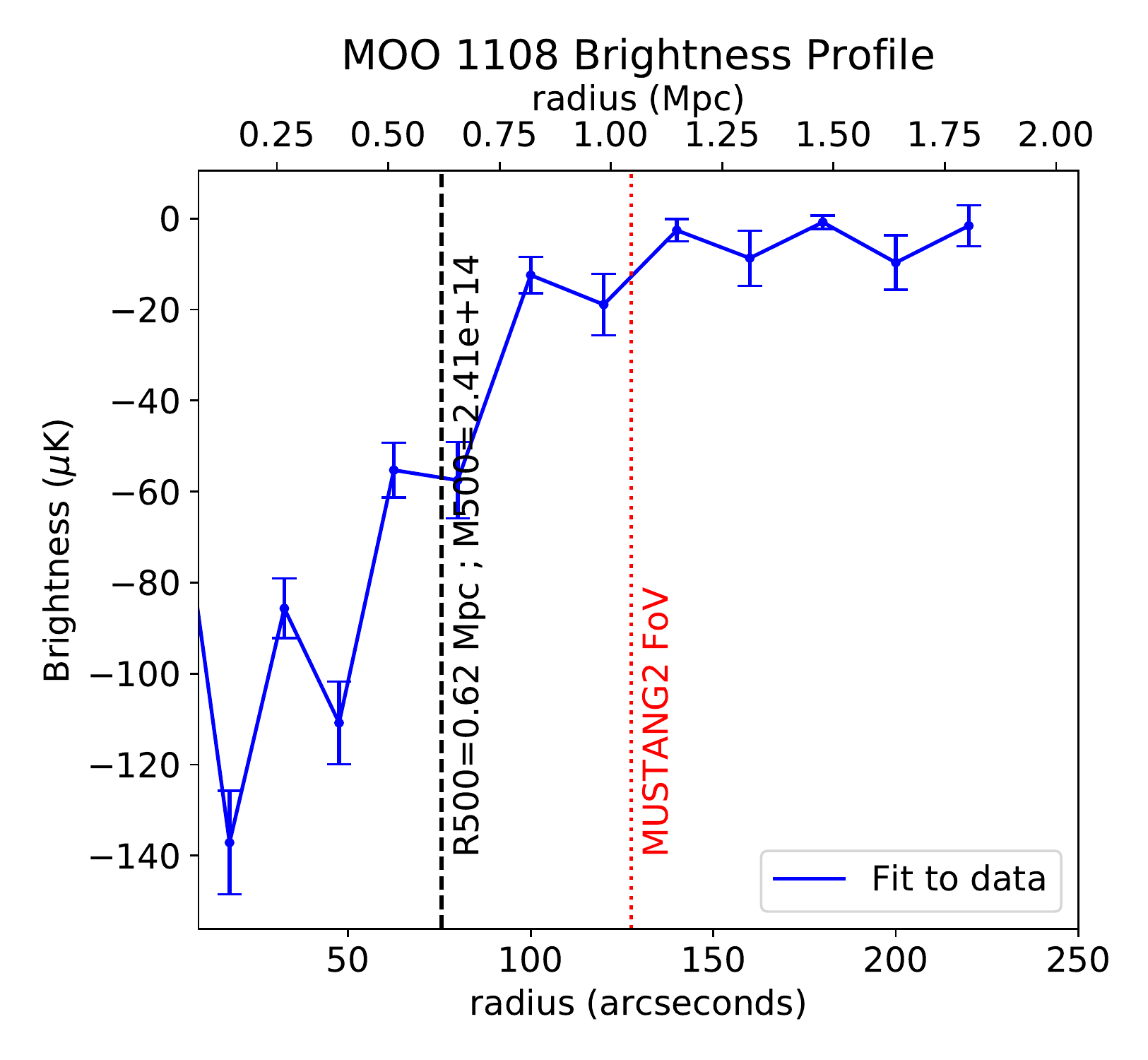} &
\includegraphics[width=0.3\textwidth]{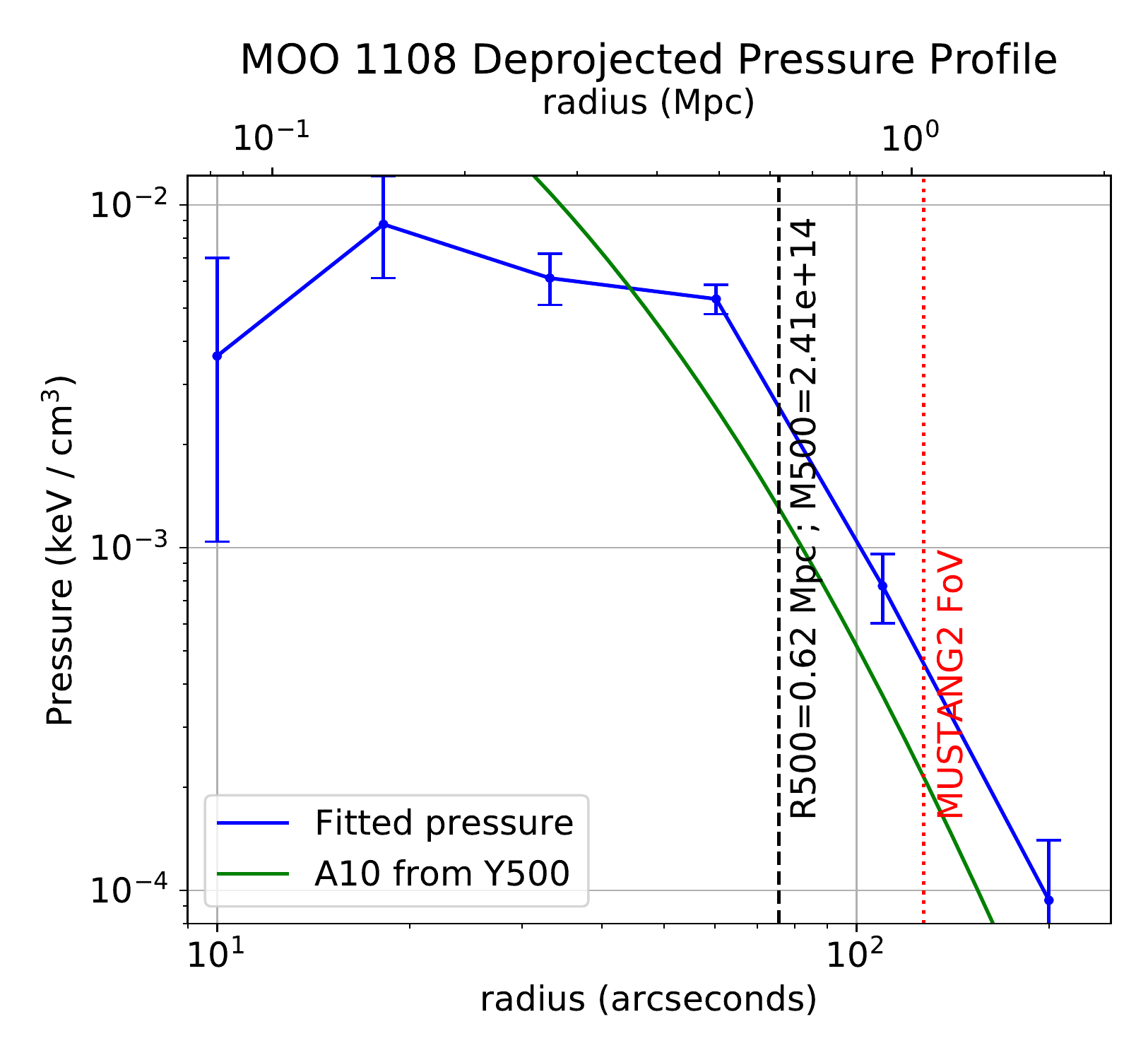}
\end{tabularx}
%{{\bf MOO 0105}}
\begin{tabularx}{\textwidth}{ccc}
\includegraphics[width=0.3\textwidth]{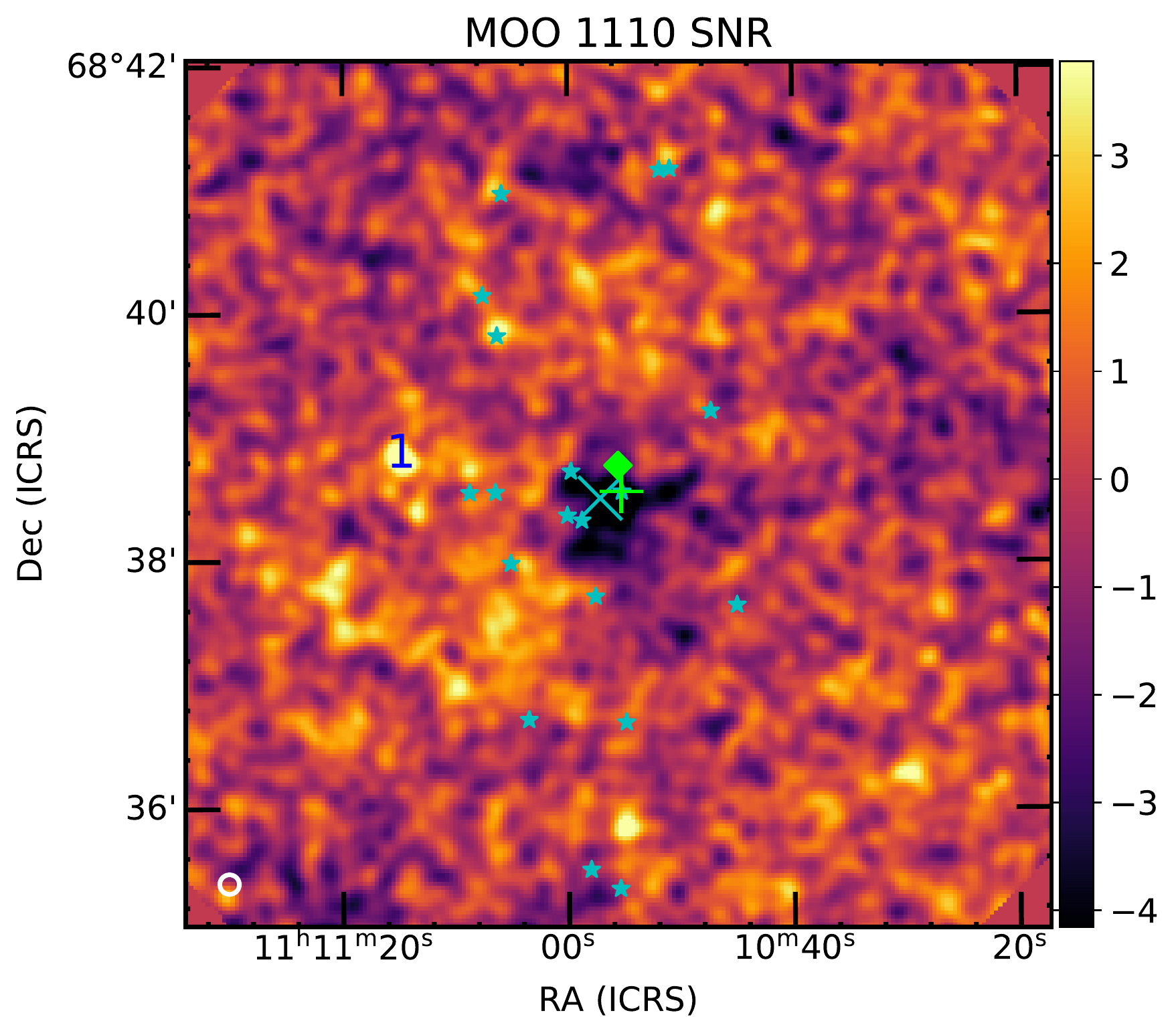} & 
\includegraphics[width=0.3\textwidth]{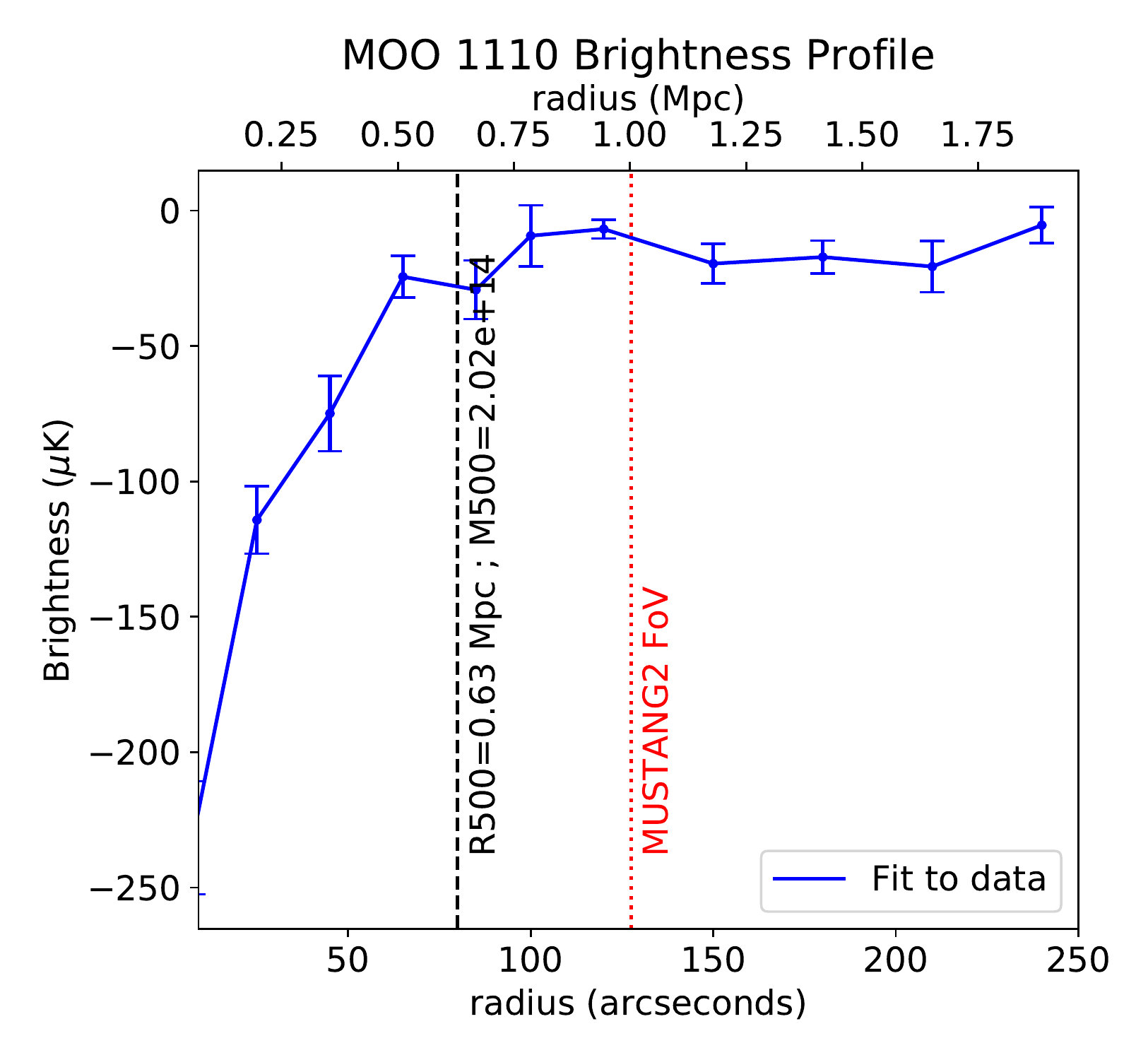} &
\includegraphics[width=0.3\textwidth]{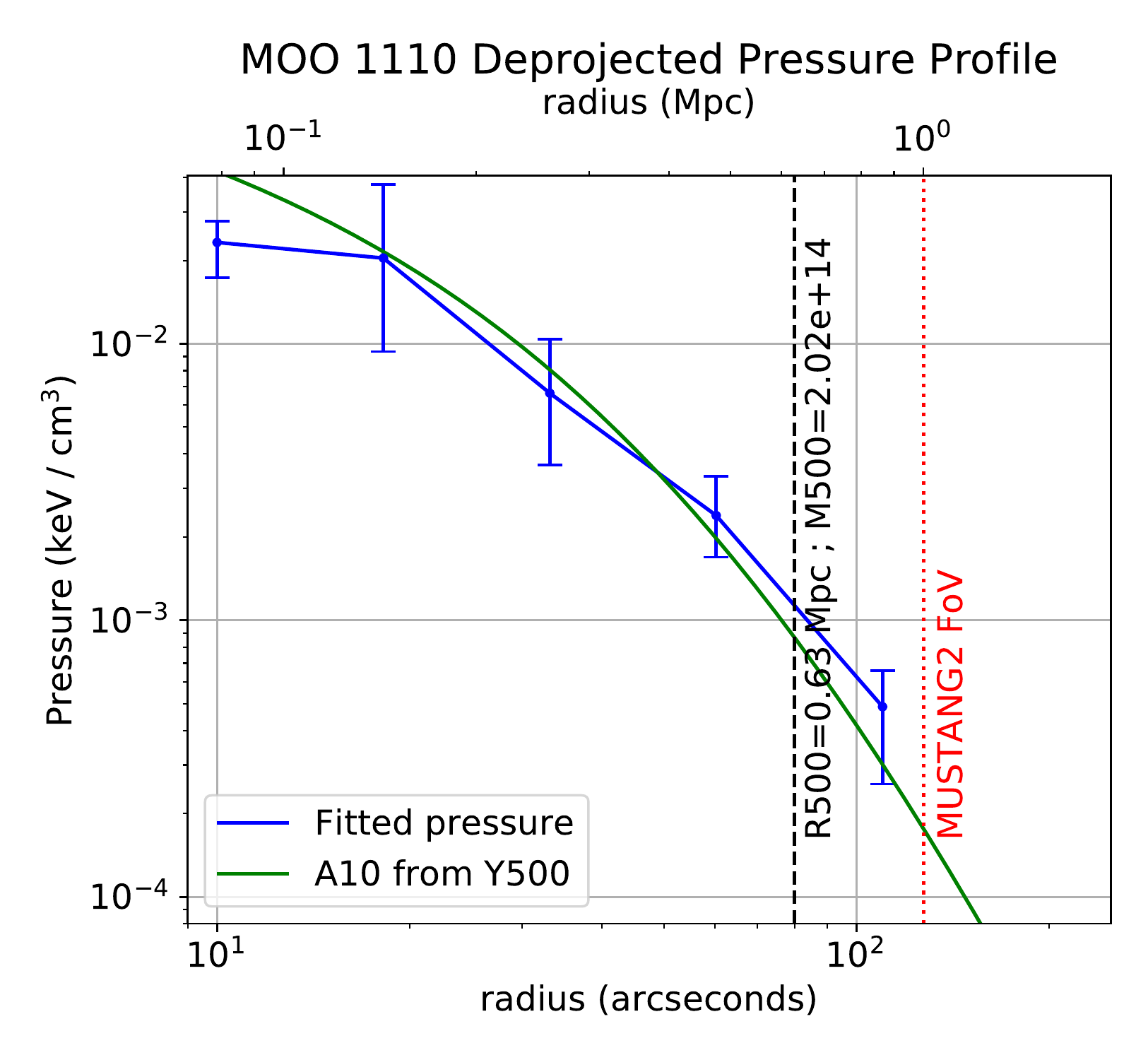}
\end{tabularx}
\begin{tabularx}{\textwidth}{ccc}
\includegraphics[width=0.3\textwidth]{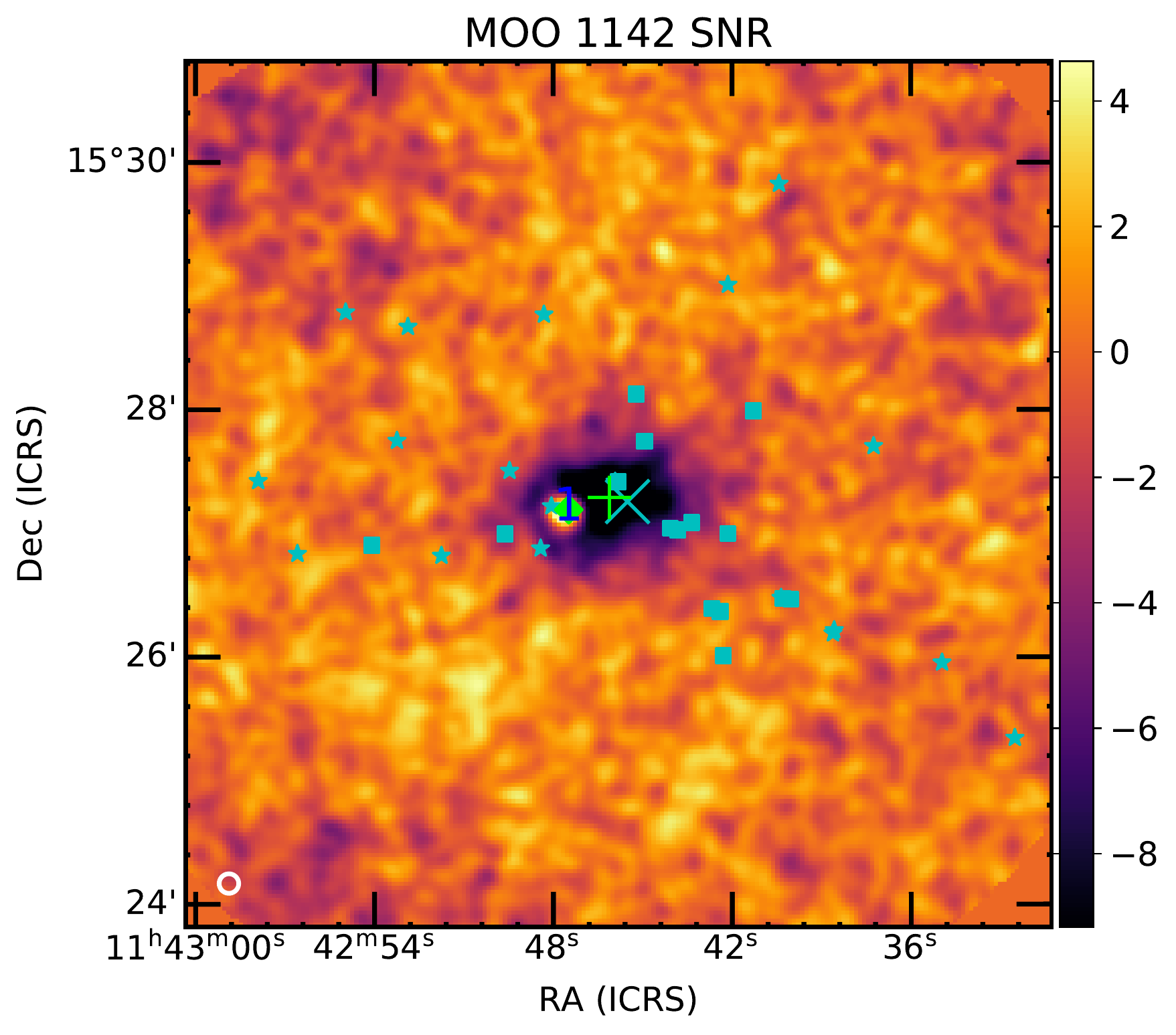} & 
\includegraphics[width=0.3\textwidth]{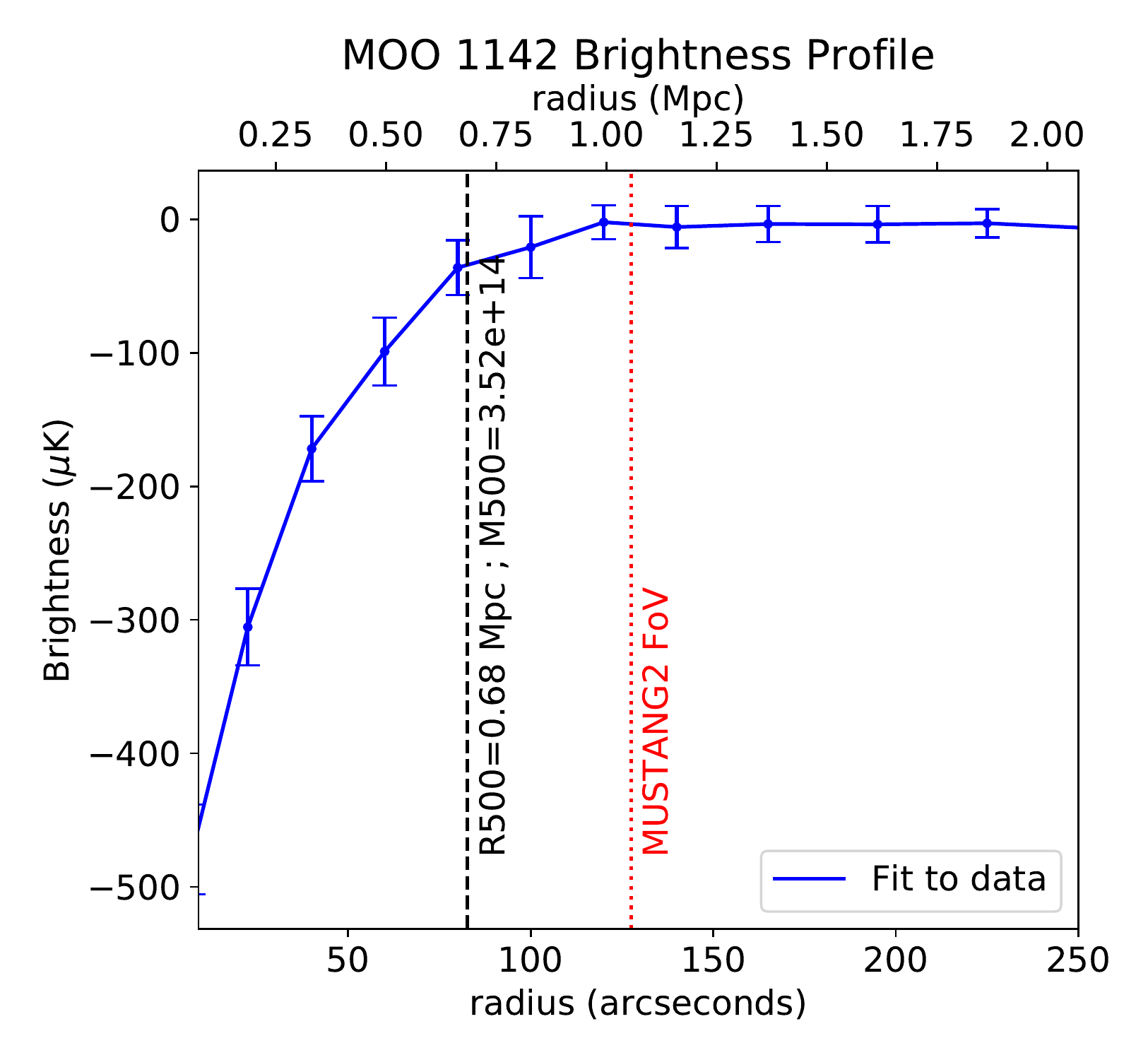} &
\includegraphics[width=0.3\textwidth]{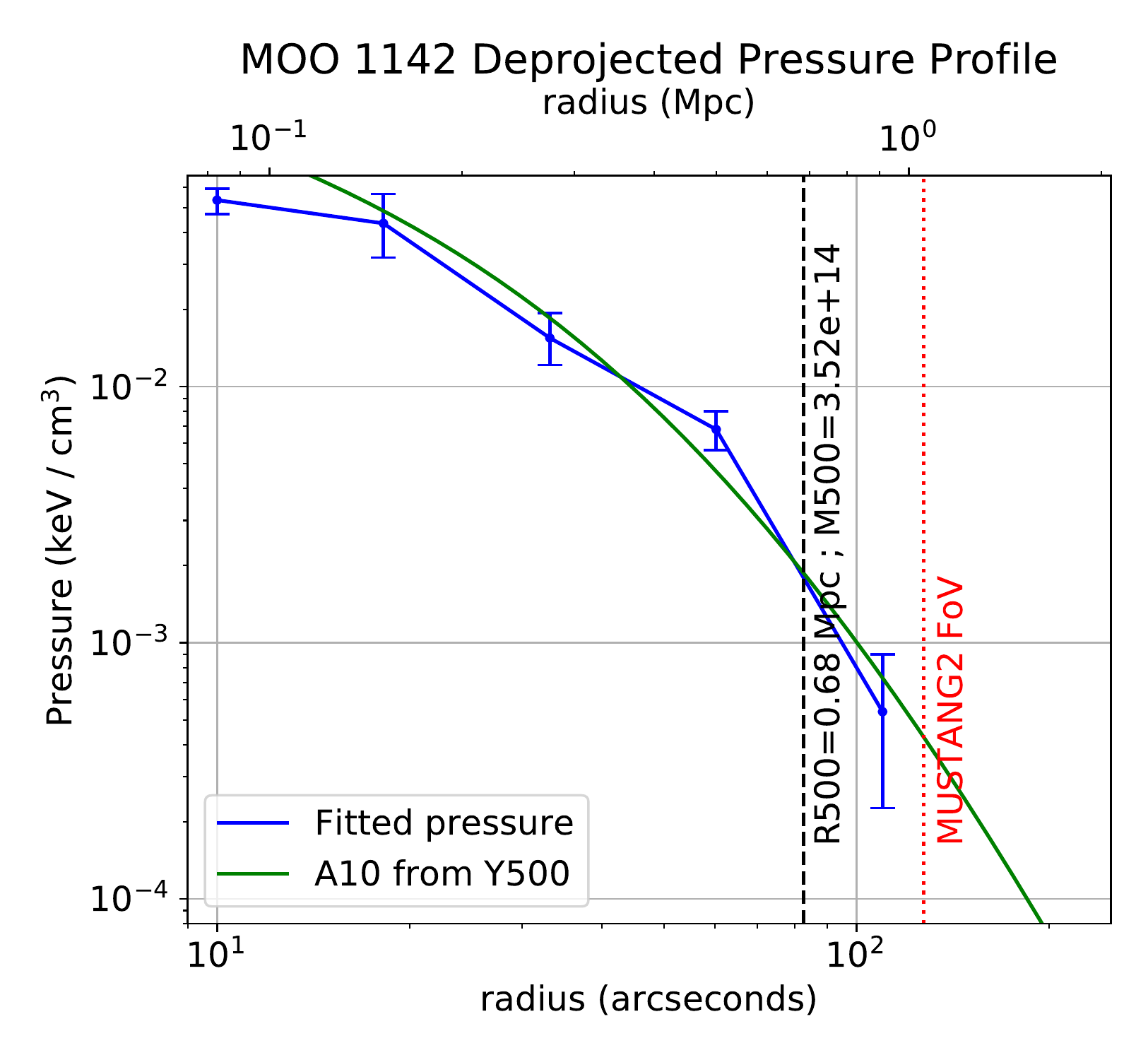}
\end{tabularx}\begin{tabularx}{\textwidth}{ccc}
\includegraphics[width=0.3\textwidth]{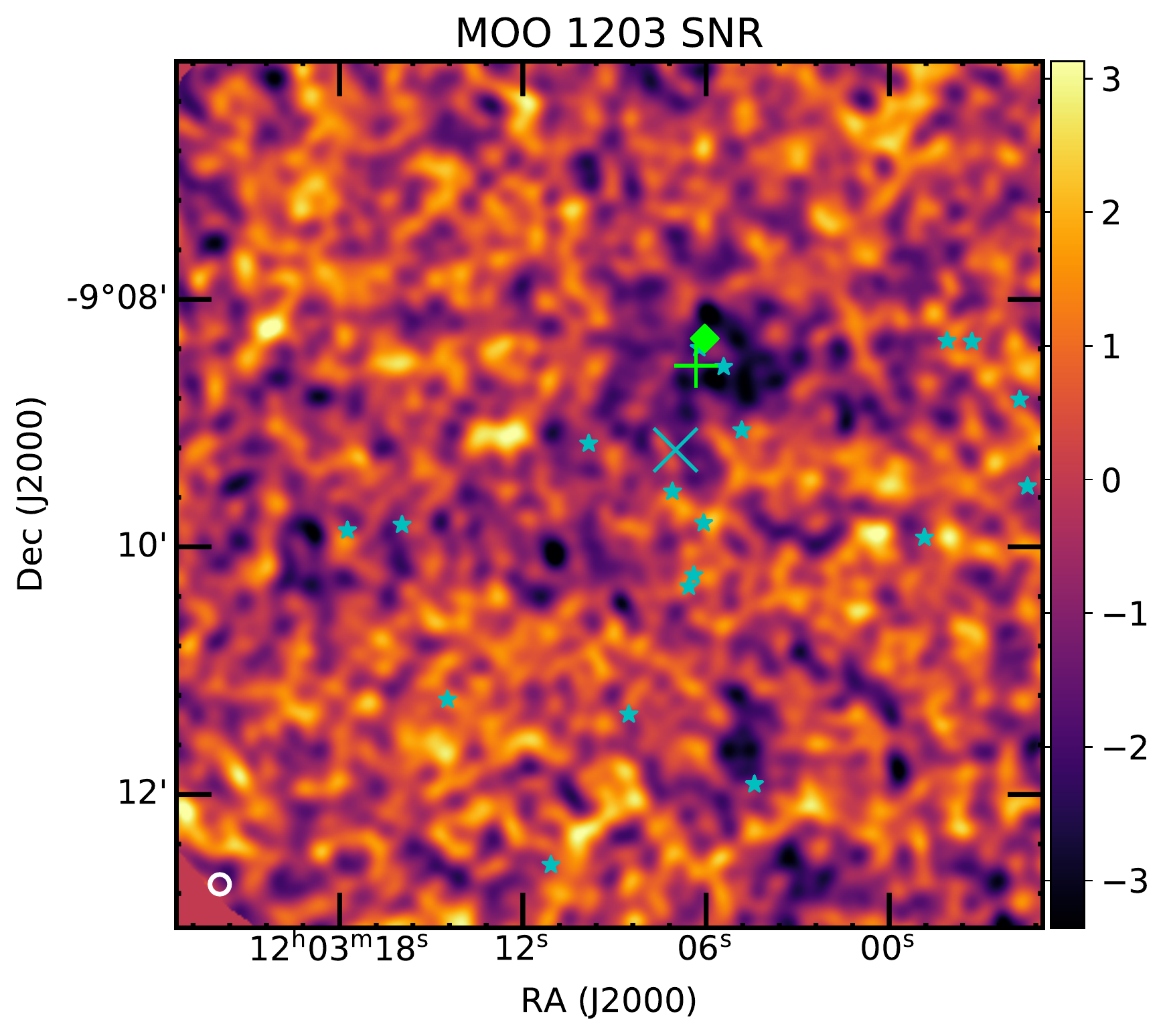} & 
\includegraphics[width=0.3\textwidth]{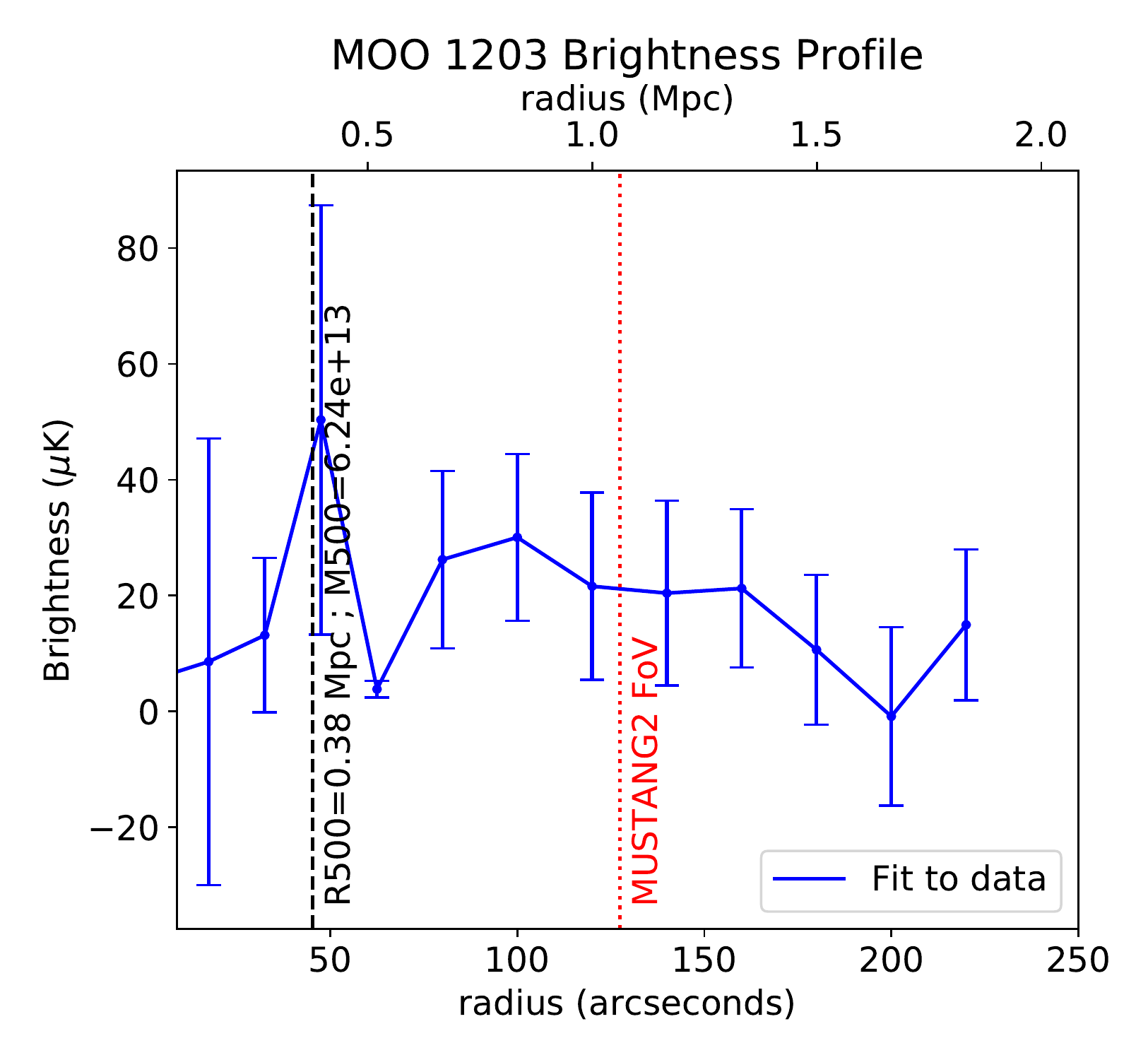} &
\includegraphics[width=0.3\textwidth]{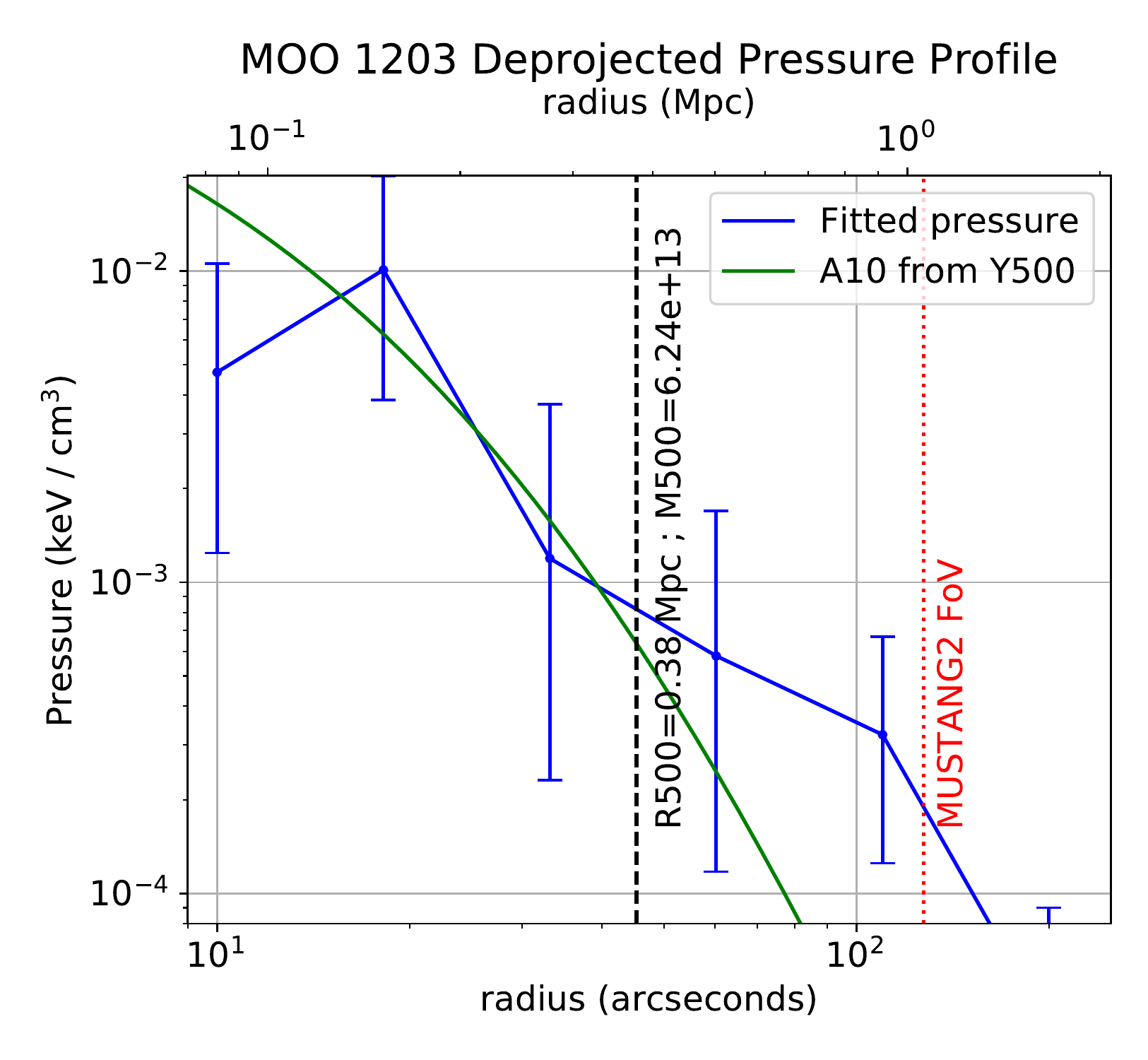}
\end{tabularx}
%{{\bf MOO 0105}}
\caption{(Continued) {\bf Left:} MUSTANG-2 images made using the MIDAS pipeline.  A cyan X marks the original center found by MaDCoWS, while the green cross marks the best fit SZE centroids.  Cyan stars and squares mark the locations of bright galaxies detected by \textit{Spitzer} and SDSS, respectively.  The BCG is marked as a green diamond and the MUSTANG2 beam is shown as a white circle on the lower left. The bright sources are clipped at $+4\sigma$ and are labeled with numbers to match Table~\ref{tab:pntSrc}. {\bf Center:} Brightness profiles of our clusters from Minkasi.\@ {\bf Right:} Pressure profiles derived from each data set.  The MUSTANG2 FoV is marked as a red line, while the black dashed line represents the $R_{500}$ for our recovered mass. \red{The A10 profile that corresponds to this mass is shown in green}.}
%\caption{Continued from Figure \ref{fig:results}.}
\end{center}
\end{figure*}

\begin{figure*}
\addtocounter{figure}{-1} %\ContinuedFloat
\begin{center}
\begin{tabularx}{\textwidth}{ccc}
\includegraphics[width=0.3\textwidth]{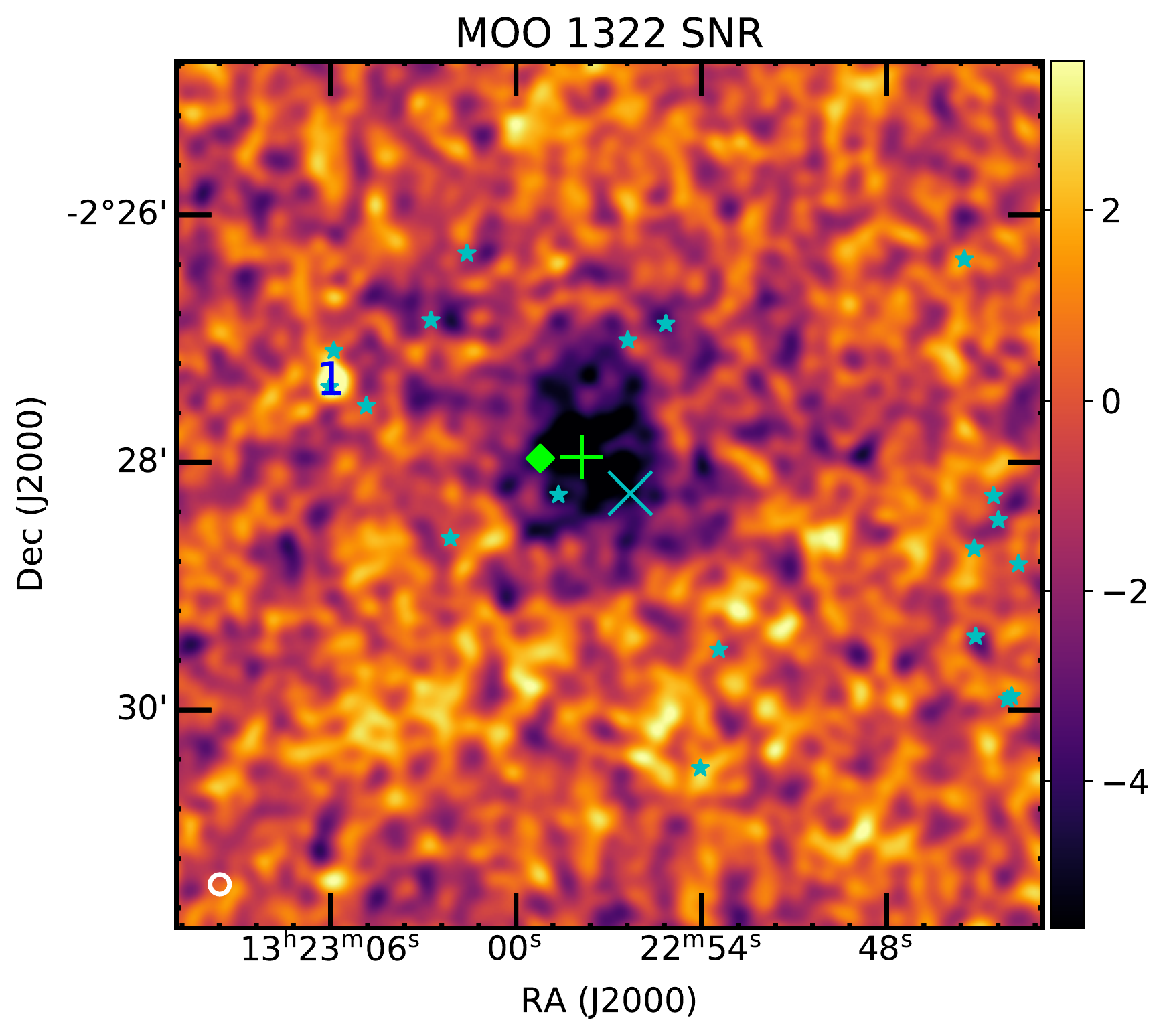} & 
\includegraphics[width=0.3\textwidth]{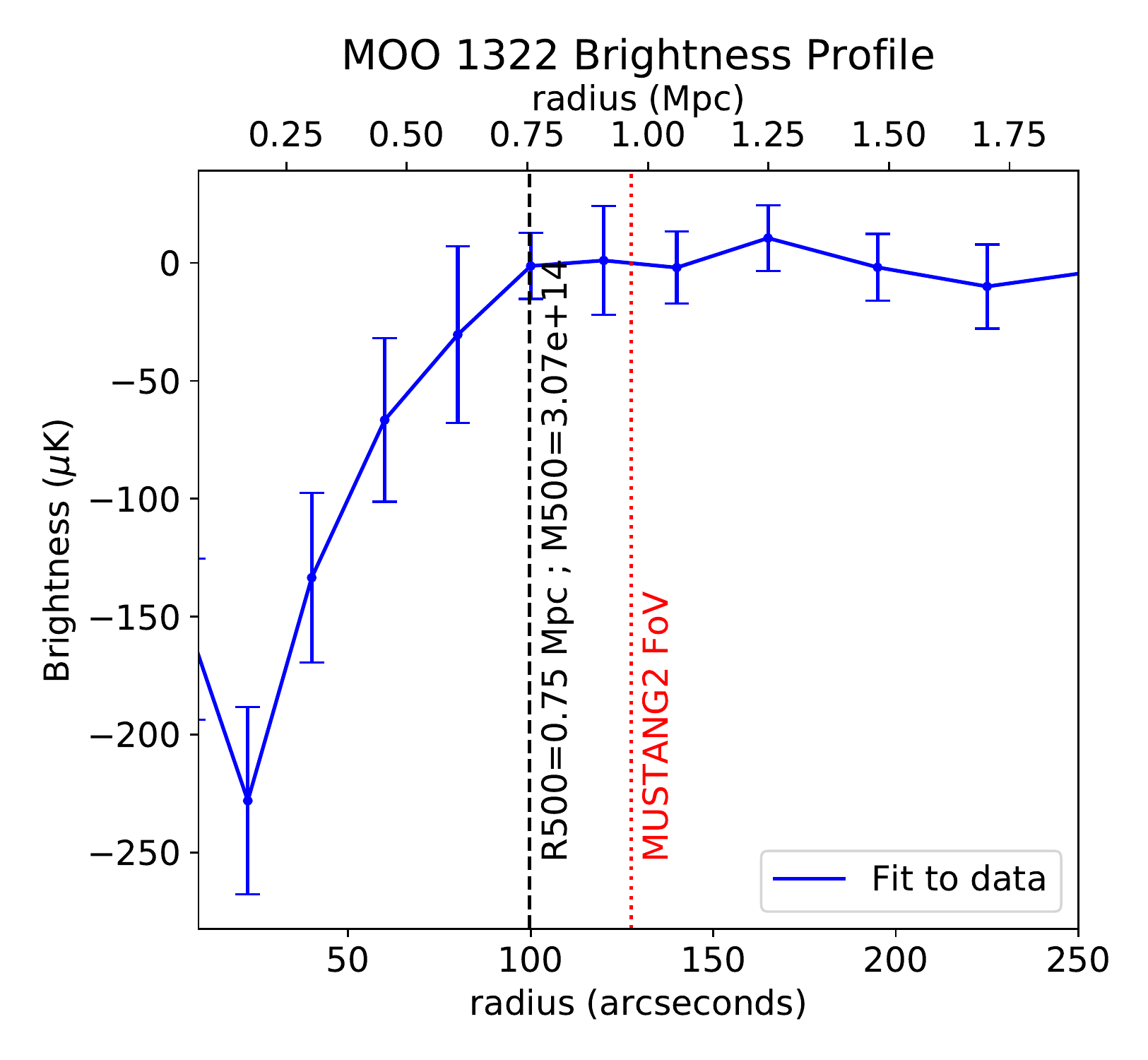} &
\includegraphics[width=0.3\textwidth]{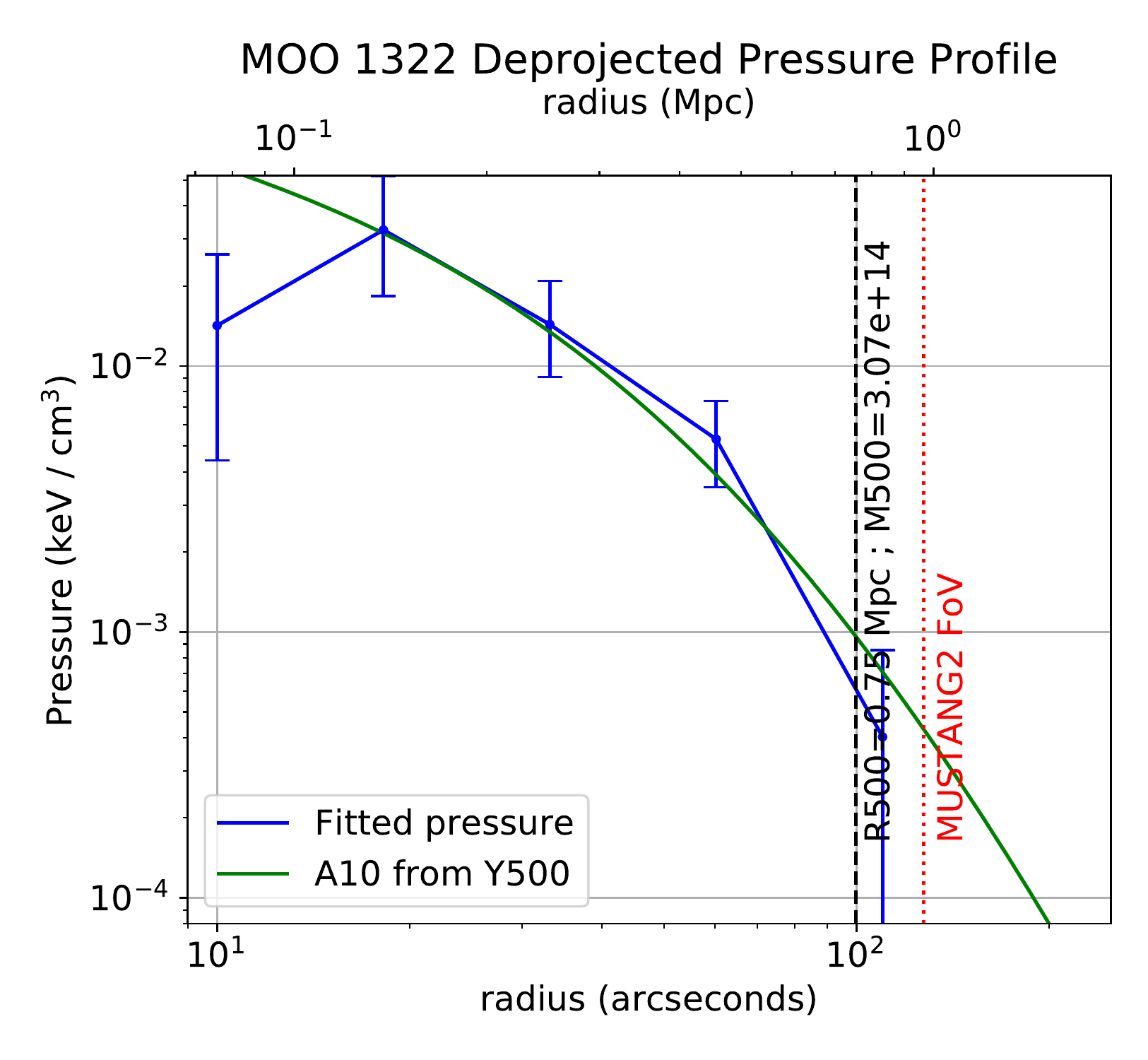}
\end{tabularx}
%{{\bf MOO 0105}}
\begin{tabularx}{\textwidth}{ccc}
\includegraphics[width=0.3\textwidth]{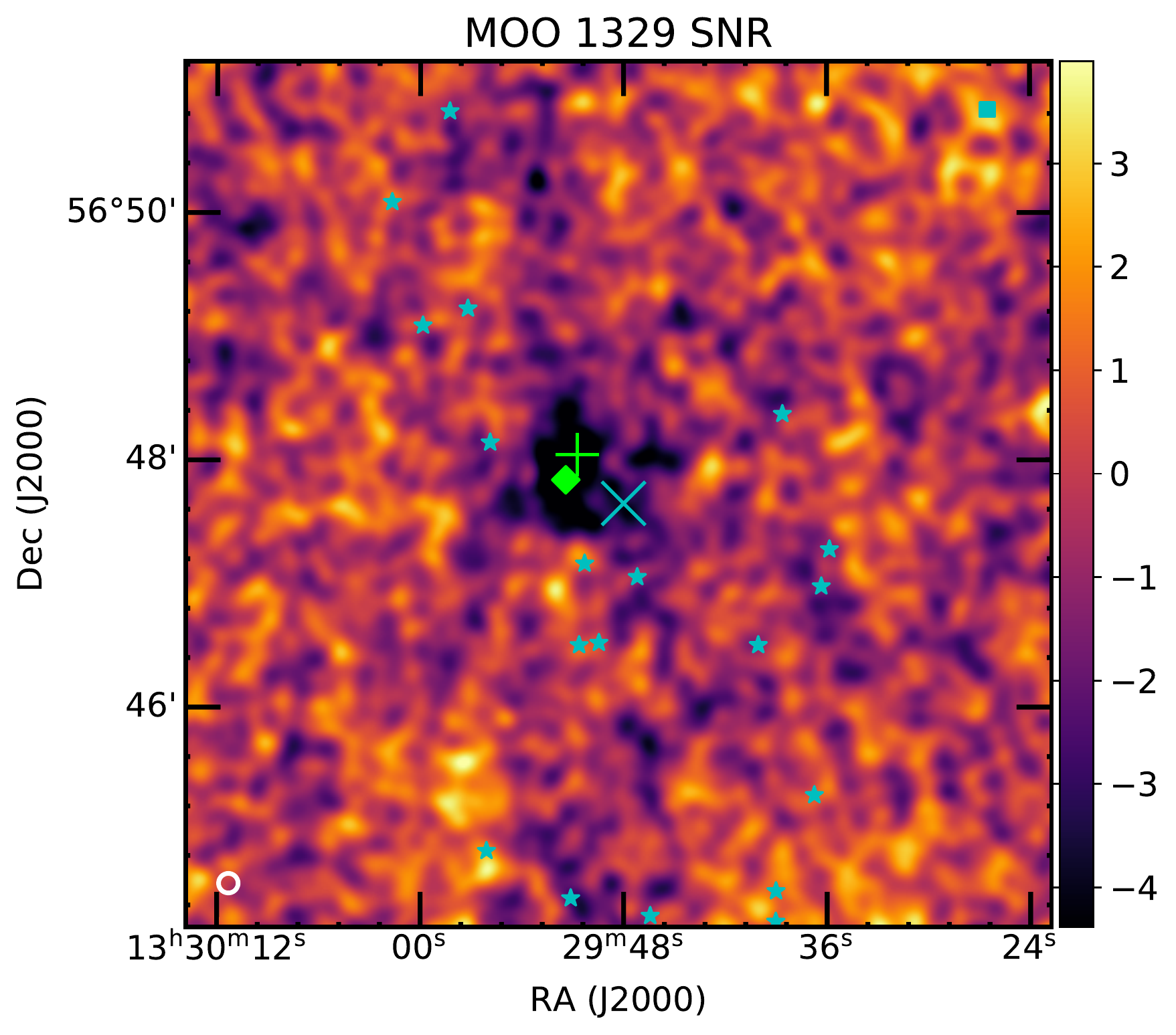} & 
\includegraphics[width=0.3\textwidth]{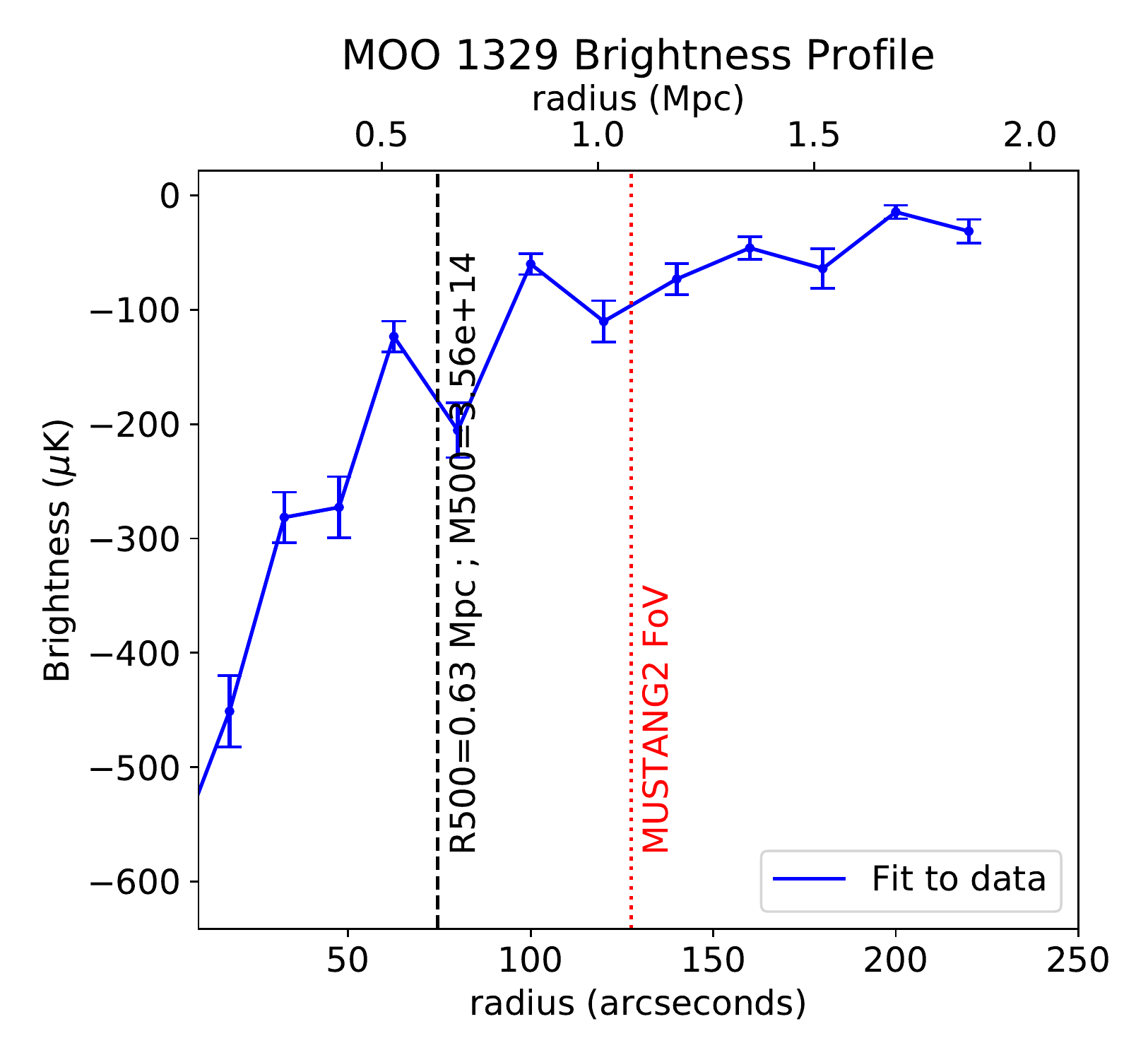} &
\includegraphics[width=0.3\textwidth]{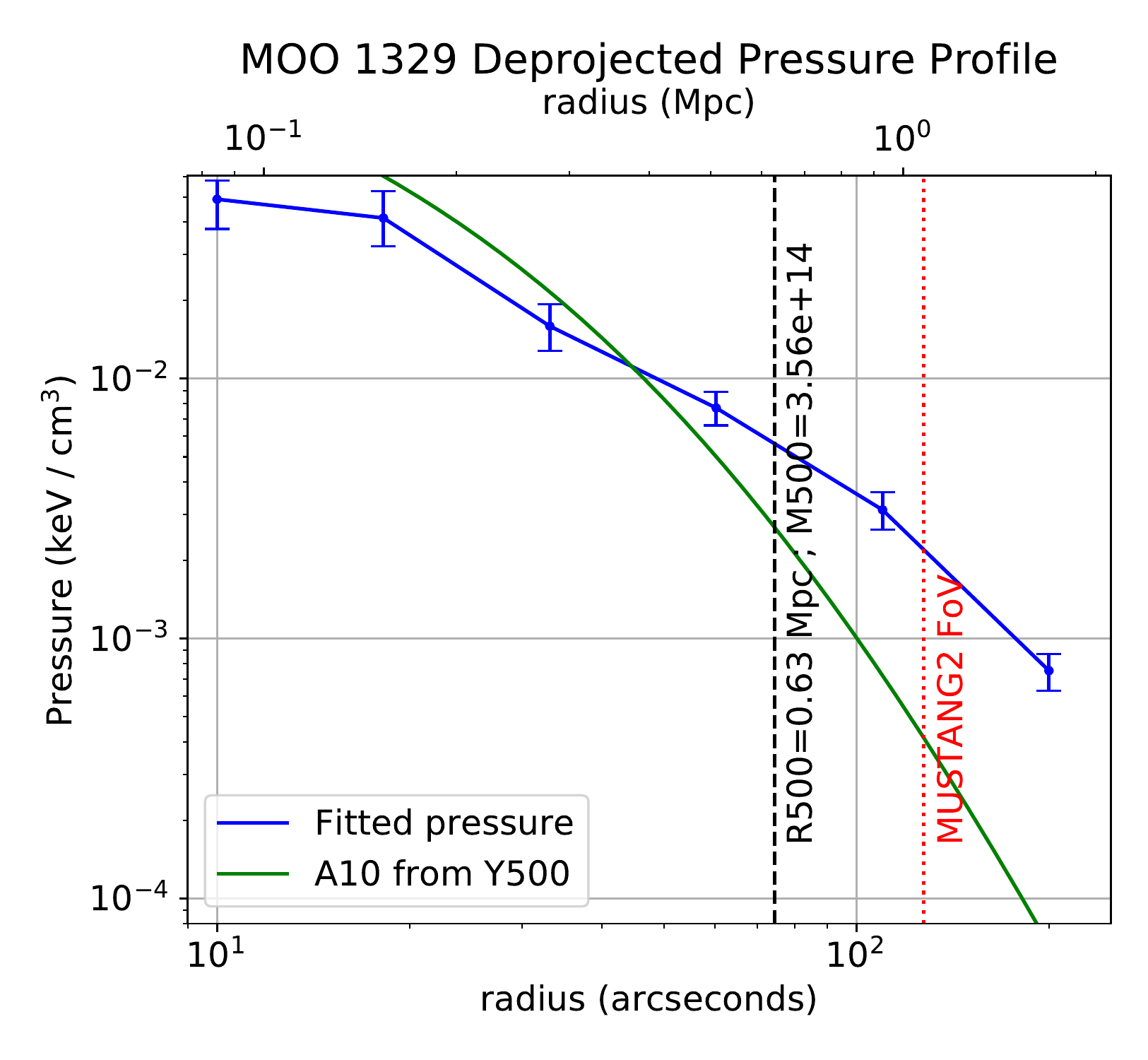}
\end{tabularx}
\begin{tabularx}{\textwidth}{ccc}
\includegraphics[width=0.3\textwidth]{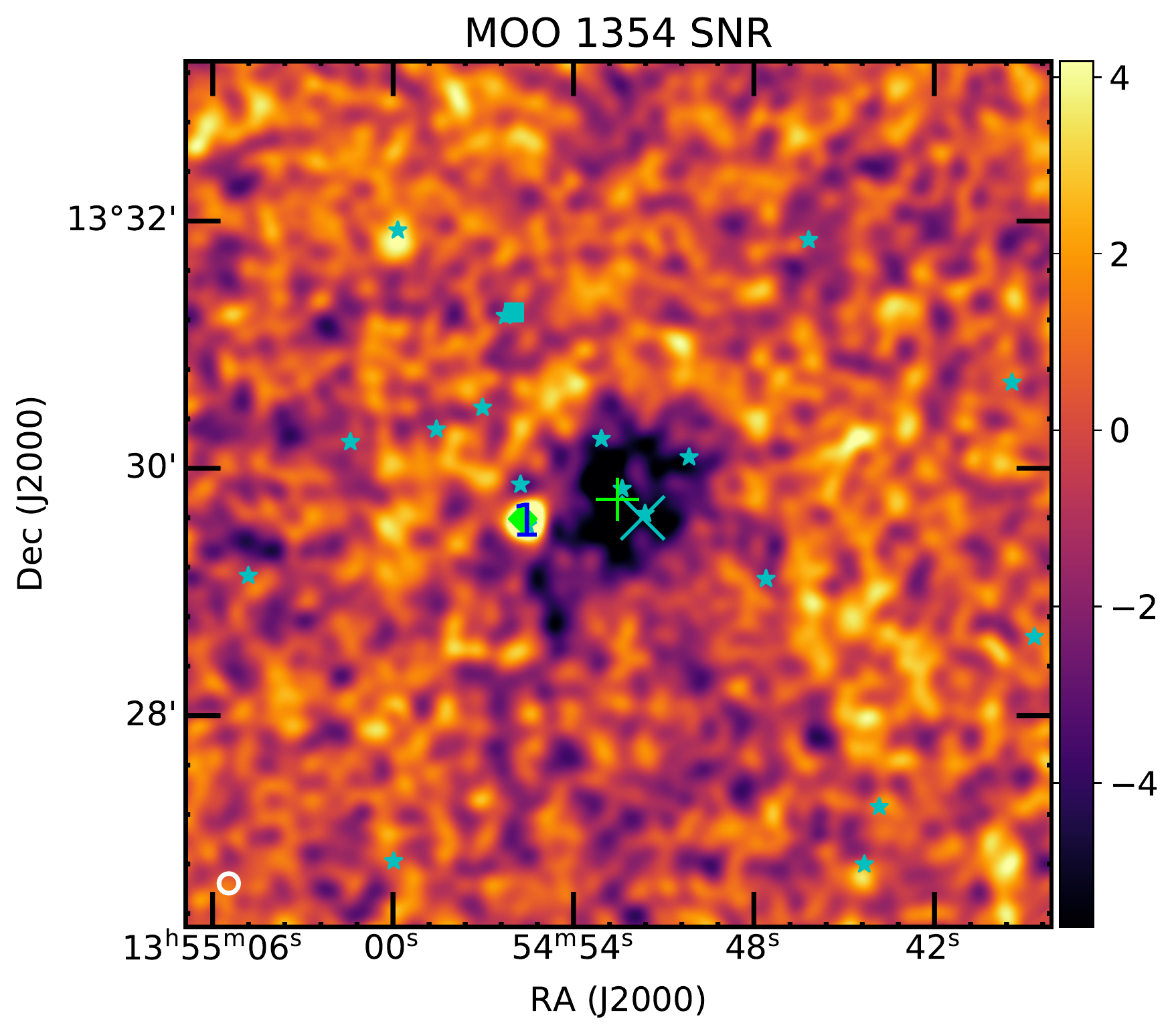} & 
\includegraphics[width=0.3\textwidth]{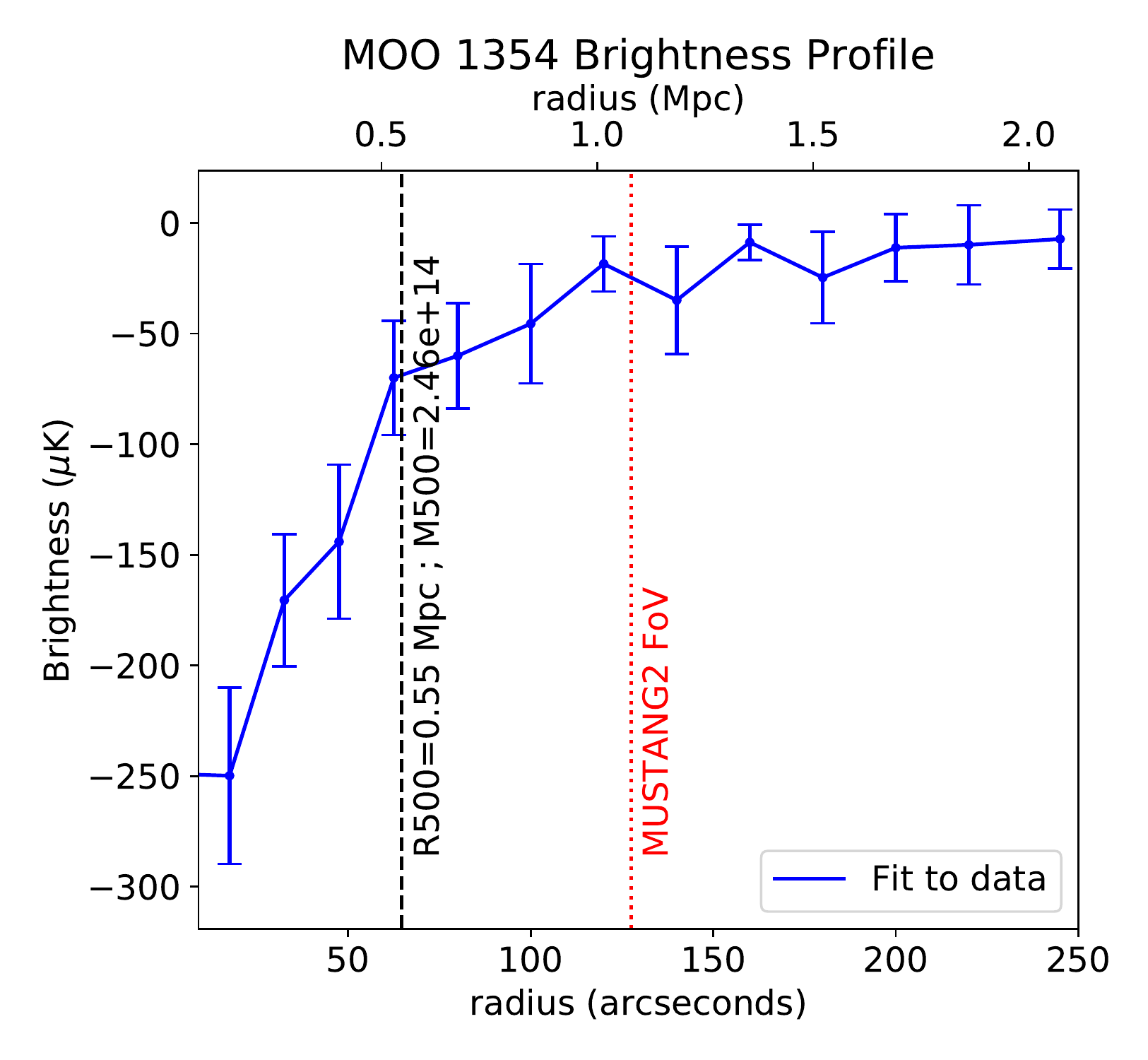} &
\includegraphics[width=0.3\textwidth]{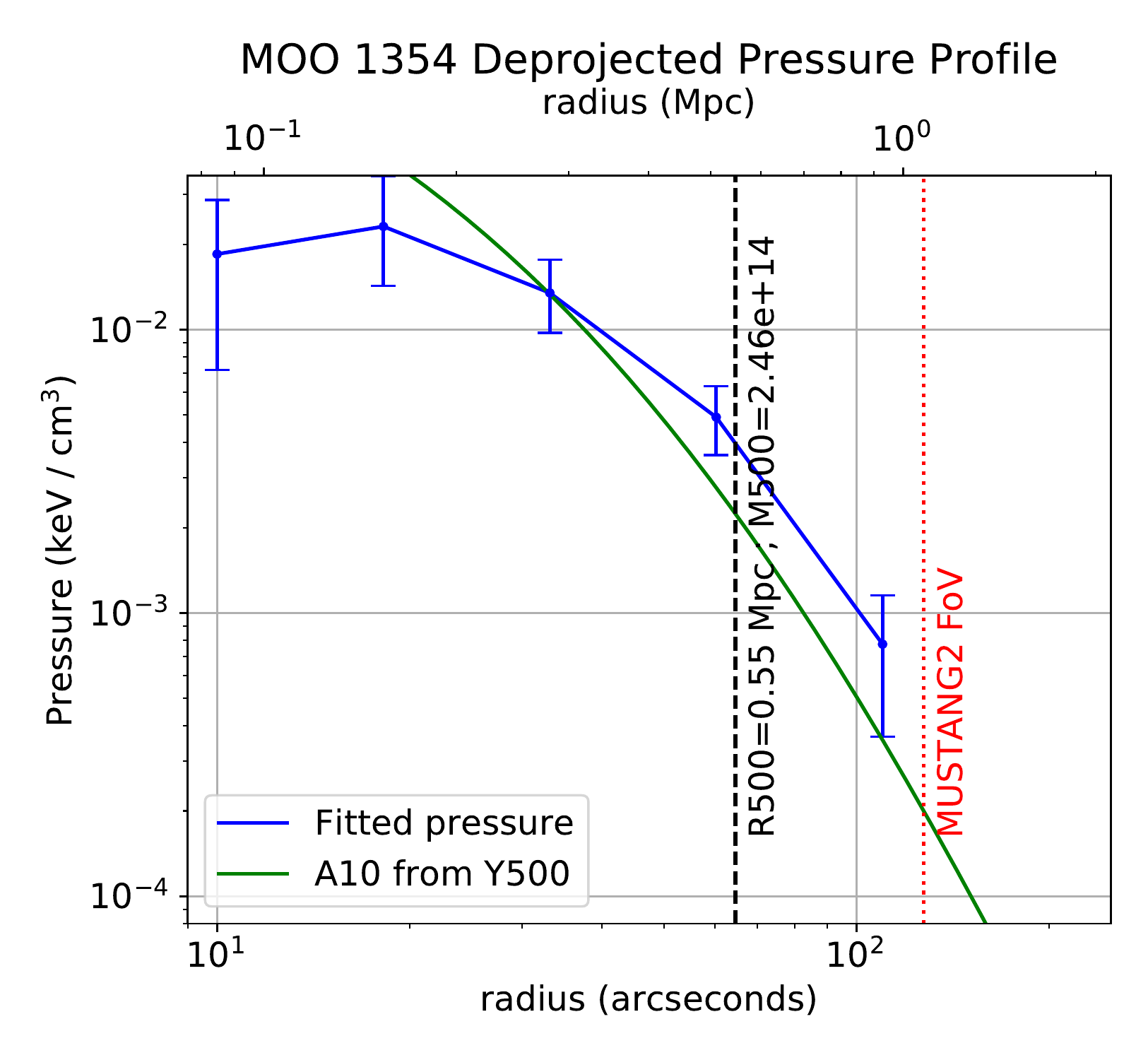}
\end{tabularx}
\begin{tabularx}{\textwidth}{ccc}
\includegraphics[width=0.3\textwidth]{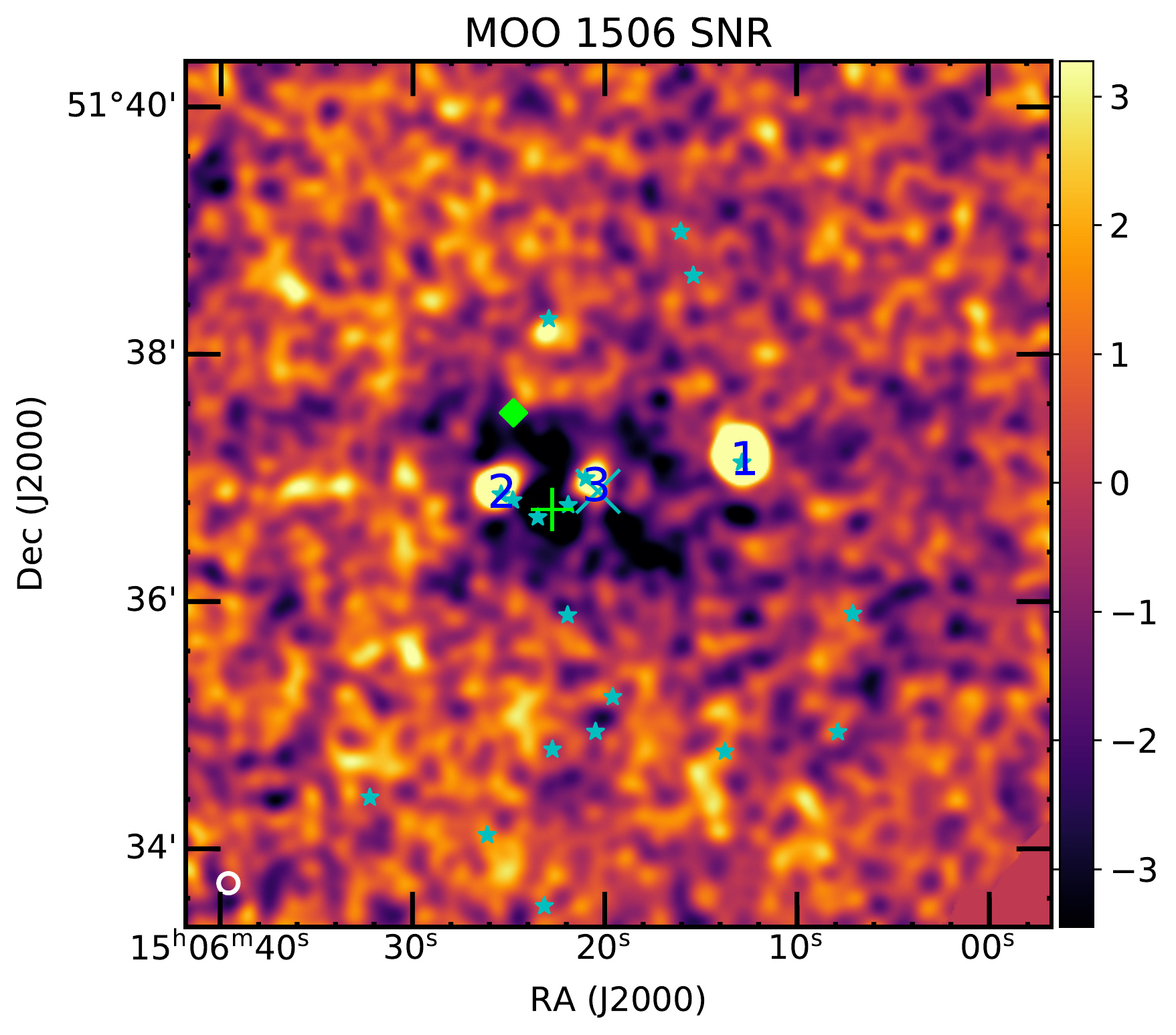} & 
\includegraphics[width=0.3\textwidth]{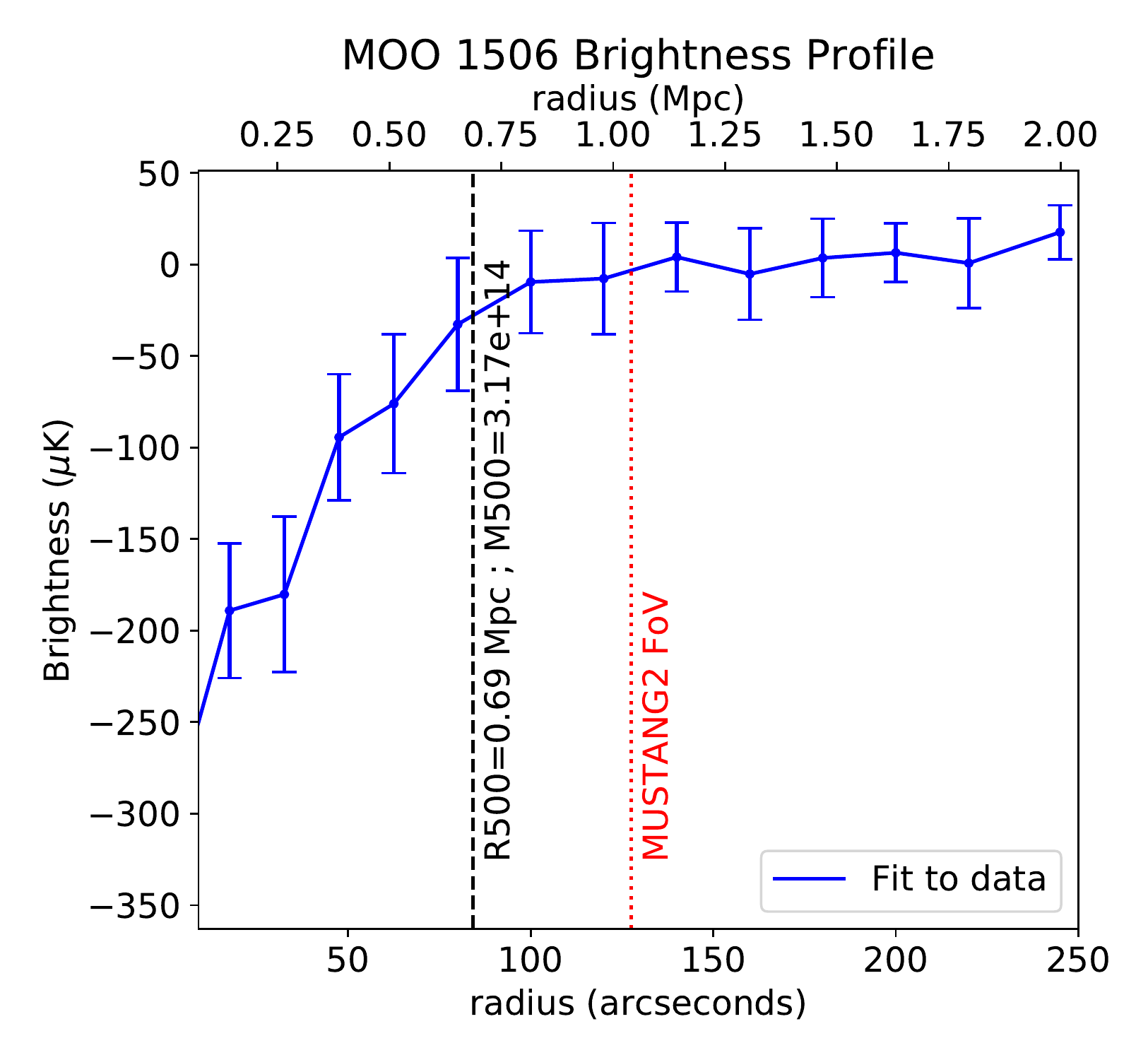} &
\includegraphics[width=0.3\textwidth]{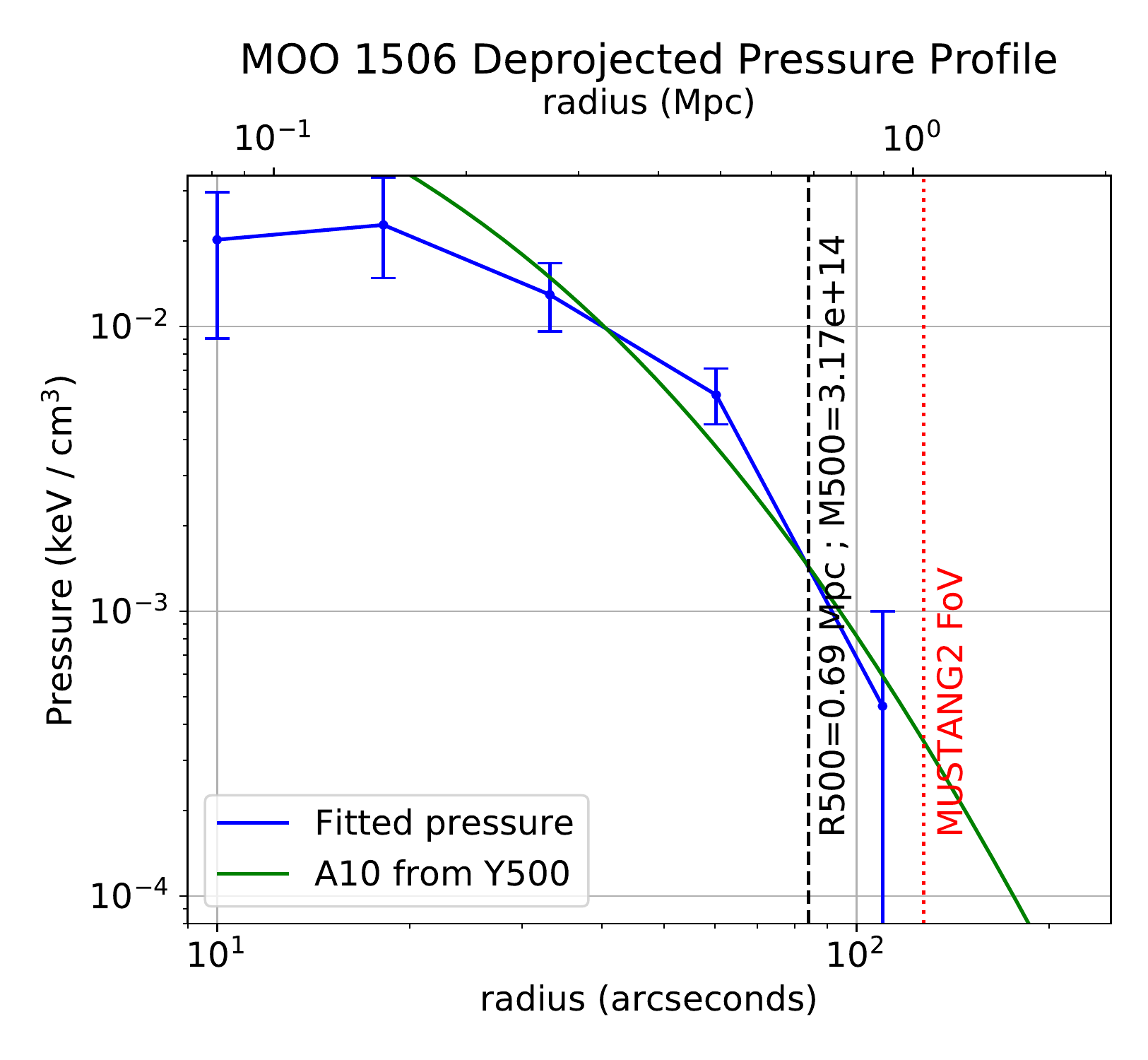}
\end{tabularx}
%{{\bf MOO 0105}}
\caption{(Continued)  {\bf Left:} MUSTANG-2 images made using the MIDAS pipeline.  A cyan X marks the original center found by MaDCoWS, while the green cross marks the best fit SZE centroids.  Cyan stars and squares mark the locations of bright galaxies detected by \textit{Spitzer} and SDSS, respectively.  The BCG is marked as a green diamond and the MUSTANG2 beam is shown as a white circle on the lower left. The bright sources are clipped at $+4\sigma$ and are labeled with numbers to match Table~\ref{tab:pntSrc}. {\bf Center:} Brightness profiles of our clusters from Minkasi.\@ {\bf Right:} Pressure profiles derived from each data set.  The MUSTANG2 FoV is marked as a red line, while the black dashed line represents the $R_{500}$ for our recovered mass. \red{The A10 profile that corresponds to this mass is shown in green}.}
%\caption{Continued from Figure \ref{fig:results}.}
\end{center}
\end{figure*}

%fitted point sources
\begin{table}[htb]
    \begin{center}
    \caption{Fitted point source locations \& amplitudes.}% Fitting was carried out simultaneously using the Minkasi pipeline.}
    \label{tab:pntSrc}
    \begin{tabular}{cccc}
     \hline\hline  Source ID   & RA & Dec (J2000) & Flux/mJy\\ \hline %Kelvin values
    MOO 0105:1  &	01:05:34.19     & +13:23:06.7	&   $0.493\pm0.087$ \\\hline %$0.704\pm0.124$	\\\hline
    MOO 1031:1  &	10:32:04.10     & +62:53:29.5	&   $0.387\pm0.038$ \\\hline %$0.580\pm0.057$	\\\hline
    MOO 1052:1  &	10:52:14.05     & +08:24:54.2	&   $0.511\pm0.064$ \\\hline %$0.759\pm0.095$	\\\hline
    MOO 1054:1  &   10:54:59.33     & +05:01:09.0	&   $5.097\pm0.277$ \\ %$7.488\pm0.260$	\\
    MOO 1054:2  &	10:54:40.61     & +05:07:36.3	&   $0.679\pm0.150$ \\\hline %$0.997\pm0.220$	\\\hline
    MOO 1108:1  &	11:08:53.50     & +32:45:04.9	&   $0.121\pm0.026$ \\\hline    %$0.166\pm0.036$	\\\hline
    MOO 1110:1  & 	11:11:14.79     & +68:38:51.9	&   $0.173\pm0.029$  \\\hline    %$0.251\pm0.042$	\\\hline
    MOO 1142:1  & 	11:42:47.48     & +15:27:12.4	&   $4.051\pm0.035$ \\\hline %$0.568\pm0.049$	\\\hline
    MOO 1322:1  &	13:23:05.98     & $-$02:27:21.2	&   $0.353\pm0.096$ \\\hline %$0.492\pm0.133$	\\\hline
    MOO 1354:1  &   13:54:55.57     & +13:29:33.1	&   $0.585\pm0.068$ \\\hline    %$0.923\pm0.107$	\\\hline
    MOO 1506:1  &   15:06:12.72     & +51:37:07.7	&   $3.489\pm0.064$ \\  %$4.815\pm0.089$	\\
    MOO 1506:2  &	15:06:25.34     & +51:36:51.5	&   $0.729\pm0.052$ \\  %$0.993\pm0.073$	\\
    MOO 1506:3  &	15:06:20.40     & +51:36:55.1	&   $0.553\pm0.063$ \\  %$0.763\pm0.087$	\\
    MOO 1506:4  &	15:05:55.55     & +51:36:23.8	&   $1.284\pm0.149$ \\\hline    %$1.772\pm0.205$	\\\hline
    \end{tabular}
    \end{center}
    NOTE: The number after each source ID refers to the source number designation in Figure~\ref{fig:results}.  Source identifications are grouped by cluster field.  A few sources are close to the edge of the maps and are not plotted in Figure~\ref{fig:results}. 
\end{table}

%fitted cluster centers.
\begin{table}[htb]
\begin{center}
    \caption{Fitted SZE centers.}% Fitting was carried out simultaneously using the Minkasi pipeline.}
    \label{tab:centers}
    \begin{tabular}{cccl}
     \hline\hline Cluster ID   & RA (J2000) & Dec (J2000) & SZE Offset ($''$) \\ \hline
    MOO 0105     &	01:05:31.30 &  $+$13:24:00.1	& (12.8,-1.8)	\\
    MOO 0135	 &	01:35:02.44 &  $+$32:07:41.4	& (-28.0,14.2)	\\
    MOO 1014     &	10:14:07.18 &  $+$00:38:18.2	& (-4.7,-12.0)	\\
    MOO 1031     &	10:32:01.56 &  $+$62:54:53.9	& (200.0,-36.6)	\\
    MOO 1046    &	10:46:52.63 &  $+$27:58:05.0	& (-2.8,2.1)	\\
    MOO 1052    &	10:52:20.31 &  $+$08:22:11.4	& (75.1,-101.6)	\\
    MOO 1054    &	10:54:58.18 &  $+$05:05:34.8	& (32.6,-4.2)	\\
    MOO 1059    &	10:59:52.21 &  $+$54:55:15.2	& (20.7,16.8)	\\
    MOO 1108    &	11:08:47.56 &  $+$32:43:43.3	& (-6.6,7.5)	\\
    MOO 1110    &	11:10:55.29 &  $+$68:38:33.9	& (-27.9,3.2)	\\
    MOO 1142    &	11:42:46.12 &  $+$15:27:17.4	& (9.1,2.0)	\\
    MOO 1203    &	12:03:06.33 &  $-$09:08:32.2	& (-10.0,40.8)	\\
    MOO 1322    &   13:22:57.87 &  $-$02:27:57.4	& (23.5,17.6)	\\
    MOO 1329    &	13:29:50.73 &  $+$56:48:02.6	& (41.0,23.6)	\\
    MOO 1354    &	13:54:52.54 &  $+$13:29:44.9	& (12.6,8.9)	\\
    MOO 1506    &	15:06:22.74 &  $+$51:36:44.9	& (35.8,-8.7)	\\\hline
    \end{tabular}\end{center}
    NOTE: Offsets are given separately in RA and Dec from the MaDCoWS galaxy overdensity center.
\end{table}

%cluster masses and richnesses
\begin{table*}[htb]
    \begin{center}
    \caption{\label{tab:mass_richness}Cluster richness, masses, and size}
    \begin{tabular}{llcccccl}\hline\hline
        Cluster ID  & MaDCoWS ID & Richness\tablenotemark{a} & $M_{500_c}$ (CARMA)\tablenotemark{a} & $M_{500_c}$ (M2)\tablenotemark{b} & $Y_{500} $ & $R_{500}$ & Notes \\ % d=disturbed n=not disturbed l=lowsnr
         && $\lambda_{15}$& ($10^{14}\mbox{M}_\odot$) & ($10^{14}\mbox{M}_\odot$)    & $(\mbox{Mpc}^{2}$)     & (Mpc) \\ \hline
        MOO 0105 & MOO J0105+1323   &   $87\pm 10$ & $3.9\pm 0.5~$   & $3.83^{+0.23,0.13,0.28}_{-0.24,0.12,0.26}$    &$2.77\times 10^{-5}$&  0.72 & merger\tablenotemark{c}\\ % d 
        MOO 0135 & MOO J0135+3207   &   $39\pm 6~$  & \ldots         & $1.82^{+0.31,0.07,0.13}_{-0.31,0.07,0.12}$    &$0.74\times 10^{-5}$&  0.50 \\ % r
        MOO 1014 & MOO J1014+0038   &   $44\pm 7~$  & $3.2\pm0.35$   & $3.12^{+0.16,0.10,0.23}_{-0.15,0.10,0.24}$    &$1.93\times 10^{-5}$&  0.65 \\ % r ACT 3.1+-0.4
        MOO 1031 & MOO J1031+6255   &   $50\pm 7~$  & \ldots         & $0.67^{+0.26,0.05,0.05}_{-0.26,0.05,0.05}$    &$0.13\times 10^{-5}$&  0.37 & low SNR\\ % l
        MOO 1046 & MOO J1046+2757   &   $52\pm 7~$  & \ldots         & $2.00^{+0.21,0.07,0.14}_{-0.23,0.07,0.14}$    &$0.87\times 10^{-5}$&  0.57    & flat profile\\ % r
        MOO 1052 & MOO J1052+0823   &    $42\pm 6~$  & \ldots         & $1.90^{+0.31,0.07,0.14}_{-0.35,0.07,0.13}$    &$0.80\times 10^{-5}$&  0.51 \\% r
        MOO 1054 & MOO J1054+0505   &    $42\pm 6~$  & \ldots         & $1.34^{+0.33,0.07,0.10}_{-0.34,0.06,0.09}$    &$0.43\times 10^{-5}$&  0.45 \\% r
        MOO 1059 & MOO J1059+5454   &    $57\pm 7~$  & \ldots         & $2.54^{+0.06,0.08,0.19}_{-0.06,0.08,0.17}$    &$1.34\times 10^{-5}$&  0.63 & flat profile \\ %r
        MOO 1108 & MOO J1108+3242   &    $63\pm 8~$  & \ldots         & $2.41^{+0.19,0.08,0.16}_{-0.20,0.08,0.16}$    &$1.22\times 10^{-5}$&  0.62 \\ %r
        MOO 1110 & MOO J1110+6838   &    $55\pm 7~$  & \ldots         & $2.02^{+0.16,0.07,0.15}_{-0.16,0.07,0.14}$    &$0.89\times 10^{-5}$&  0.63 \\ %r
        MOO 1142 & MOO J1142+1527   &    $58\pm 8~$  & $5.7\pm0.5~$   & $3.52^{+0.19,0.11,0.26}_{-0.19,0.11,0.24}$    &$2.39\times 10^{-5}$&  0.68 & merger \\ %r ACT 4.37 +- 0.64
        MOO 1203 & MOO J1203$-$0909   &    $56\pm 7~$  & \ldots         & $0.64^{+0.26,0.05,0.05}_{-0.26,0.05,0.05}$ &  $0.11\times 10^{-5}$ &   0.37 &low SNR\\ %l
        MOO 1322 & MOO J1322$-$0228   &    $83\pm 9~$   & \ldots        & $3.07^{+0.41,0.10,0.22}_{-0.53,0.09,0.21}$    &$1.88\times 10^{-5}$&  0.75 \\ %r ACT 2.9+-0.5
        MOO 1329 & MOO J1329+5647   &    $42\pm 6~$  & \ldots         & $3.56^{+0.20,0.12,0.26}_{-0.20,0.11,0.24}$    &$2.43\times 10^{-5}$&  0.63    & flat profile\\ %r 
        MOO 1354 & MOO J1354+1329   &    $44\pm 6~$  & \ldots         & $2.46^{+0.25,0.08,0.18}_{-0.30,0.08,0.17}$    &$1.26\times 10^{-5}$&  0.55 \\ %r ACT 1.83+-0.3
        MOO 1506 & MOO J1506+5136   &    $74\pm 8~$  & \ldots         & $3.17^{+0.29,0.10,0.22}_{-0.29,0.09,0.21}$ &$1.98\times 10^{-5}$& 0.69 & merger\tablenotemark{c}\\ %d
            \hline
    \end{tabular}
    \end{center}
    \tablenotetext{a}{Richness and mass values from \citet{Gonzalez2019}. More information on MOO~1506 can be found in \citet{Moravec2020}.}
    \tablenotetext{b}{The errors on the MUSTANG2 masses are, from left to right, the statistical errors, errors due to the Y-M relationship, and errors due to the absolute calibration of MUSTANG2 data. }
    \tablenotetext{c}{\red{Merger status of MOO~0105 \& MOO~1506 are based on x-ray morphologies in Chandra imaging, \citep{Gonzalez2019} and MOO~1142 is claimed to be a merger in \citet{Ruppin2020}.}}
    %The masses given include the corrections from our simulation pipeline, including systematic errors (Section~\ref{sec:sims}).}
    %\tablenotetext{c}{MOO 1031 and MOO 1203 are marginal detections.}
    %\textcolor{red}{TBD - fill in masses/references - sort out just what we want to go here so as to make it fit.}
    %\comment{I think we should report $Y_{500}$ here using both $r_{500}$ from the SZE fits and from the mass-richness relation.}
    %\tablenotetext{b}{where this mass came from.....}
\end{table*}

\section{Simulations}\label{sec:sims}
In order to confirm our ability to recover masses and pressure profiles, we performed simulations of our observations and analysis pipeline.  Complete end-to-end tests were carried out by creating fake cluster profiles at known redshifts between 0.7 and 1.4 (to cover the range of redshifts in our sample) and masses ranging from $M_{500}=10^{14}$ to $10^{15}~\mbox{M}_\odot$ (corresponding to our expected mass range).  Three different generalized Navarro-Frenk-White profiles %\citep{Navarro1997,nagai2007} 
were used as inputs. These were taken from \citet{arnaud2010}: an ensemble average ``universal'' profile, the average profile found to fit cool core clusters, and the average profile found to fit disturbed clusters. Each profile was convolved with a $10''$ beam, then fake timestreams were generated by sampling these maps using the real observational scanning patterns used on the MaDCoWS clusters (Figure~\ref{fig:scan}).  Noise was added by taking MUSTANG2 timestreams from other projects that had observed blank fields and the data were analyzed using the same steps described in Section~\ref{sec:dataReduction}.

Initial simulations used the telescope pointings/scans for MOO~0105. These showed good recovery of the surface brightness profiles, the cluster pressure profile, and the cluster mass for all redshifts, masses, and profiles chosen.  However, when the same tests were carried out using the telescope pointings used for the blank fields the noise timestreams were taken from, less than half the original mass values were recovered and the surface brightness profiles showed significant errors of the order of $500~\mu $K.  Further investigation showed that this was due to residual atmospheric gradients -- when timestreams from one set of scans were used as noise in another, the phase of the atmospheric signal no longer matched that of the telescope scan pattern in elevation and was easily rejected as noise. When the same timestreams were used with the original telescope scan pattern and maps made in elevation/cross-elevation then a residual atmospheric signal of the order of 100~$\mu$K remained in some maps. Due to sky rotation, maps made in RA/Dec on a single pointing center sometimes produced circular features of the order 4$'$ in diameter.  To mitigate this, a second order polynomial in the elevation direction around the scan center was fit for and subtracted from each scan, after which our simulations showed accurate mass recovery for single pointing observations. \red{These extra fit parameters (one set of second order polynomial coefficients for each scan) were added to parameters in the Minkasi surface brightness fitting procedure described in Section~\ref{sec:profiles} and used on all our real data}. To make this more robust, in later observations, instead of a single pointing centered on the cluster four separate pointing centers around each cluster were used, each offset by $\pm 1.5'$ in RA or Dec.  Although this resulted in slightly less integration time on source, the improved ability to reject atmospheric noise more than made up for this. 

As well as testing the recovery of cluster mass with different  simulated cluster shapes, our sensitivity to errors in finding the cluster centers was tested.  It was found that manually adding offsets of 30$''$ to the centers of the clusters found by the first step of our pipeline had a negligible effect on the masses recovered.  Also, changing the range of the fits between 180$''$ and 240$''$ changed the recovered masses by only a small fraction of the measurement error. % In other tests, masses for clusters with significant ($>10\sigma$) point sources were calculated without fitting for these sources.  In the case of MOO~1142 where the source was over 4~mJy this made only a 6\% difference in the recovered mass.

\section{Results}\label{sec:results}
Maps produced using the IDL pipeline are shown in Figure~\ref{fig:results} along with pressure fits from Minkasi.  Masses derived from these fits are shown in Table~\ref{tab:mass_richness} along with derived values for $Y_{500}$ and $R_{500}$.  Out of our initial sample of 16 clusters, 14 show significant detections of the ICM.  For the remaining two (MOO~1031 and MOO~1203), the noise in the maps is similar to that for the other clusters, allowing us to place strong upper limits on the masses ($\leq1.3\times10^{14}~\mbox{M}_\odot$ at $3\sigma$) and there is a weaker (3.7$\sigma$) detection of the ICM. 

Of the three clusters in this paper with both MUSTANG2 and CARMA mass measurements, two (MOO~0105 and MOO~1014) are in good agreement (see  Figure~\ref{fig:massRichness}). However the MUSTANG2 mass for MOO~1142 is almost 40\% below the CARMA value.  %\citep[which is in agreement with X-ray measurements from][]{Ruppin2020}. % We note that such discrepancies are not uncommon in the literature \citep[for example, see][]{Romero2020}.
 Major differences between the CARMA and MUSTANG2 measurements are CARMA's $37''$ resolution and that the CARMA masses were obtained by directly fitting an A10 model to the interferometric observations, while in this paper the initial fit is non-parametric.  Also \citet{Gonzalez2015} found a $41''$ offset between their SZE center and the MaDCoWS center while with MUSTANG2 this offset was less than $10''$. One possible explanation for the difference in recovered masses is that MOO~1142 has two halos, and that MUSTANG2 has only fit for the largest, though no second halo is apparent in Figure \ref{fig:results} there are hints of a bimodal distribution of the {\it WISE} galaxy densities in (Figure~\ref{fig:spizteroverlays} in Appendix~\ref{sec:scaling_app}).  \red{\citet{Ruppin2020} recently published observations of this cluster using NIKA2 (18$''$ resolution at 150~GHz). Using X-ray data and the galaxy distribution they conclude that this cluster is an early stage merger (before first core passage). A similar combined analysis of the MUSTANG2 data will be investigated in future work.}
 %another is that the lower resolution of the CARMA data and the larger offset have biased the CARMA fit. Resolution of the true reason for this discrepancy (including the X-ray data) will require further analysis, possibly including additional data sets from experiments such as ACTpol. %Deeper observations would be needed to test this hypothesis.  

The fitted SZE centers of the clusters are given in Table~\ref{tab:centers}.  Both the  fitted SZE centers from this paper and the MaDCoWS centers from \citet{Gonzalez2019} have formal errors less than $21''$.  For most of the clusters, the differences between these centers are less than $1'$ so they are consistent with each other. The exceptions are MOO~1031 (although this cluster is low SNR) and MOO~1052. Subsequent re-examination of the {\it WISE} galaxy density for MOO~1052 shows a second peak in the galaxy density closer to the fitted SZE center (see Figure~\ref{fig:spizteroverlays} in Appendix~\ref{sec:scaling_app}).  This could indicate an on-going merger or it could be the result of contamination by foreground/background galaxies.   Follow-up studies (e.g. X-ray  or deeper SZE observations, or to obtain gravitational lensing and galaxy dynamics) would be of interest.

When compared to an A10 pressure profile, many of the clusters in Figure~\ref{fig:results} show a shallower slope with the pressures being systematically higher above a radius of 100$''$.   At masses around $10^{14}~\mbox{M}_\odot$ (where detections are marginal), the simulations typically recovered shallower profiles.  However, at larger masses, the simulations showed that our pipeline recovered the correct profiles.  Of our observed clusters with a SNR over $6\sigma$, MOO~1046, MOO~1059, and MOO~1329 have a significantly shallower profile than A10, possibly indicating disturbance in the ICM or a possible merger.  

Another feature in some of our recovered profiles is significantly lower pressure in the central $r \sim 25''$. MOO~0105, MOO~0135, MOO~1108, MOO~1354, and MOO~1506 are the five strongest examples. Again, this was not seen in recovered profiles from simulations and tests to see if using the incorrect beam size to recover the profiles could not reproduce this effect.  The only way it was reproduced in simulations was by the introduction of large ($>30''$) errors in the fitted cluster centers. The maps in Figure~\ref{fig:results} clearly show that the SZE centers match up to the decrements far better than this. However, looking at these maps, these clusters also show features at their centers.  Active AGNs, low significance sources of any type, or a disturbed profile due to an ongoing merger could all cause such features.  Comparison with other data sets will be explored in future papers.% but our tests in Section~\ref{sec:sims} showed that disturbed profiles or point sources do not significantly affect the recovered cluster masses.  

\section{The mass richness relationship}\label{sec:mass-richness}

%\begin{figure*}[tbh]
%    \centering
%    \includegraphics[clip, trim=0.50cm 0.05cm 0.85cm 0.49cm, width=0.496\textwidth]{Mass_richness/new/test_1_must2+carma+aca_no-merger_no-nosign_big_lab.pdf}
%    \includegraphics[clip, trim=0.50cm 0.05cm 0.85cm 0.49cm, width=0.496\textwidth]{Mass_richness/new/test_2_must2+carma+aca_wt-merger_wt-nosign_big.pdf}
%    \caption{SZE-inferred masses, using data from three instruments/observatories that have targeted samples of 10 or more MaDCoWS clusters, plotted against the richness values reported in \citet{Gonzalez2019}. The gray, yellow and blue points are from CARMA, the ACA, and MUSTANG2 (this paper) respectively. Low significant points are indicated by triangles, and the known mergers are depicted as squares. The circles represent all other clusters. Points included in the fits on each plot are solid while those that are not are empty.  All error bars are $1\sigma$. The best fit relations are shown as dashed lines and the shaded areas represent the errors in the fits. Grey shaded areas are fits to just the CARMA points while the blue shaded areas include all filled symbols on the plots.  In the {\bf left panel} the fit excluded known mergers and low-significance ($<4\sigma$) detections while in the {\bf right panel} these points are included.  Plots of other fits and the fit parameters can be found in Appendix~\ref{sec:scaling_app}.}
%    \label{fig:massRichness}
%\end{figure*}

%%%%%%%%%%%%%%%%%%%%%%%%%%%%%%%%%%%%%%%%%%%%%%%%%%%%%%%%%%%%
% new plots

\begin{figure*}[tbh]
    \centering
    \includegraphics[width=\textwidth]{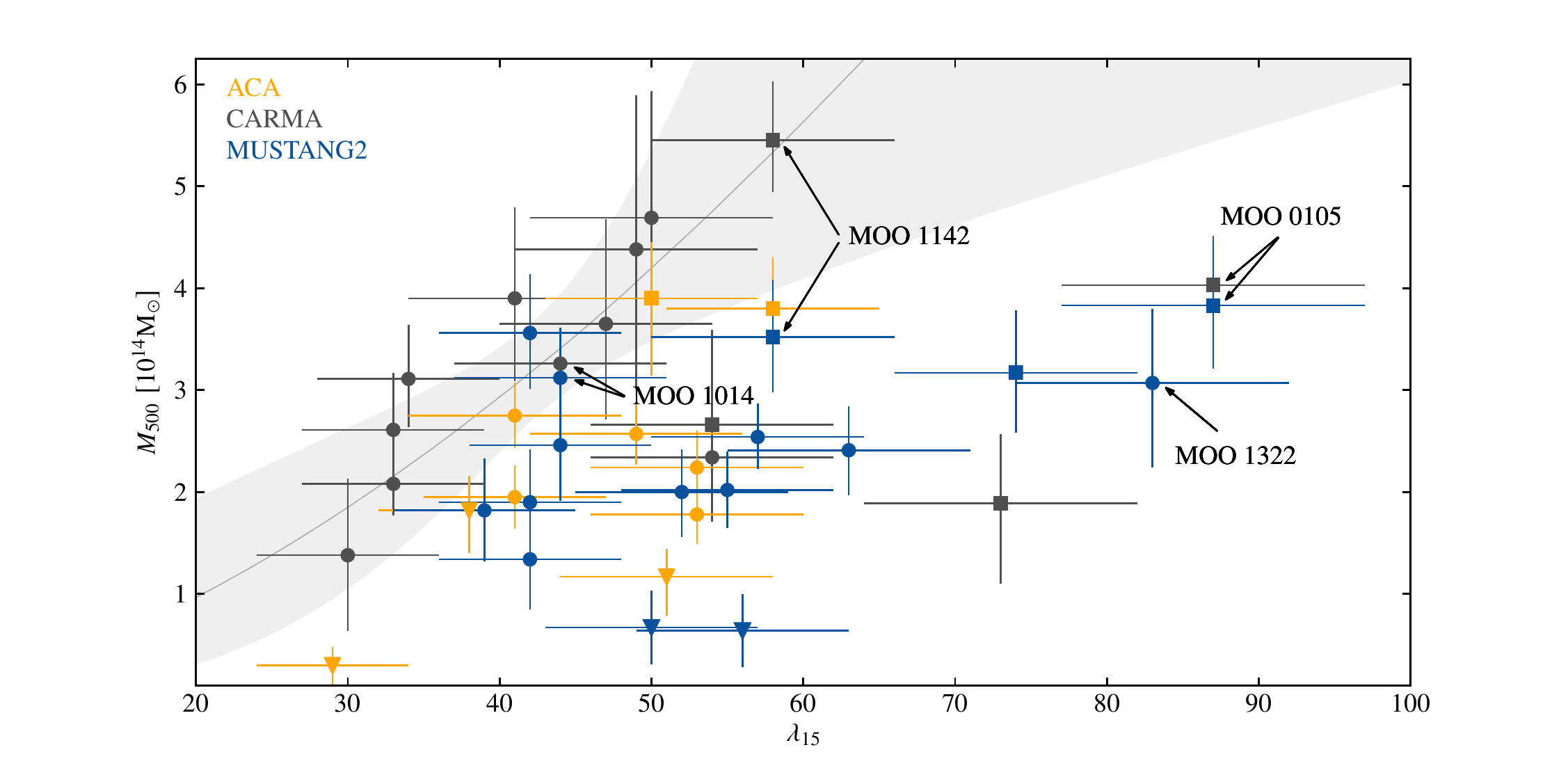}
    \caption{SZE-inferred masses from three instruments/observatories that have targeted samples of 10 or more MaDCoWS clusters plotted against their richness reported in \citet{Gonzalez2019}. The black, yellow and blue points are from CARMA, the ACA, and MUSTANG2 (this paper) respectively. The mass-richness relation (with errors) that was fitted to the CARMA data in \citet{Gonzalez2019} is shown as the gray line and shaded region. Low significant points are indicated by triangles, and the known mergers \red{(from \citealt{Gonzalez2019,Ruppin2020})} are depicted as squares. The circles represent all other clusters. All errors are $1\sigma$.}
    \label{fig:massRichness}
\end{figure*}

\begin{figure}[tbh]
    \centering
    \includegraphics[width=\columnwidth]{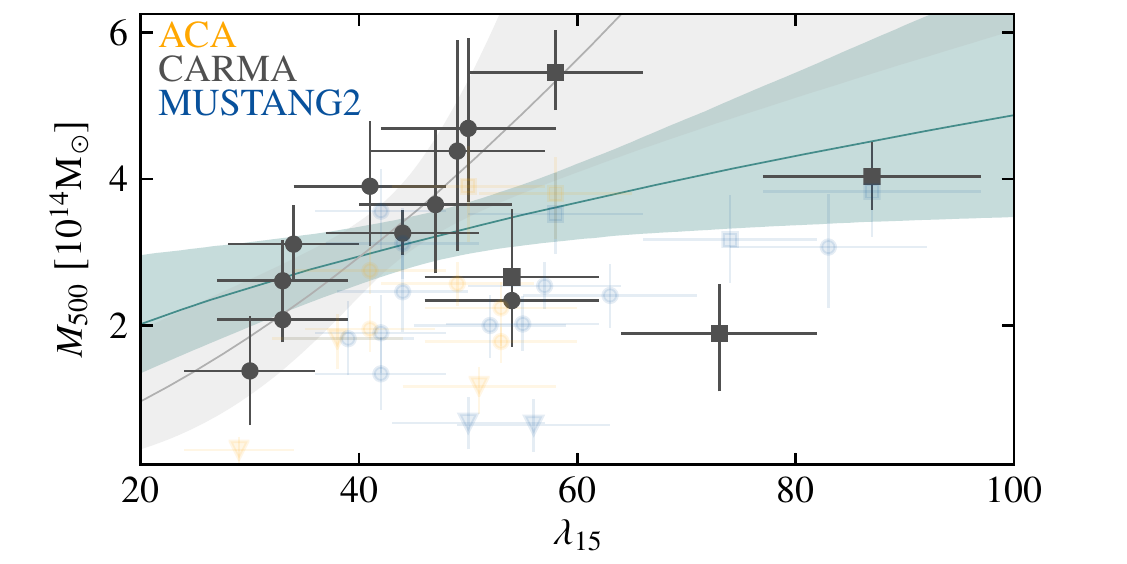}\\
    \includegraphics[width=\columnwidth]{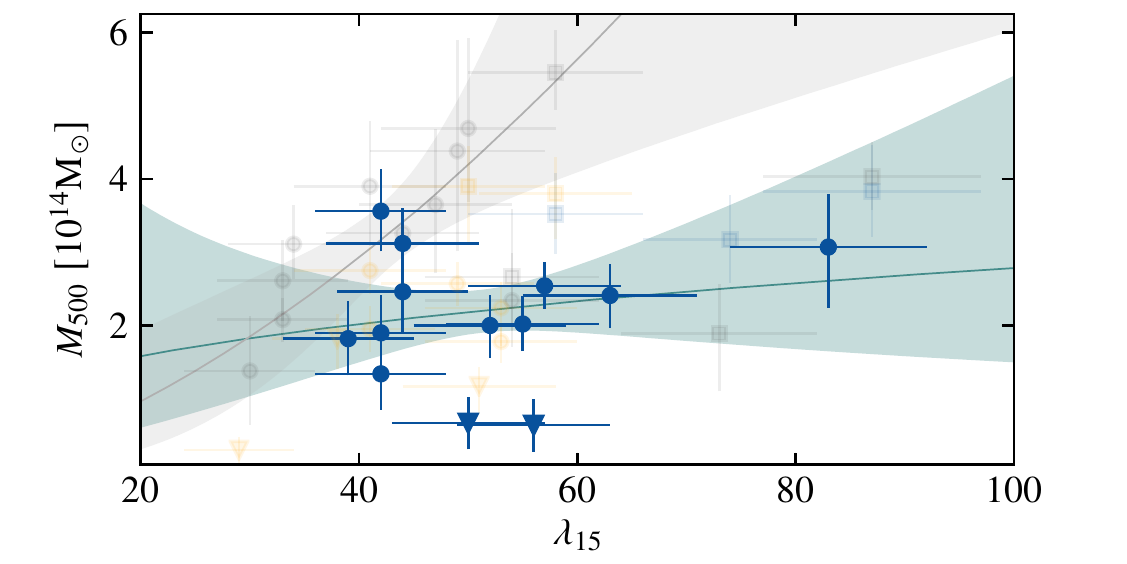}\\
    \includegraphics[width=\columnwidth]{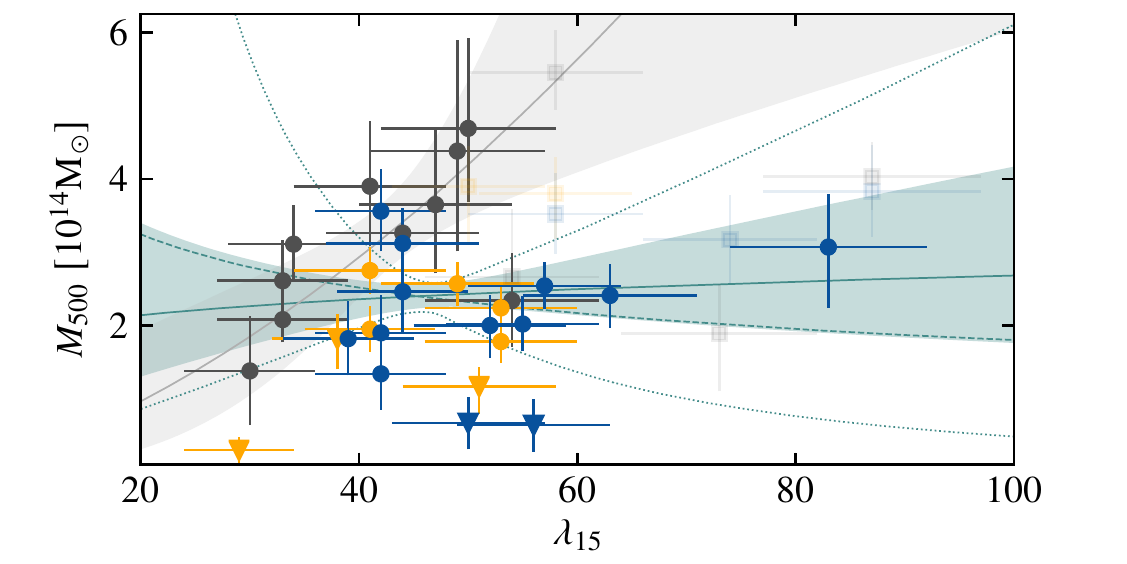}
    \caption{Mass-richness scaling relations obtained from different data sets. The data included/excluded from each of the fits are denoted with bold/faint markers (we refer to Figure~\ref{fig:massRichness} for more details about the notation). The best fit relations are shown as green solid lines and the shaded areas represent the errors in the fits. As in Figure~\ref{fig:massRichness}, the gray line and shaded region refer to the mass-richness relation reported in \citealt{Gonzalez2019}. {\bf Top:} The effects of including known mergers in the CARMA fit. {\bf Middle:}  A fit to the MUSTANG2 data excluding known mergers.  The dashed and dotted lines denote the best-fit relation and corresponding credibility interval obtained when excluding MOO 1322 (the richest non-merging cluster in the MUSTANG2 MaDCoWS pilot sample). {\bf Bottom:} A fit to all the data excluding known mergers.}
    \label{fig:massRichness2}
\end{figure}

%%%%%%%%%%%%%%%%%%%%%%%%%%%%%%%%%%%%%%%%%%%%%%%%%%%%%%%%%%%%

Optical surveys for clusters of galaxies often employ some form of richness measure as a proxy for cluster mass \citep[e.g.][]{abell1958,Rykoff2012,Andreon2015,Saro2015,Geach2017,Simet2017,Rettura2018,Chiu2020}.  For MaDCoWS, \citet{Gonzalez2019} defined the measure $\lambda_{15}$ to be the excess number density of galaxies selected by \textit{Spitzer} color cuts as possible cluster members with a brightness cut-off of $15~\mu$Jy.  This flux cut-off corresponds to a stellar mass of $\sim 5\times 10^{10}~\mbox{M}_\odot$.  An aperture of 1~Mpc in radius was used for the calculation.  To calibrate the $M_{500}-\lambda_{15}$ scaling relation, masses determined using 14 SZE observations from CARMA were used.  The CARMA sample included MOO~0105 and MOO~1142, which are also present in the sample studied here using MUSTANG2.  A fit for the relationship between $\lambda_{15}$ and $M_{500}$ with the assumed form:
\begin{equation*}
    \log_{10}\frac{M_{500}}{10^{14}\mbox{M}_\odot} = \alpha \log_{10} \lambda_{15} + \beta + \eta 
\end{equation*}
was made where $\alpha$ and $\beta$ are the slope and intercept of the relation, respectively. The term $\eta$ takes into account any scatter inherent to the data points with respect to the reconstructed linear relation (i.e.\ intrinsic scatter), and is assumed to be a Gaussian random variable with variance $\sigma^2_{\log M |\lambda}$ and a null expectation value.
Clusters with twice as many galaxies can be expected to be approximately twice as massive so values of $\alpha$ close to unity are expected \citep[e.g.][]{AndreonHurn2010,Chiu2020}. The Bayesian linear regression algorithm \texttt{linmix} \citep{Kelly2007} was used for the fit \red{in \citet{Gonzalez2019}. Although \texttt{linmix} does not fully account for scatter in the x parameter (and so can produce biased results), } for consistency, we use the same method but include the MUSTANG2 SZE-inferred masses presented in this paper and those inferred from ACA observations \citep{DiMascolo2020}.

A plot of the mass-richness data can be seen in Figure~\ref{fig:massRichness} and plots of some of the fits can be seen in Figure~\ref{fig:massRichness2}. More plots of fits along with a table of parameters can be found in Appendix~\ref{sec:scaling_app}. \citet{Gonzalez2019} excluded known mergers from their fits and found best fit values of $\alpha=1.65^{+1.45}_{-0.96}$ and $\beta=-2.16^{+1.57}_{-2.38}$ with a large scatter of $\sigma_{\log M |\lambda}=0.12$. % (a significant amount of which comes from MOO~J0037+3306).  
Adding in data from the ACA lowers the slope only slightly (to $\alpha=1.41^{+2.42}_{-1.70}$) but the change in the offset is larger (to $\beta=-1.88^{+2.81}_{-4.01}$) and as noted in \citet{DiMascolo2020}, the scatter increases.  This is consistent with the fact that ACA clusters that are not known mergers or upper limits are spread over a relatively narrow range of richness ($40<\lambda_{15}<55$) so they cannot constrain the slope well.  They also lie below the CARMA data (Figure~\ref{fig:massRichness}). 

A fit to the MUSTANG2 clusters (excluding known mergers MOO~0105, MOO~1142, and MOO~1506 and low SNR clusters MOO~1031, and MOO~1203) has a significantly shallower slope ($\alpha=0.30^{+0.76}_{-0.68}$), a higher intercept ($\beta=-0.13^{+1.16}_{-1.30}$), but comparable scatter to the CARMA/ACA fits.  The inclusion of low SNR clusters has only a small effect on the fitted slopes with the main effect to be an increase in the scatter.  Although including upper limits in the fits increases the scatter, it is important to do so if one wants to avoid biasing results -- for example if a significant number of high richness clusters had far lower SZE masses than expected and were not detected then only fitting the detections would bias the slope high.  Information on non-detections in the CARMA observations is not given in \citet{Gonzalez2019} but the authors caution that a more complete analysis of the CARMA fits to include the non-detections is needed.  If some of these non-detections were high richness clusters with a much lower than expected flux, the exclusion of these data points would bias a fit to the CARMA data to a steeper slope.% than the more complete MUSTANG2 sample.

Figure~\ref{fig:massRichness} shows hints that the population of galaxy clusters could exhibit a break or bimodality, with non-mergers following the steeper CARMA fit, and mergers falling significantly below that relation and closer to the flatter MUSTANG2 fit.  To first order, the number of galaxies above a given flux would be unchanged during a merger so as soon as the two merging clusters are within the line-of-sight radius in which richness is being measured then the richness value should increase to close to the value for the new combined cluster. Richness, including $\lambda_{15}$ used in this paper, should be relatively unaffected by the dynamical state of a post-merger cluster.  The intrinsic scatter in mass-richness relations is less certain with values between 15\% and 50\% being quoted depending on the exact definition of richness \citep[for examples see][]{AndreonHurn2010,Andreon2015}.  The intrinsic scatter in the Y-M relation for the SZE is known to be lower at around 10\%.  However, during a merger individual clusters can vary  by more than this.  Simulations show a brief enhancement in the SZE signal during the first core passage and then less SZE signal by up to 40\% until the gas in the ICM has virialized or thermalized in the merged gravitational potential \citep[][]{Wik2008, Marrone2012}. Empirical studies have generally confirmed this \citep[e.g.][]{Bocquet2015,Hilton2018}. Consequently one would expect that including merging clusters when using the SZE to calibrate a mass-richness scaling relation would result in a more shallow slope \citep[see in addition the discussion of mergers in][]{Moravec2020}.
% but with recent, major mergers appearing below undisturbed clusters and smaller mergers that are partly virialized being between.

To test if undetected mergers are driving the differences between the MUSTANG2 and CARMA slopes, the data were refit to include known mergers.  The result is that the slope of all data sets became similar (0.5--0.9) but still slightly steeper than the slope from MUSTANG2 excluding mergers (0.30). 
We also note that known mergers dominate the high richness end of Figure~\ref{fig:massRichness}, with the exception of MOO~1322, which has a median mass $M_{500}\approx 3.1\times10^{14}~\rm M_\odot$ and MaDCoWS richness $\lambda_{15}=83$. Data points in this region of the plot are driving the fits towards flatter slopes.
From the MUSTANG2 data alone (Figure~\ref{fig:results}), there is no conclusive evidence that MOO~1322 is a merger.  However, in Figure~\ref{fig:spizteroverlays} in Appendix~\ref{sec:opt_images}, we provide \textit{Spitzer}/IRAC color-selected galaxy density maps with the MUSTANG2 SZE decrement contours over-plotted.  Although MOO~1322 has the second highest richness in our sample, it appears not to have a strong galaxy concentration when applying this color cut.  This may be due to the galaxy members being bluer than expected for a virialized system, which would be consistent with the low SZE signal and likely imply this is an unvirialized, line of sight merger. We tested the effect of excluding MOO~1322 along with the known mergers from the scaling relation fits.  Due to the narrow range of richness of the remaining data, this results in much poorer constraints on the slope of the fit to the MUSTANG2 data alone (Figure~\ref{fig:massRichness2}).  Follow-up observations such as optical spectroscopy to verify if this cluster is a merger would be of interest and planned observations of more MaDCoWS clusters using MUSTANG2 over a wider range of richnesses (to fill in this region) will better constrain the scaling relation with or without this cluster. Given how much the MUSTANG2 fit is affected by this one cluster then until such observations are obtained the scaling relationship derived from MUSTANG2 data should be considered preliminary.

\begin{figure*}[th]
    \centering
    \includegraphics[clip, trim=0.0cm 0.2cm 1.5cm 0.3cm, width=0.95\columnwidth]{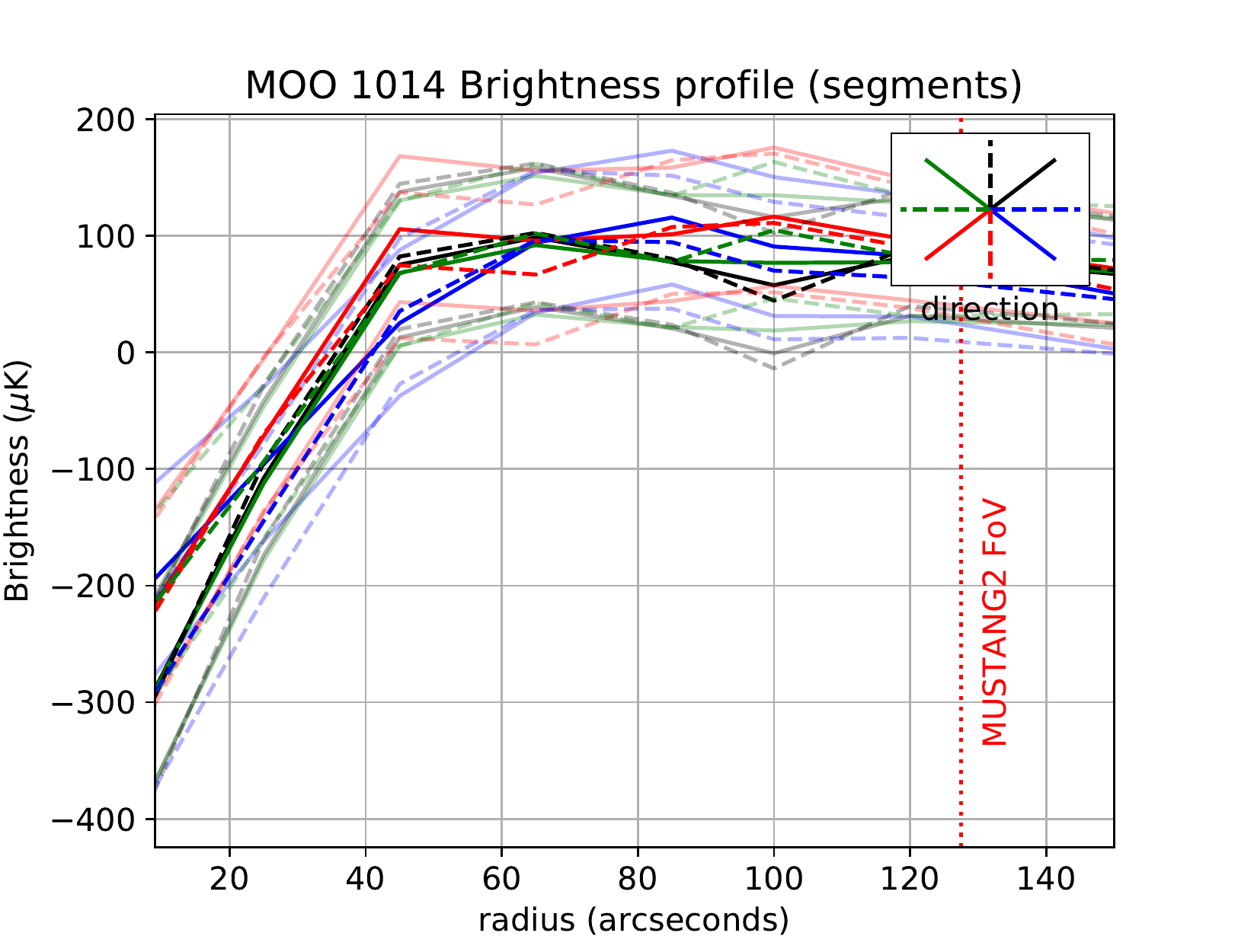}  
    \includegraphics[clip, trim=0.0cm 0.2cm 1.5cm 0.3cm, width=0.95\columnwidth]{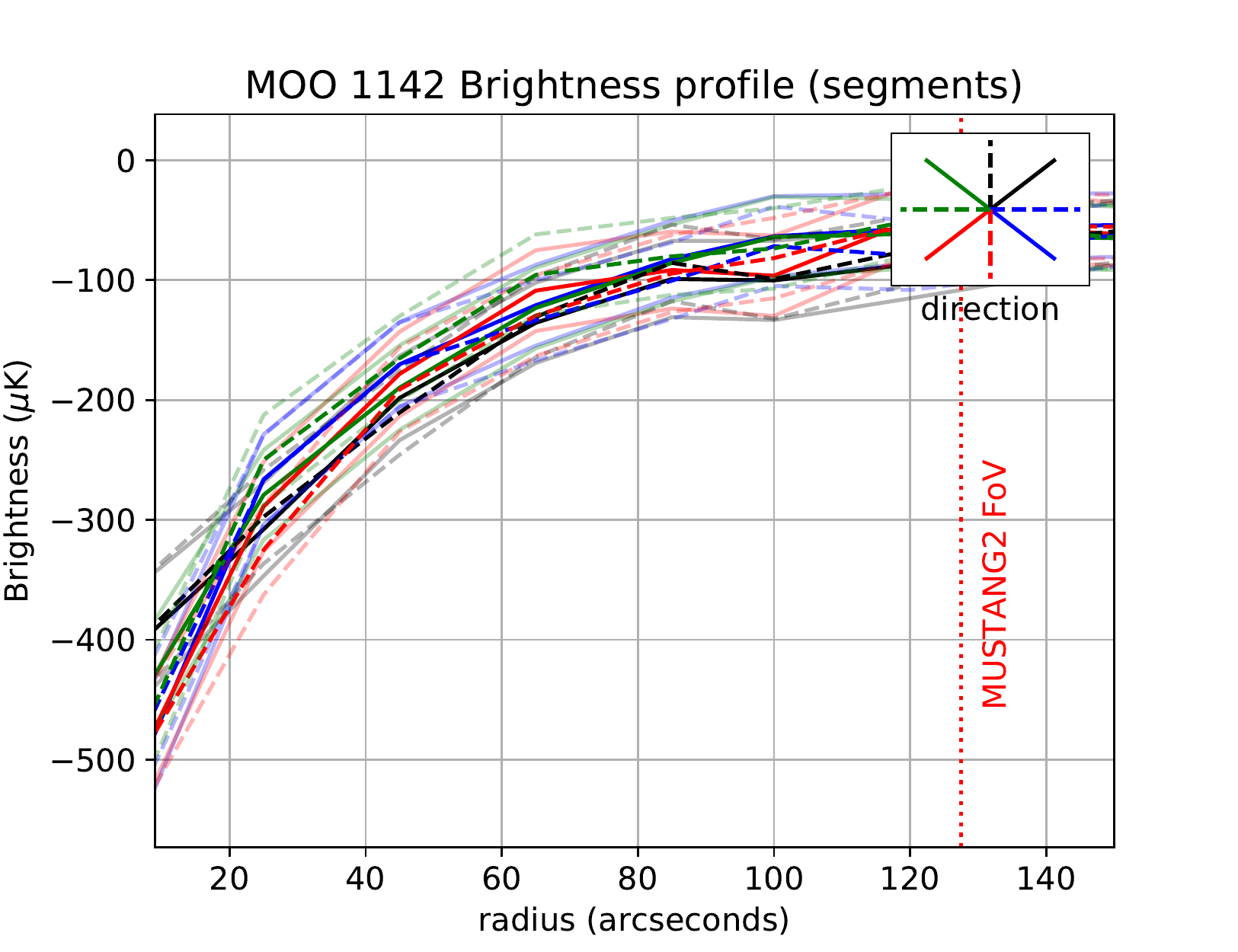}  
    \caption{Cluster profiles fit over 90\dg\ slices. The solid and dashed lines represent the 90\dg\ slice directions shown in the top right of each plot while the fainter lines on either side show the $3\sigma$ error bars corresponding to that color and line style.}
    \label{fig:sym}
\end{figure*}

Mergers per unit time are expected to be more common at high redshift \citep{Fakhouri2010} so it is possible that several of the systems in this sample not currently thought to be mergers are in fact mergers. Ongoing mergers could explain why  MOO~1031 and MOO~1203 have similar richnesses to many robustly-detected clusters in the sample but are relatively weakly detected by MUSTANG2. The correct identification of which clusters are merging could become the limiting factor when finding an accurate mass-richness scaling relation using SZE observations.  MUSTANG2's ability to resolve structure within clusters could be of some use here - with the planned larger sample it will be possible to group clusters by their profile shapes and carry out searches for signs of shocks in the ICM. Other signs of mergers include double peaks in galaxy distributions (either spatially for mergers in the plane of the sky or spectroscopically for mergers along the line of sight) and offsets between the SZE center and the center of the galaxies (for example MOO~1052).  When available, comparison with deep X-ray data could be used in a joint analysis (to obtain better temperature constraints than either data sets alone) and X-ray masses compared to SZE masses. 

%\change{(need to combine this with paragraph 2 down :) We note that the $Y_{\rm SZ}-M_{500}$ scaling relation expected under the assumption of self-similarity, has a relatively steep slope of 5/3 \citep[e.g.][]{Morandi2007, Poole2007,Rasia2011,Battaglia2012}.    This means that the SZE signal from a virialized, post-merger system of a given mass is higher than the sum of the signals from the individual pre-merger components.  Fitting the total SZE flux from unvirialized systems in the same field as a single halo would therefore result in an underestimate of the mass with respect to the true total.  }

\section{Segmented fits}\label{sec:segments}
The cluster profiles presented in Section~\ref{sec:results} assume symmetry. However, through examination of the maps, we see some evidence that the cluster morphology is asymmetric (\eg\ the center of MOO~1142), and at least three of the clusters are known mergers (where one might expect an asymmetric profile).  The analysis in Section~\ref{sec:profiles} was repeated but with the rings broken into four segments of 90\dg. An asymmetrical cluster with one axis aligned within the segments chosen (in the first round of analysis 0\dg, 90\dg, 180\dg, and 270\dg\ east of north) would show up as different profiles in some segments.  To fully test for asymmetry this process was repeated for angles of 45\dg, 135\dg, 225\dg, and 315\dg.  Examples of some of the fits are shown in Figure~\ref{fig:sym} -- within the errors no significant derivations from spherical symmetry could be detected in the brightness profiles and the masses recovered from each segment were the same to within the statistical error.

As a further check, the symmetrical profiles from Section~\ref{sec:profiles} were subtracted from the maps (along with any point sources).  In all cases the residual signal was consistent with zero.  While this does not mean there is no asymmetry in the clusters, deeper maps would be needed to detect it.  Asymmetry is a sign of possible mergers and mergers could have a dramatic effect on the mass-richness scaling relation (Section~\ref{sec:mass-richness}) so joint analysis using other data sets on the complete MUSTANG2 sample is planned.
Other possible future avenues include using matched filters to search for small scale features such as shock fronts in the MUSTANG2 maps.

\section{Conclusions}\label{sec:concl}
In this paper we have presented initial results of using MUSTANG2 to follow up a pilot sample of clusters detected by the MaDCoWS project \citep[for the full catalog, see][]{Gonzalez2019}.  With integration times between two and seven hours on 16 clusters, we were able to detect, through the SZE effect, fourteen of our targets and obtain robust estimates of their masses as well as measurements of their SZE brightness and pressure profiles. Strong upper limits on the masses of the other two were obtained.  MUSTANG2's 9$''$ resolution (10$''$ after smoothing) makes the identification and removal of point sources from the observations relatively straightforward compared to smaller single dish and survey instruments.  The cluster profiles and the offsets between the best fit centers for the SZE compared with the MaDCoWS centers give indications of the dynamical states of our fourteen detections. In general the fitted SZE centers agreed well with those identified in \citet{Gonzalez2019}, and no strong evidence could be found for asymmetry in any of the clusters.  However in the case of MOO~1052 there is a significant ($>2'$) offset between the SZE and MaDCoWS centers, making this a good target for follow-up work to investigate whether is it undergoing a merger.  

The mass-richness relationship derived using the MUSTANG2 MaDCoWS pilot sample alone, excluding known mergers but including upper limits, exhibits similar high scatter as those from CARMA and ACA data.  However, the best-fit scaling relations to any data sets that include MUSTANG2 data are all significantly flatter than those derived for CARMA or CARMA+ACA, but still consistent with $\alpha=1$.  We note that the SZE observations with MUSTANG2 and ACA are in general deeper and higher resolution than those performed with CARMA, and posit that we may be probing previously unexplored parts of the MaDCoWS cluster population. We also note that the CARMA data excluded a number of weak and non-detections lying below their sensitivity limit, which could biases the inferred scaling relation in \cite{Gonzalez2019} to a steeper slope. Conversely the slope of the fit to the MUSTANG2 data is strongly leveraged by the inclusion of a single cluster, MOO~1322.  The dynamical state of this cluster is uncertain and it could be biasing the inferred scaling relation from MUSTANG2 data to a flatter value -- including mergers in the fits to ACA and CARMA data dramatically flattens these slopes as well.  Further studies, such as the already-approved, upcoming MUSTANG2 observations, will be necessary to resolve the discrepancies in mass-richness relationships. In particular it will be important to include a greater number of MaDCoWS clusters spanning a wider range of richnesses and to ensure only those clusters known to be relaxed are used in the fit. 

\acknowledgments
MUSTANG2 is supported by the NSF award number 1615604 and by the Mt.\ Cuba Astronomical Foundation. 
The National Radio Astronomy Observatory is a facility of the National Science Foundation operated under cooperative agreement by Associated Universities, Inc. GBT data was taken under the project IDs AGBT18B\_215 and AGBT19B\_200.
EM acknowledges the support of the EU-ARC.CZ Large Research Infrastructure grant project LM2018106 of the Ministry of Education, Youth and Sports of the Czech Republic.
The work of T.C. and D.S. %P.E., and D.S. 
was carried out at the Jet Propulsion Laboratory, California Institute of Technology, under a contract with NASA. 
This paper made use of APLpy, an open-source plotting package for Python \citep{APLpy}.\\
\facilities{GBT (MUSTANG2)}
\software{astropy (\url{https://www.astropy.org/}), APLpy (\url{https://aplpy.github.io/})}

%\clearpage
%\vspace{1mm}

%\clearpage
%\vspace{4cm}

%\newline
\bibliographystyle{aasjournal}
\bibliography{M2MaDCoWS}

\appendix 
\section{Galaxy number densities inferred from \textit{Spitzer}/IRAC data}\label{sec:opt_images}
\begin{figure*}
%\begin{center}
\includegraphics[clip,trim=1.75cm 1.75cm 1.75cm 1.75cm,height=0.23\textwidth]{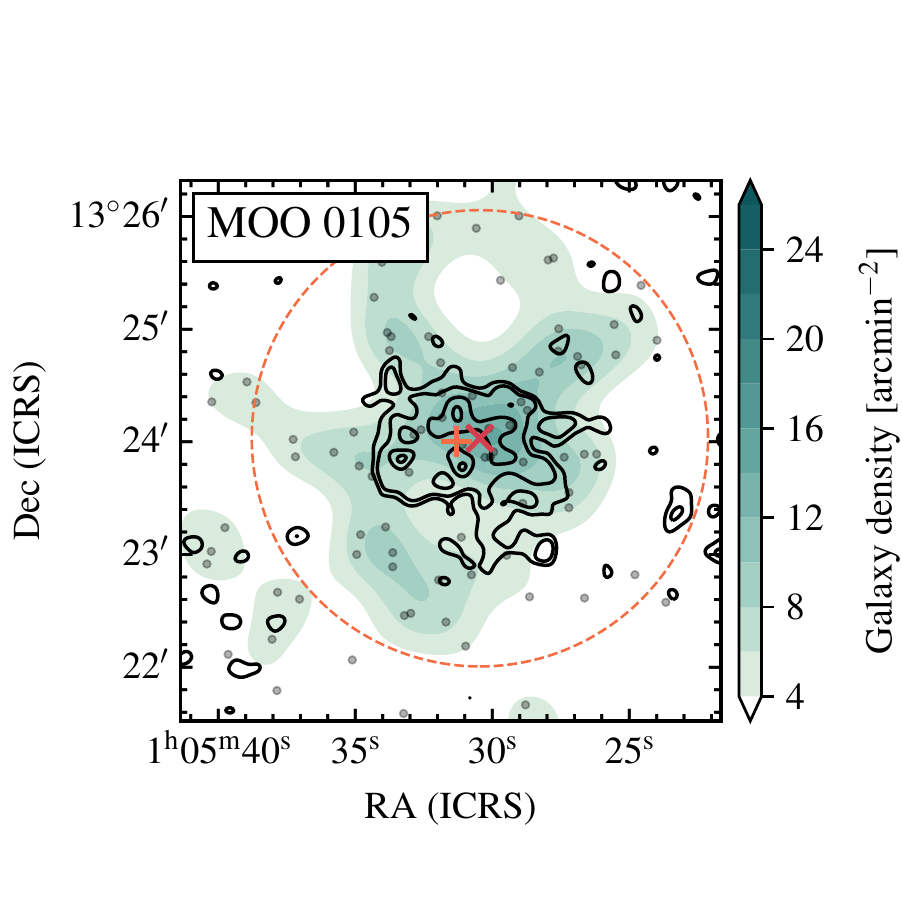}
\includegraphics[clip,trim=1.75cm 1.75cm 1.75cm 1.75cm,height=0.23\textwidth]{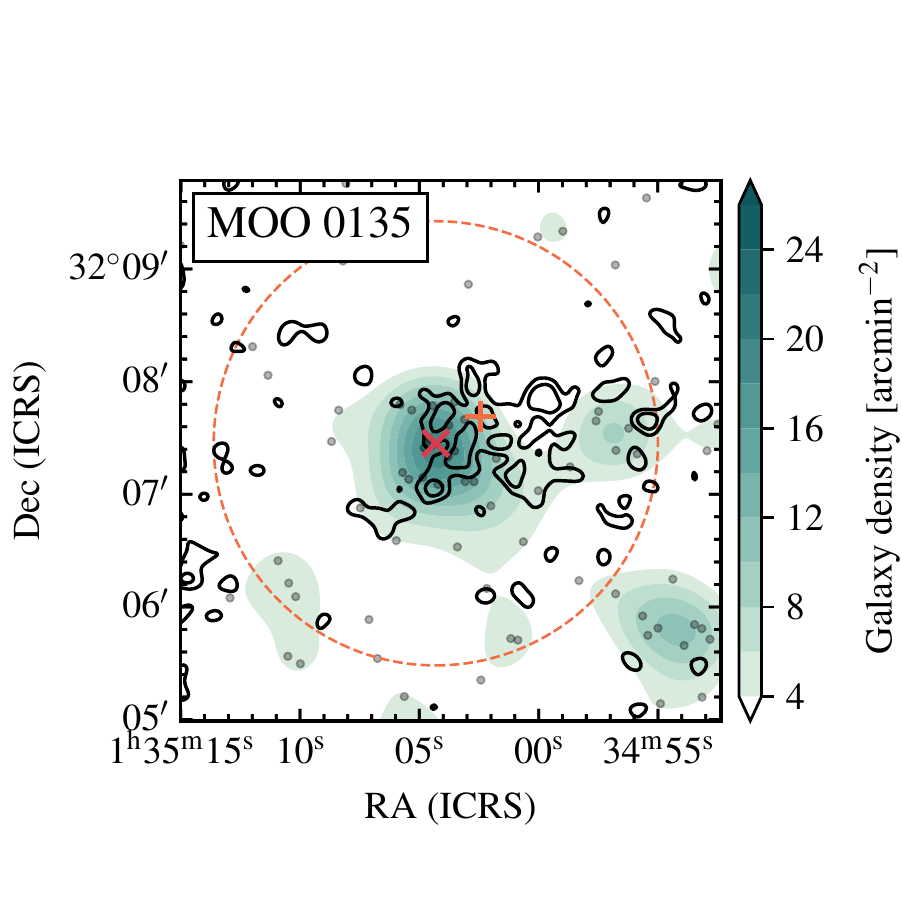}
\includegraphics[clip,trim=1.75cm 1.75cm 1.75cm 1.75cm,height=0.23\textwidth]{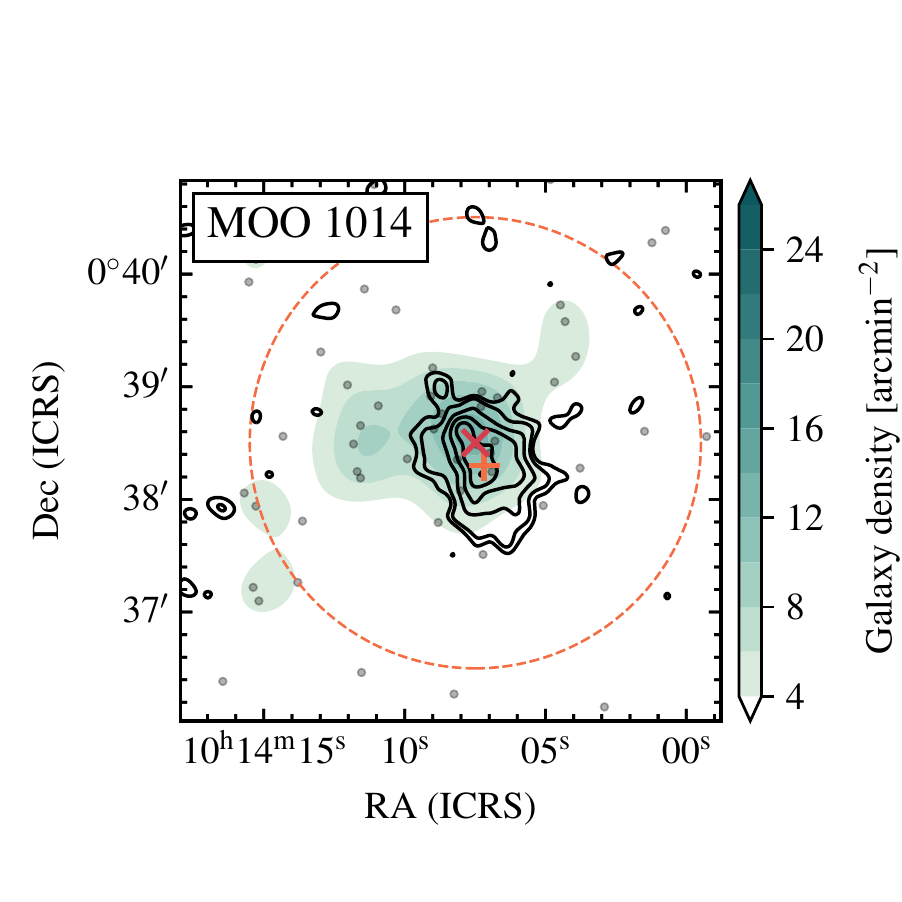}
\includegraphics[clip,trim=1.75cm 1.75cm 0.00cm 1.75cm,height=0.23\textwidth]{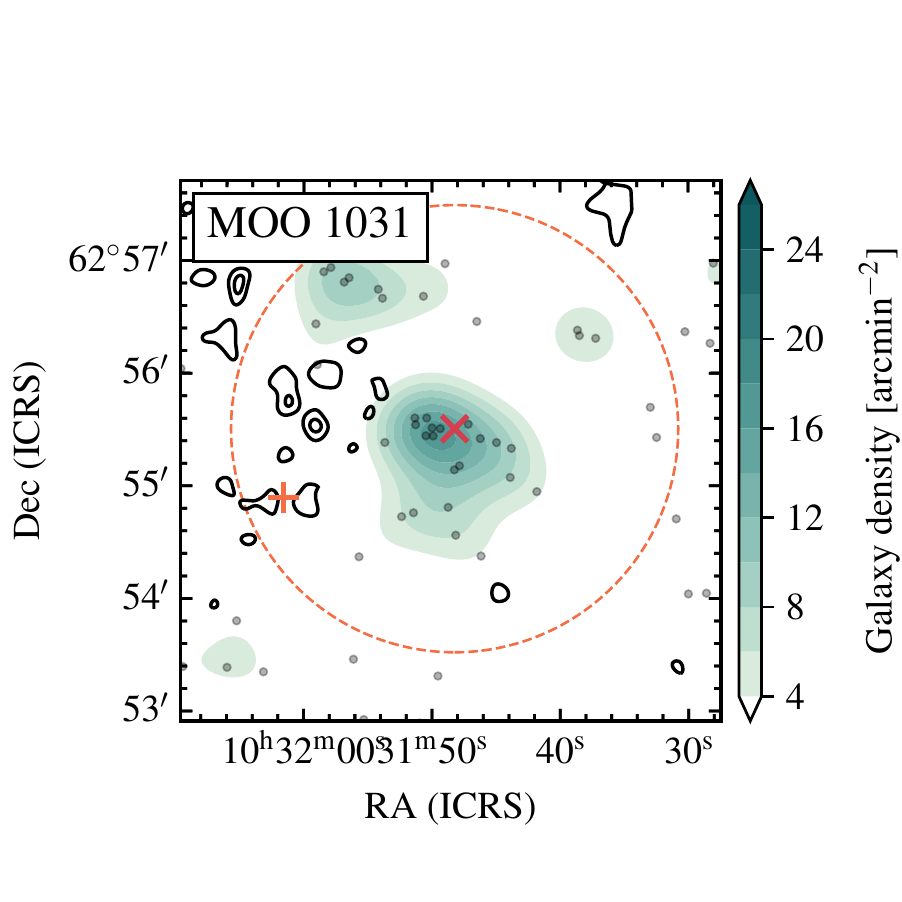}
\includegraphics[clip,trim=1.75cm 1.75cm 1.75cm 1.75cm,height=0.23\textwidth]{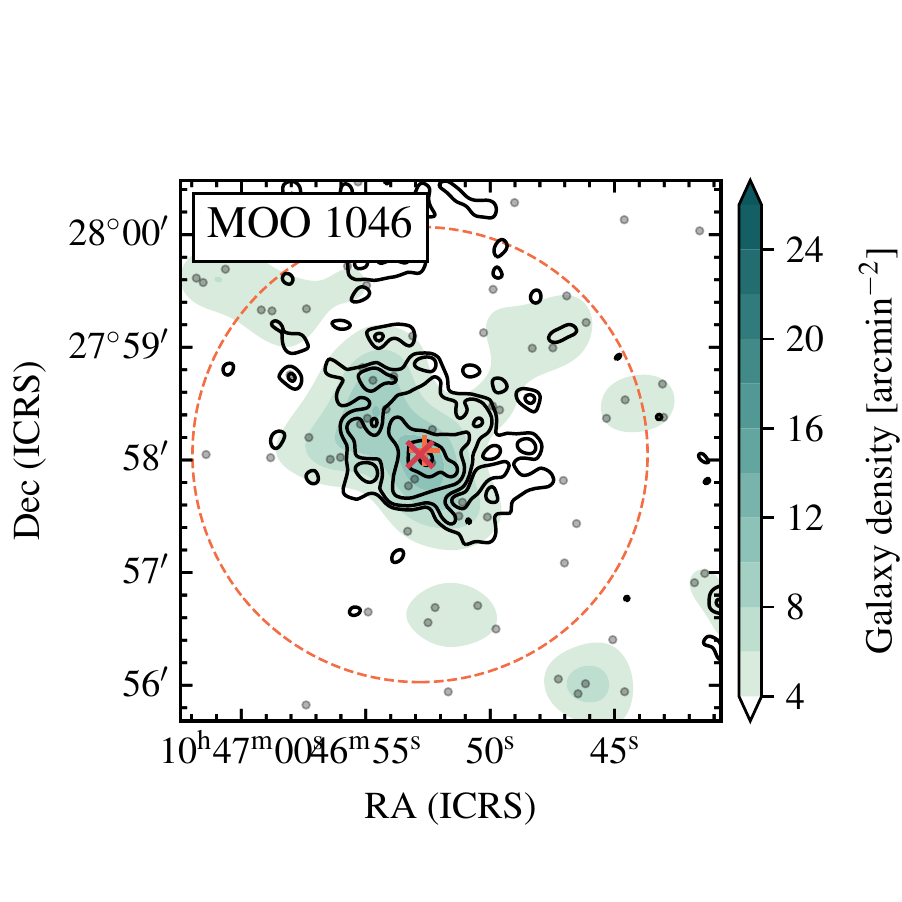}
\includegraphics[clip,trim=1.75cm 1.75cm 1.75cm 1.75cm,height=0.23\textwidth]{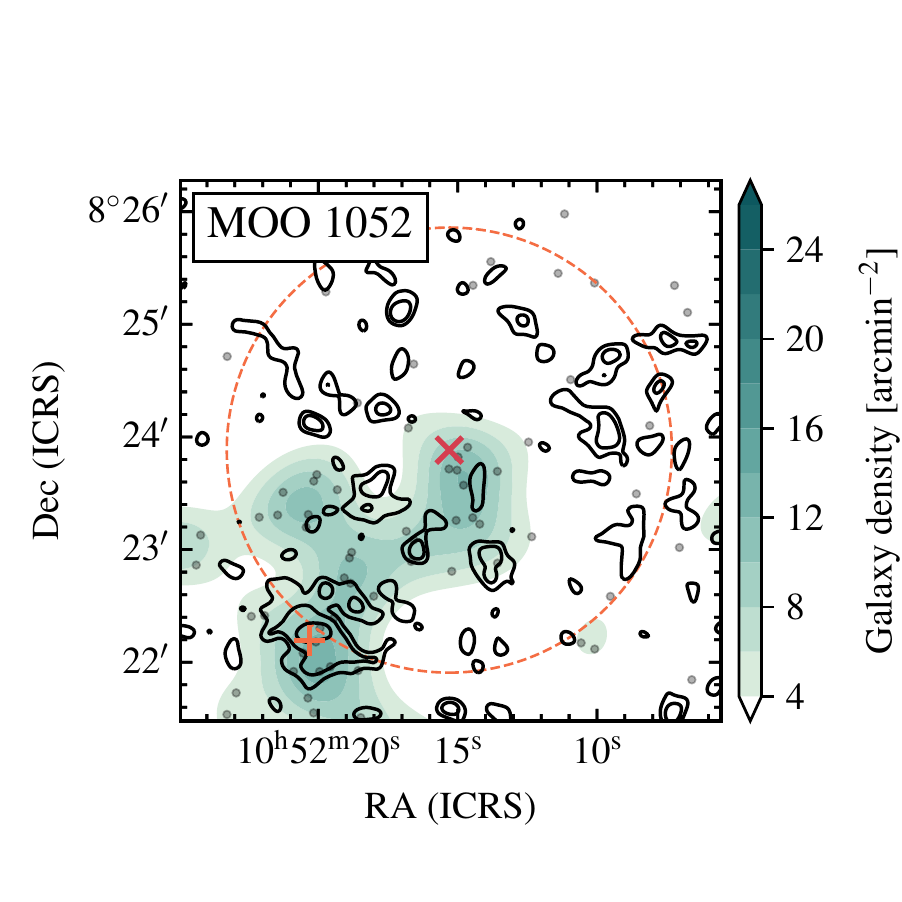}
\includegraphics[clip,trim=1.75cm 1.75cm 1.75cm 1.75cm,height=0.23\textwidth]{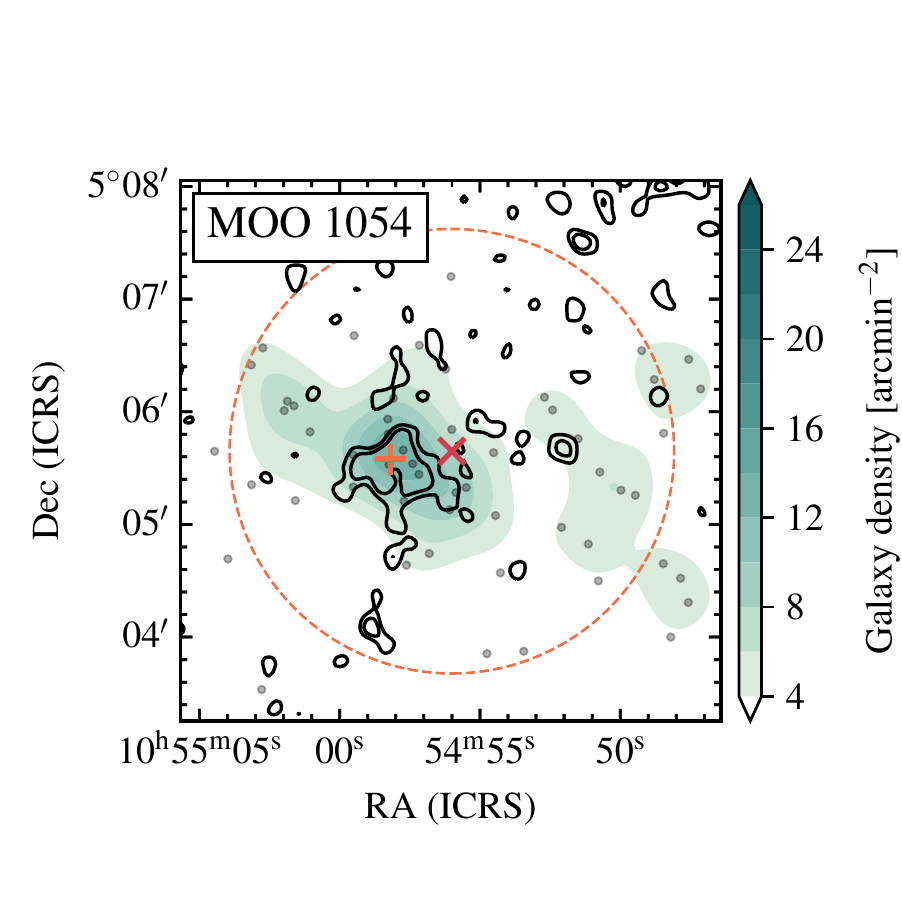}
\includegraphics[clip,trim=1.75cm 1.75cm 0.00cm 1.75cm,height=0.23\textwidth]{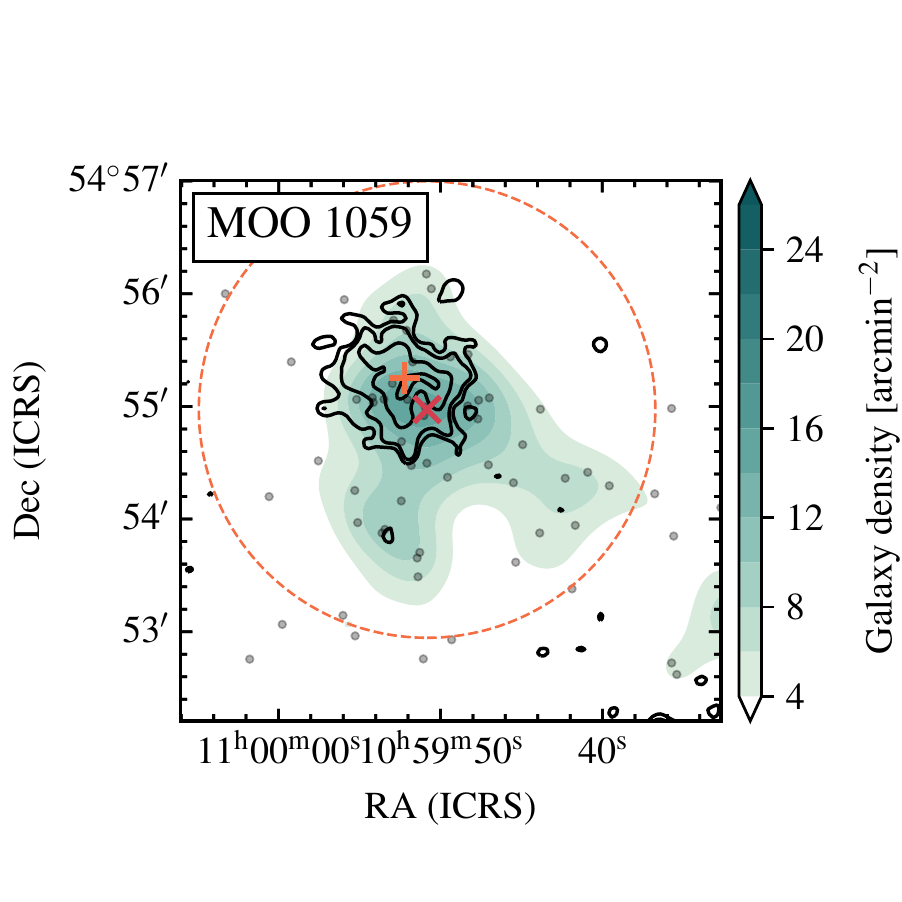}
\includegraphics[clip,trim=1.75cm 1.75cm 1.75cm 1.75cm,height=0.23\textwidth]{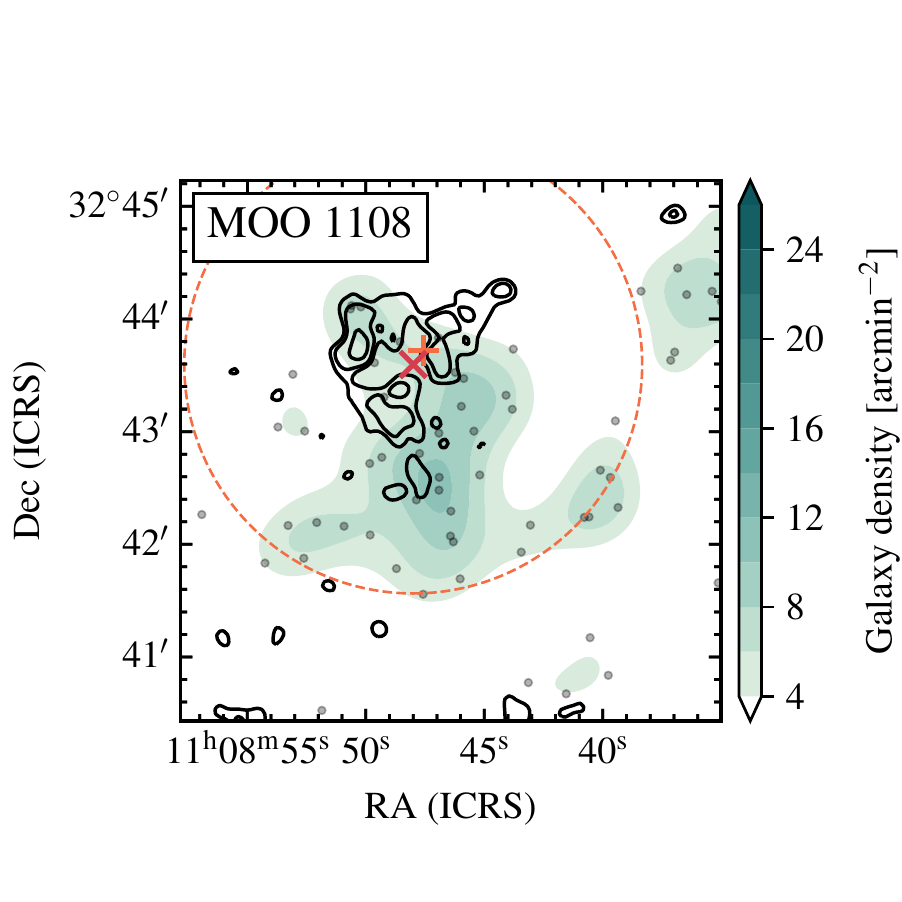}
\includegraphics[clip,trim=1.75cm 1.75cm 1.75cm 1.75cm,height=0.23\textwidth]{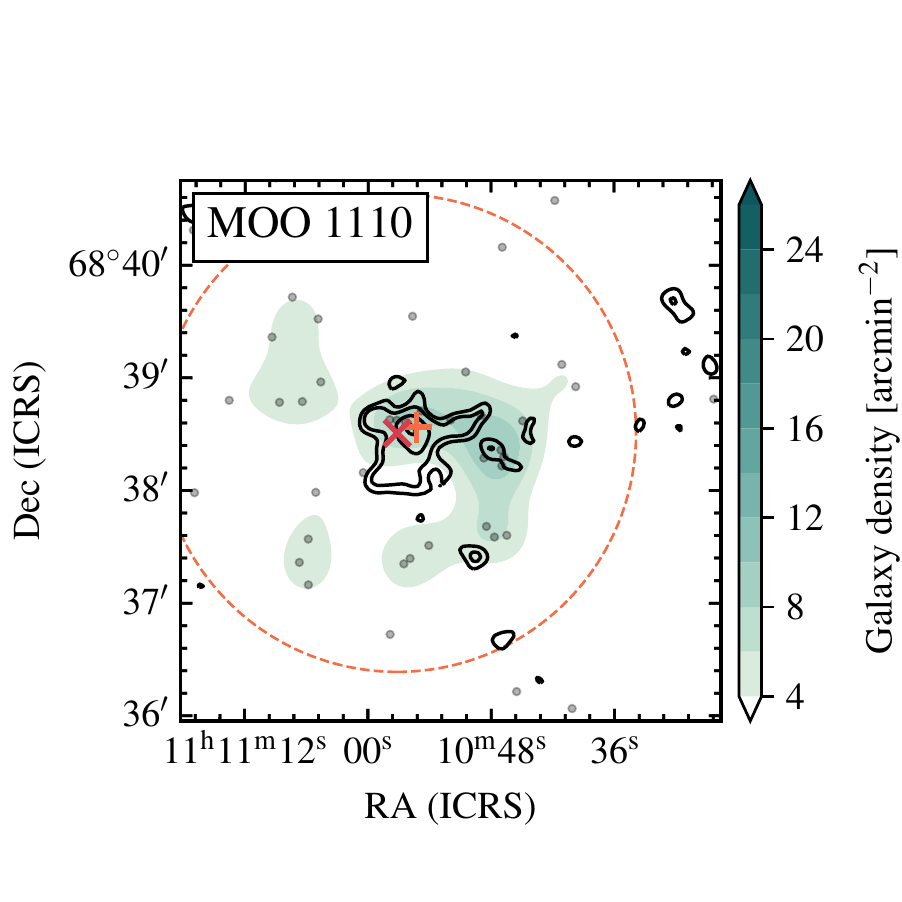}
\includegraphics[clip,trim=1.75cm 1.75cm 1.75cm 1.75cm,height=0.23\textwidth]{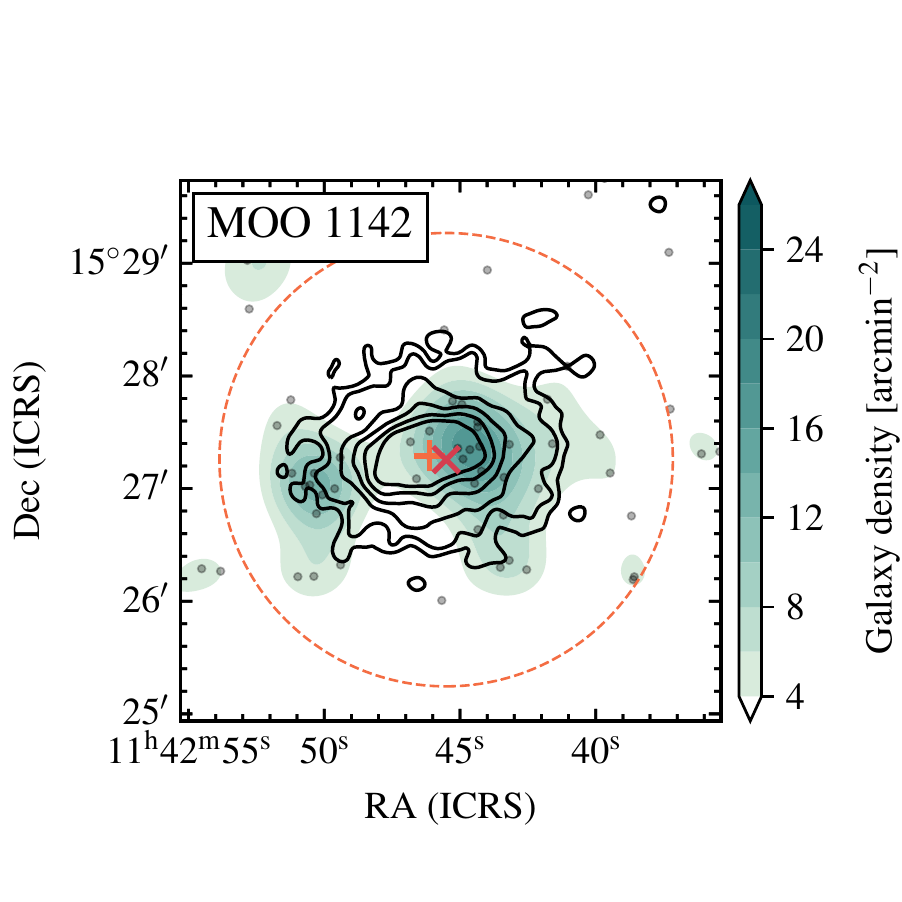}
\includegraphics[clip,trim=1.75cm 1.75cm 0.00cm 1.75cm,height=0.23\textwidth]{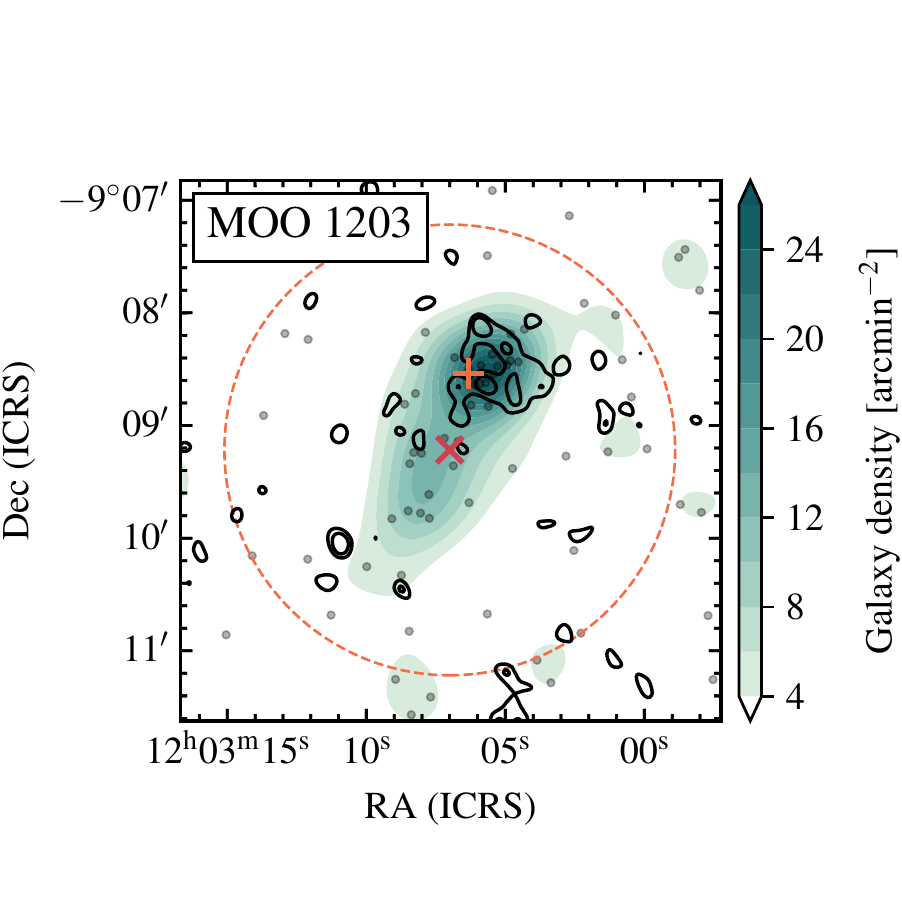}
\includegraphics[clip,trim=1.75cm 1.75cm 1.75cm 1.75cm,height=0.23\textwidth]{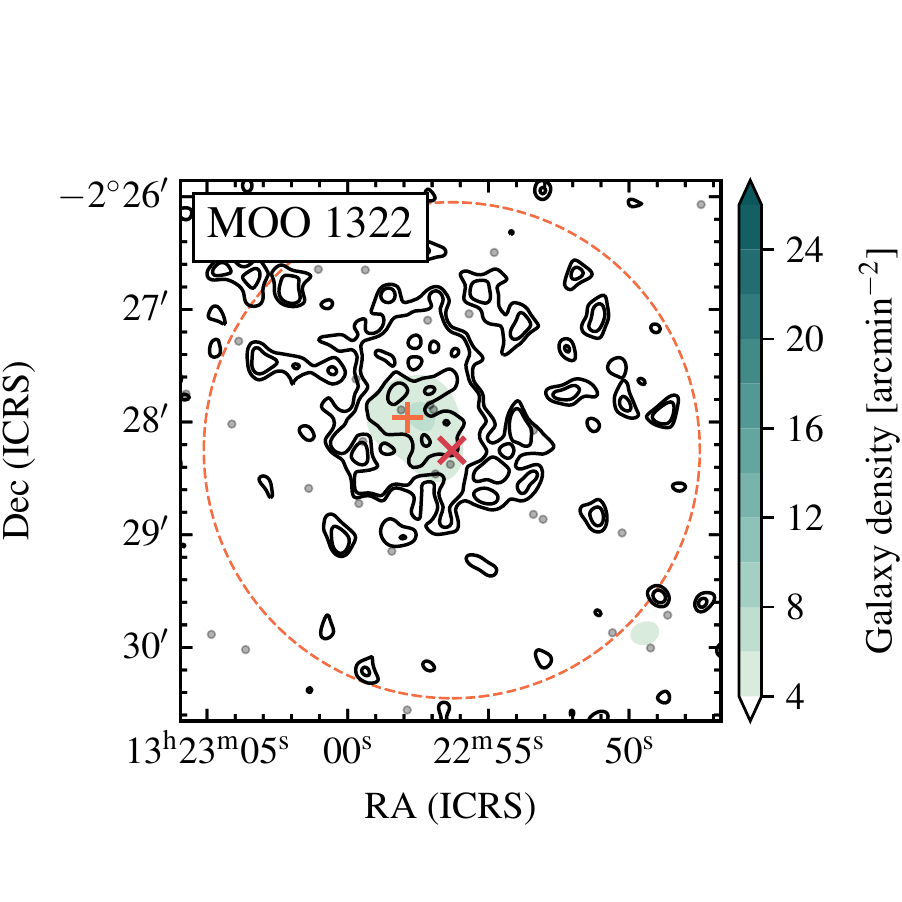}
\includegraphics[clip,trim=1.75cm 1.75cm 1.75cm 1.75cm,height=0.23\textwidth]{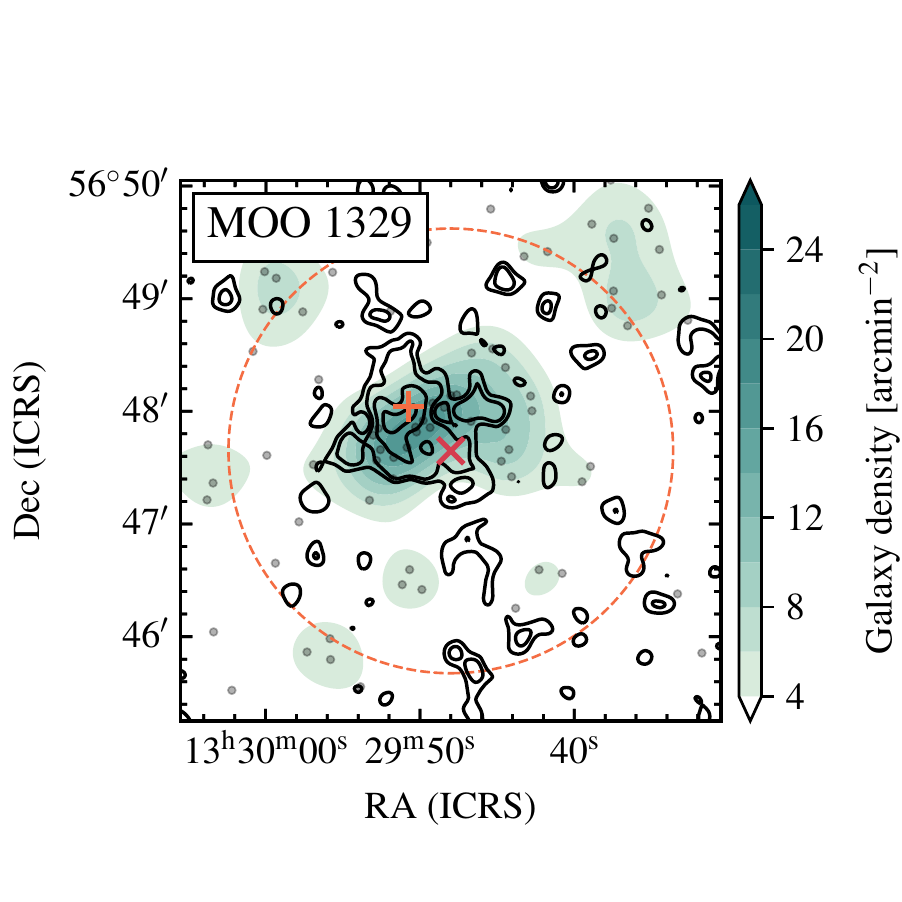}
\includegraphics[clip,trim=1.75cm 1.75cm 1.75cm 1.75cm,height=0.23\textwidth]{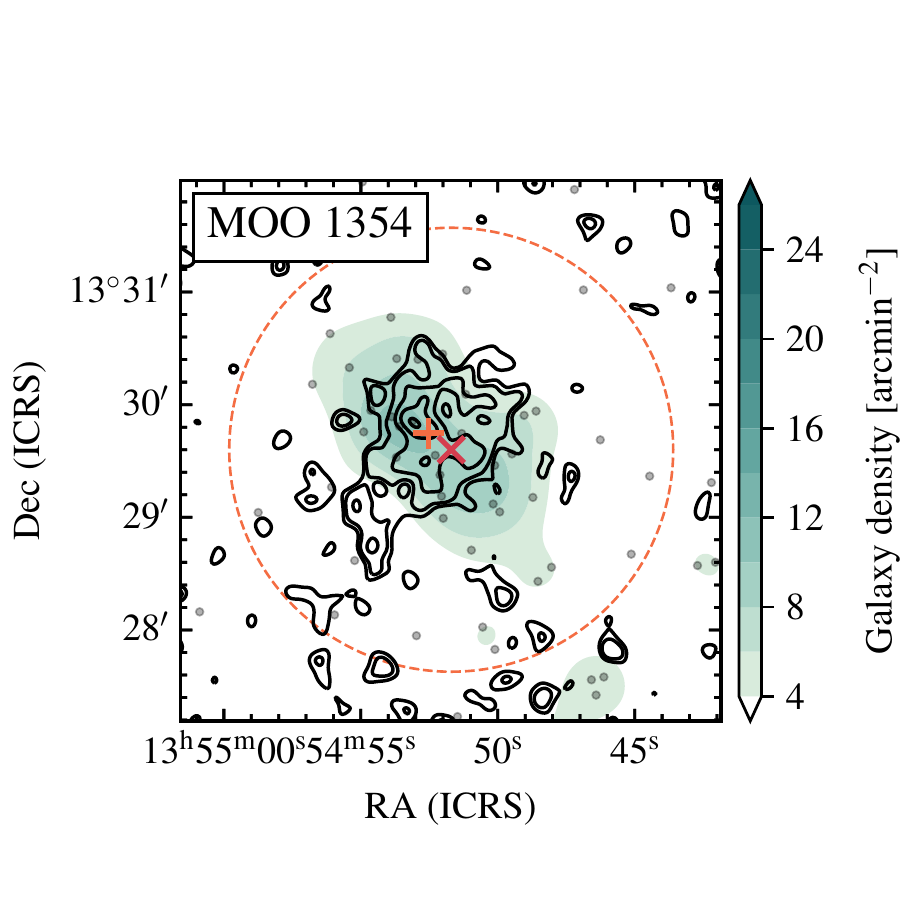}
\includegraphics[clip,trim=1.75cm 1.75cm 0.00cm 1.75cm,height=0.23\textwidth]{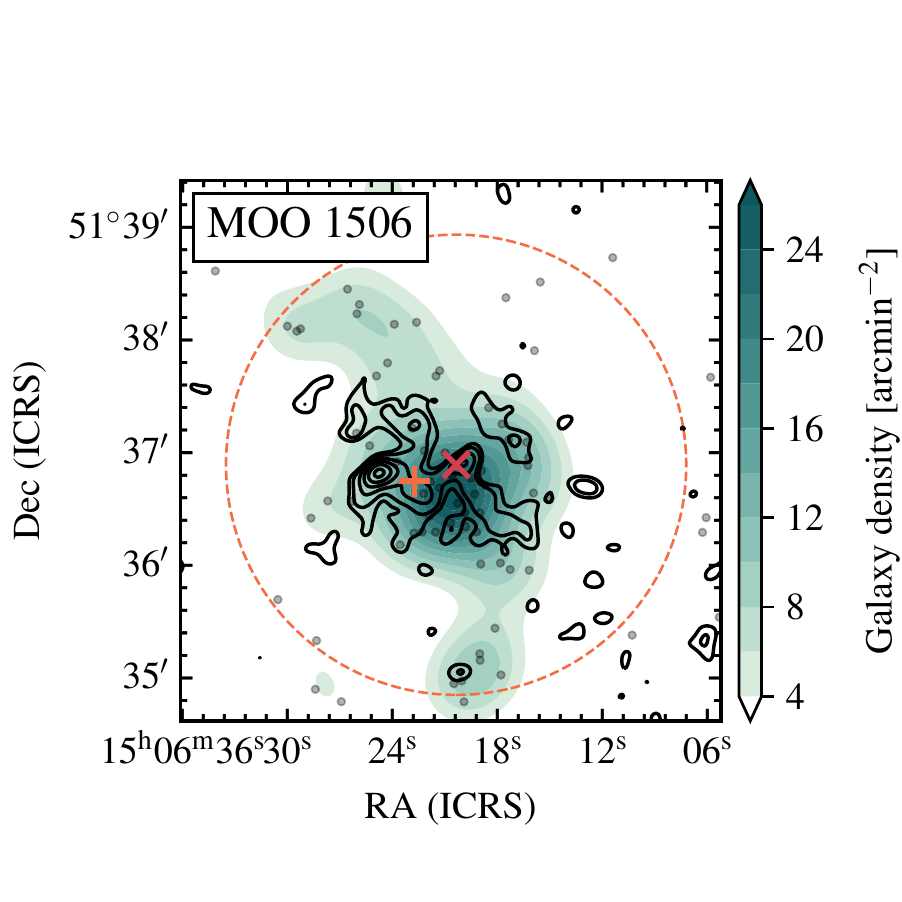}
\caption{\label{fig:spizteroverlays} \textit{Spitzer}/IRAC galaxy density maps.  Galaxies are selected by color cuts to be preferentially in the redshift range 0.7 to 1.4 \citep[see][for details]{Wylezalek2014,Gonzalez2019}. The galaxy number densities are smoothed with a 40\arcsec\ FWHM Gaussian kernel. Each plot is 4.8$'\times$4.8$'$. The MaDCoWS position (Table \ref{tab:obs}) is shown as a red X and the dashed red lines represent the 1~Mpc radius circle used to calculate $\lambda_{15}$.  Point source subtracted contours from the MUSTANG2 SNR maps are over-plotted in black, with contour levels of -2, -3, -5, -7, -9, and -11$\sigma$. The fitted SZE centers are marked with an orange $+$.}
%\end{center}
\end{figure*}

In Figure \ref{fig:spizteroverlays}, we present overlays of the SZE decrements on galaxy density maps inferred from \textit{Spitzer}/IRAC observations.  The galaxies are selected to be  preferentially at the MaDCoWS redshifts using the same color cuts described in \citet{Gonzalez2019} and to be brighter than 15~mJy at a wavelength of 4.5~$\mu$m. Overall there is good agreement between the optical and SZE images.  Exceptions include our low SNR clusters, MOO~1031 (where there is very little sign of any SZE signal where the galaxy density is) and MOO~1203 (where the SZE signal is only detected on an off-centered peak in the bulk \textit{Spitzer} galaxy overdensity).   The SZE signal for MOO~1052 is centered on a second peak in the galaxy distribution, away from the MaDCoWS-identified center, indicating a possible merger or potential contamination from foreground/background galaxies.

\section{Further considerations for the SZE Mass vs Richness scaling relations}\label{sec:scaling_app} 
\begin{figure*}[th!]
    \centering
    
    \includegraphics[clip,trim=0.50cm 0 0.50cm 0,width=0.49\textwidth]{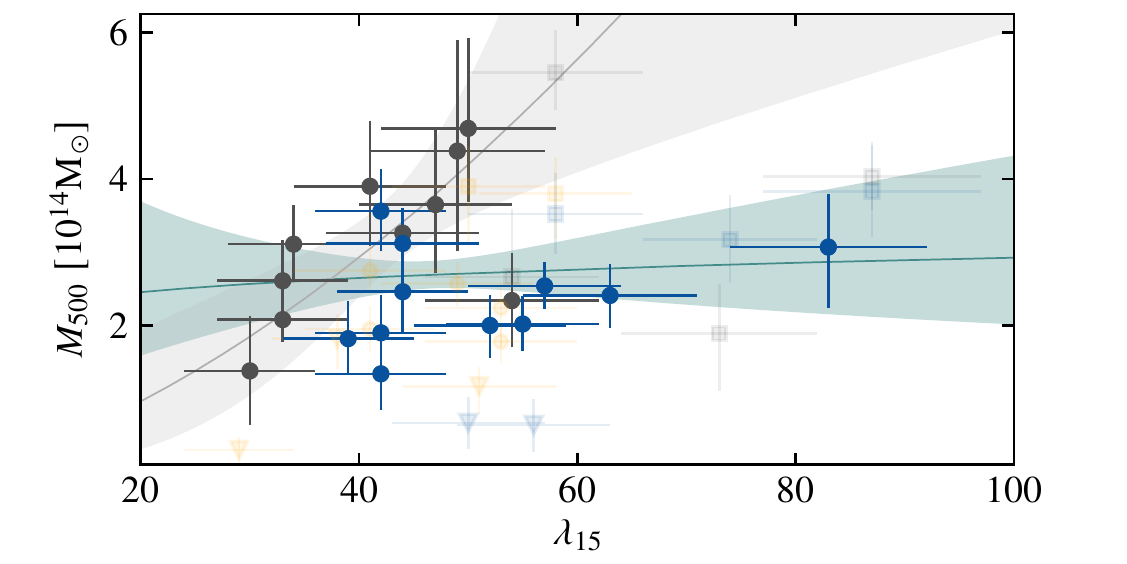}
    \includegraphics[clip,trim=0.50cm 0 0.50cm 0,width=0.49\textwidth]{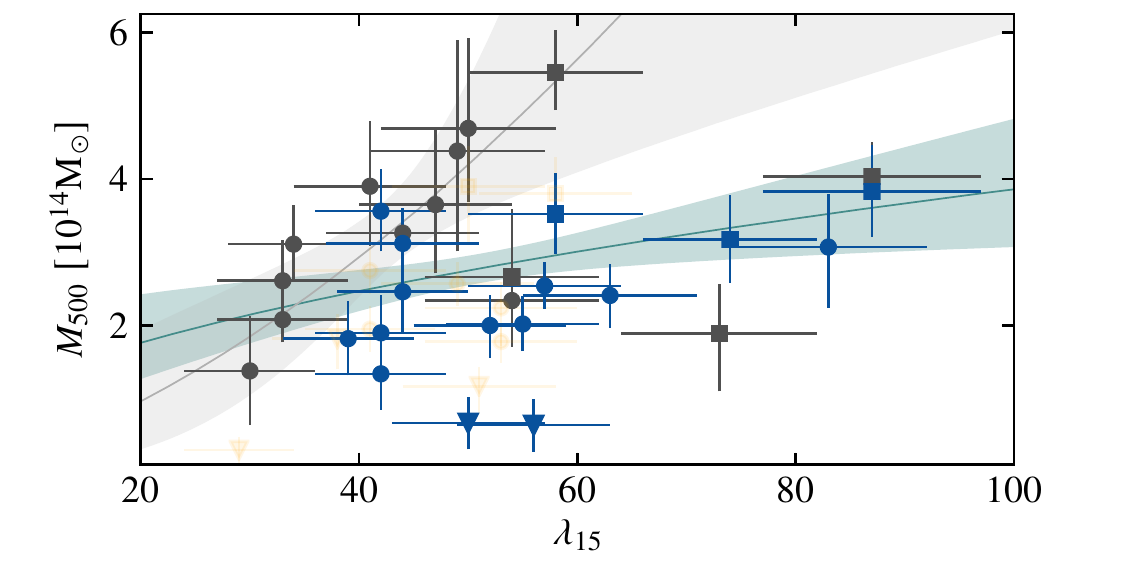}\\
    \includegraphics[clip,trim=0.50cm 0 0.50cm 0,width=0.49\textwidth]{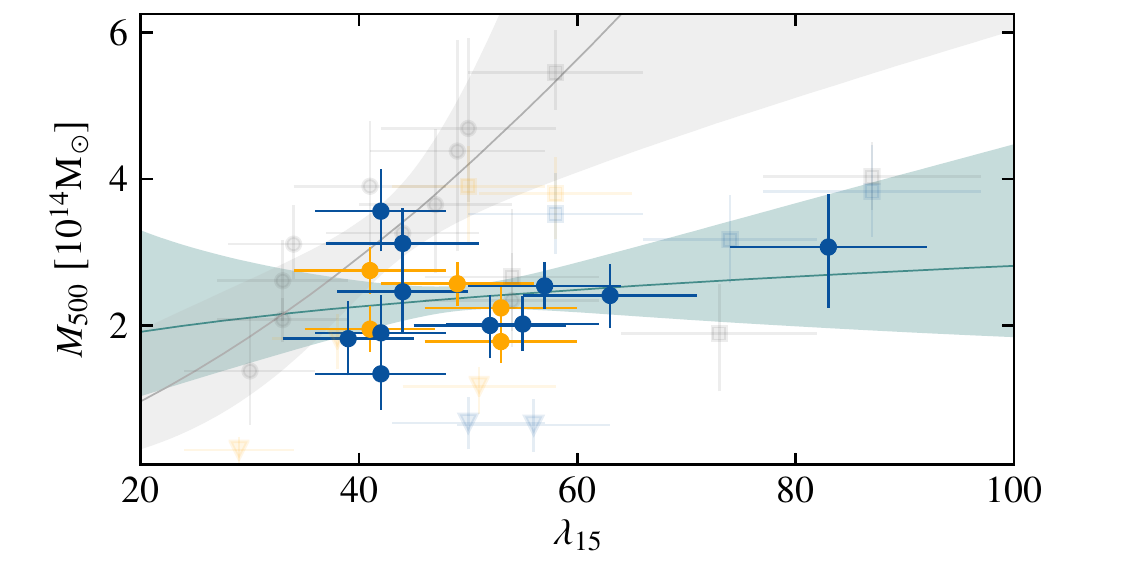}
    \includegraphics[clip,trim=0.50cm 0 0.50cm 0,width=0.49\textwidth]{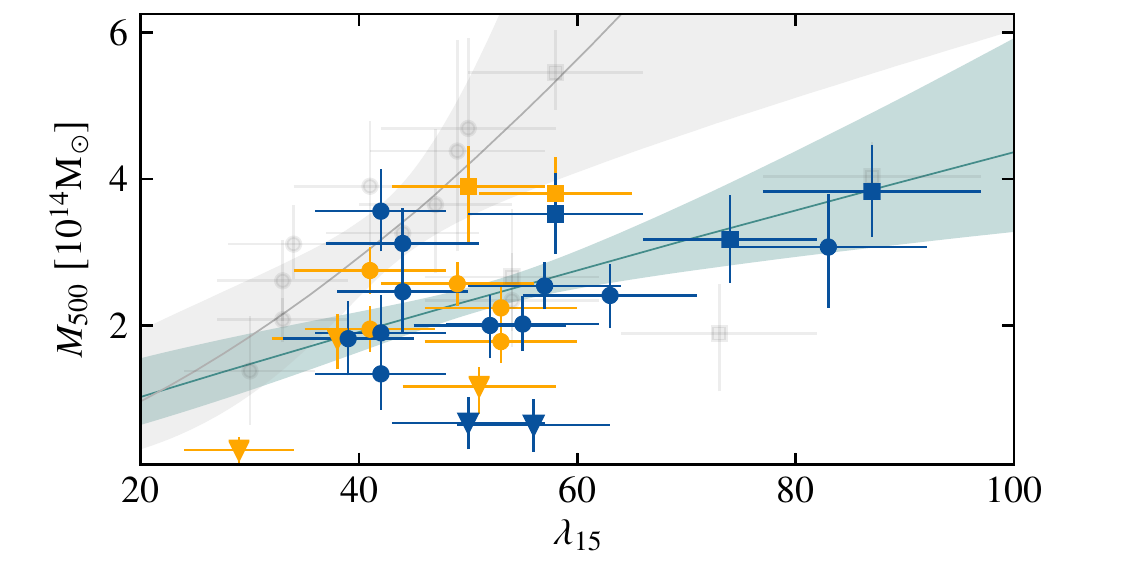}\\
    \includegraphics[clip,trim=0.50cm 0 0.50cm 0,width=0.49\textwidth]{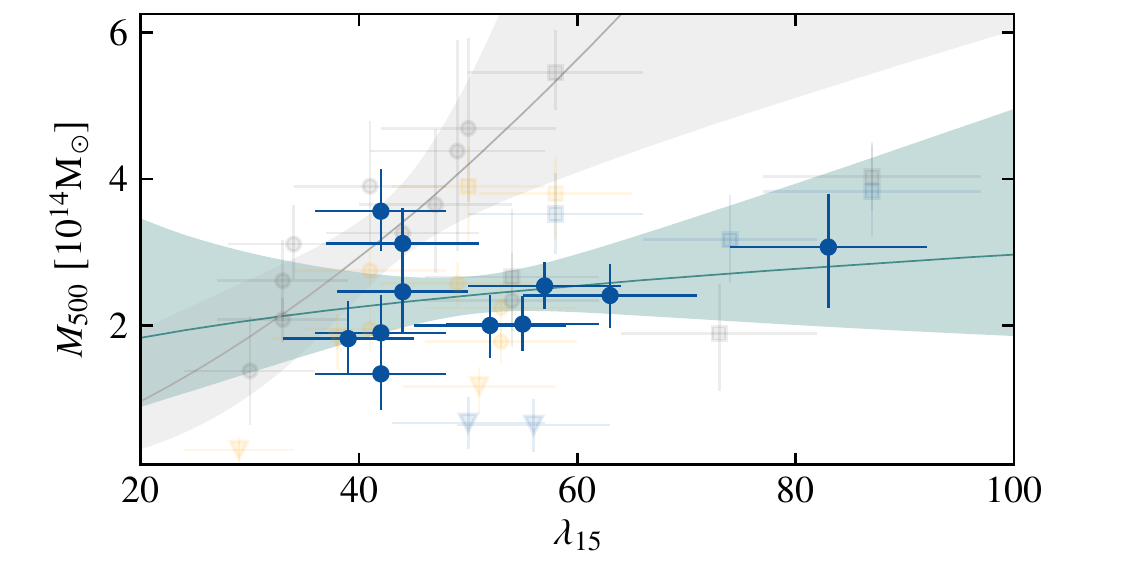}
    \includegraphics[clip,trim=0.50cm 0 0.50cm 0,width=0.49\textwidth]{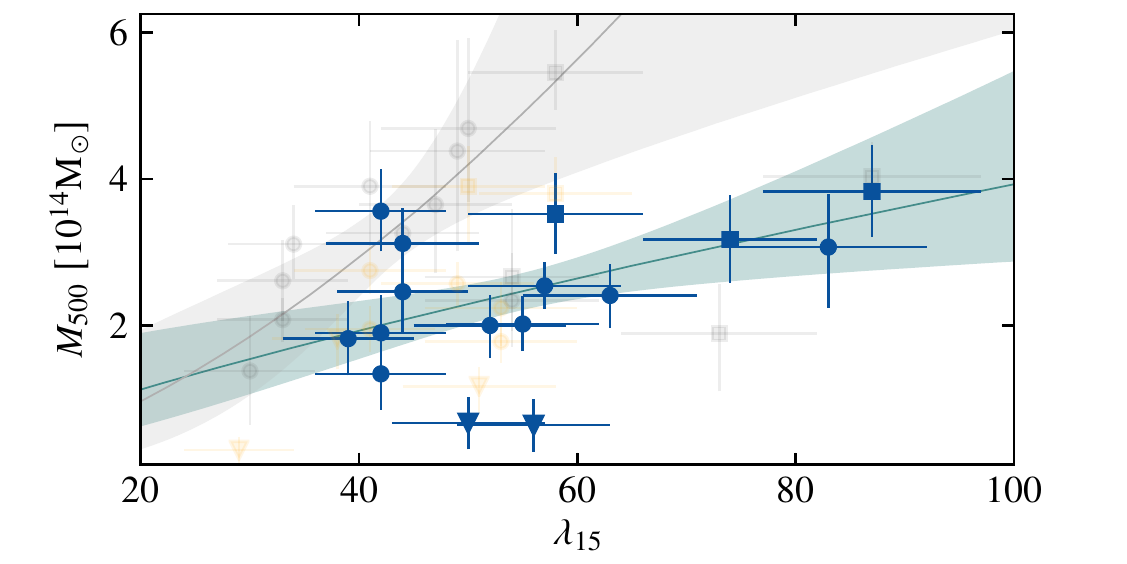}
    
    \caption{ \label{fig:massRichness_app} Plots of selected scaling relations reported in Table~\ref{tab:scalingrels}, including or excluding known mergers, weak or non-detections, and including or excluding the CARMA and ACA data (from \citealt{Gonzalez2019} and \citealt{DiMascolo2020} respectively).  The shaded gray region in each plot is the fit to the CARMA points only (including 1-$\sigma$ error bars), while each color region shows the mass-richness scaling relation appropriate for the data points in bold, which are: 
    \newline $\bullet$
    {\bf Upper left:} Fit to MUSTANG2+CARMA data, excluding known mergers and weak/non-detections. 
    \newline $\bullet$
    {\bf Upper right:} Fit to MUSTANG2+CARMA data, including known mergers and weak/non-detections. 
    \newline $\bullet$
    {\bf Middle left:} Fit to MUSTANG2+ACA data, excluding known mergers and weak/non-detections. 
    \newline $\bullet$
    {\bf Middle right:} Fit to MUSTANG2+ACA data, including known mergers and weak/non-detections. 
    \newline $\bullet$
    {\bf Lower left:} Fit to MUSTANG2 data alone, excluding known mergers and weak/non-detections. 
    \newline $\bullet$
    {\bf Lower right:} Fit to MUSTANG2 data alone, including known mergers and weak/non-detections.
    \newline
    The exclusion of known mergers and low significance detection and non-detections does not reconcile the scaling relations when including MUSTANG2 data with those found fitting the CARMA detections alone.}
\end{figure*}

In Figure~\ref{fig:massRichness_app} and Table \ref{tab:scalingrels}, we present results for the mass--richness ($M_{500}-\lambda_{15}$) scaling relations that we infer when including or excluding various SZE observational data sets, known mergers, and weak or non-detections (i.e.\ upper limits).
%fitted scaling relations
\begin{table}[tbh]
\caption{\label{tab:scalingrels} Fits for mass-richness scaling relations.}
\begin{center}
\begin{tabular}{lccc}
\hline\hline SZE Data Used           &        $\alpha$        &         $\beta$         &    $\sigma_{\log M |\lambda}$\\\hline
\multicolumn{4}{c}{Excluding known mergers and non-detections}\\
CARMA              & $ 1.65^{+1.45}_{-0.96}$ & $-2.16^{+1.57}_{-2.38}$ & $ 0.12$                \\
CARMA+ACA          & $ 0.07^{+0.51}_{-0.50}$ & $ 0.29^{+0.83}_{-0.86}$ & $ 0.15^{+0.01}_{-0.00}$\\
MUSTANG2 (M2)      & $ 0.30^{+0.76}_{-0.68}$ & $-0.13^{+1.16}_{-1.30}$ & $ 0.14^{+0.03}_{-0.01}$\\
M2+CARMA           & $ 0.11^{+0.51}_{-0.48}$ & $ 0.26^{+0.79}_{-0.86}$ & $ 0.16^{+0.02}_{-0.00}$\\
M2+ACA             & $ 0.24^{+0.67}_{-0.59}$ & $-0.02^{+1.01}_{-1.13}$ & $ 0.12^{+0.02}_{-0.01}$\\
M2+CARMA+ACA\!\!\! & $ 0.06^{+0.52}_{-0.50}$ & $ 0.30^{+0.83}_{-0.88}$ & $ 0.15^{+0.02}_{-0.00}$\\
\multicolumn{4}{c}{Excluding known mergers including non-detections}\\
CARMA              &         \ldots          &         \ldots          &         \ldots         \\
CARMA+ACA          & $ 0.13^{+0.58}_{-0.56}$ & $ 0.16^{+0.94}_{-0.97}$ & $ 0.27^{+0.01}_{-0.01}$\\
MUSTANG2 (M2)      & $ 0.35^{+0.99}_{-0.89}$ & $-0.26^{+1.52}_{-1.69}$ & $ 0.27^{+0.02}_{-0.00}$\\
M2+CARMA           & $ 0.01^{+0.58}_{-0.56}$ & $ 0.40^{+0.94}_{-0.96}$ & $ 0.24^{+0.01}_{-0.00}$\\
M2+ACA             & $ 0.51^{+0.76}_{-0.67}$ & $-0.53^{+1.13}_{-1.29}$ & $ 0.28^{+0.02}_{-0.01}$\\
M2+CARMA+ACA\!\!\! & $ 0.15^{+0.57}_{-0.55}$ & $ 0.14^{+0.91}_{-0.95}$ & $ 0.27^{+0.01}_{-0.01}$\\
\multicolumn{4}{c}{Including known mergers excluding non-detections}\\
CARMA              & $ 0.54^{+0.44}_{-0.42}$ & $-0.40^{+0.71}_{-0.73}$ & $ 0.17^{+0.02}_{-0.00}$\\
CARMA+ACA          & $ 0.55^{+0.30}_{-0.28}$ & $-0.49^{+0.49}_{-0.52}$ & $ 0.15^{+0.01}_{-0.00}$\\
MUSTANG2 (M2)      & $ 0.63^{+0.45}_{-0.41}$ & $-0.68^{+0.72}_{-0.78}$ & $ 0.12^{+0.02}_{-0.00}$\\
M2+CARMA           & $ 0.47^{+0.29}_{-0.29}$ & $-0.34^{+0.50}_{-0.51}$ & $ 0.16^{+0.01}_{-0.00}$\\
M2+ACA             & $ 0.69^{+0.42}_{-0.38}$ & $-0.78^{+0.66}_{-0.73}$ & $ 0.12^{+0.01}_{-0.00}$\\
M2+CARMA+ACA\!\!\! & $ 0.56^{+0.30}_{-0.30}$ & $-0.50^{+0.52}_{-0.51}$ & $ 0.15^{+0.01}_{-0.00}$\\
\multicolumn{4}{c}{Including known mergers including non-detections}\\
CARMA              &         \ldots          &         \ldots          &         \ldots         \\
CARMA+ACA          & $ 0.65^{+0.33}_{-0.32}$ & $-0.68^{+0.54}_{-0.56}$ & $ 0.25^{+0.01}_{-0.00}$\\
MUSTANG2 (M2)      & $ 0.77^{+0.56}_{-0.50}$ & $-0.95^{+0.88}_{-0.98}$ & $ 0.25^{+0.01}_{-0.00}$\\
M2+CARMA           & $ 0.49^{+0.33}_{-0.32}$ & $-0.39^{+0.56}_{-0.56}$ & $ 0.23^{+0.01}_{-0.00}$\\
M2+ACA             & $ 0.90^{+0.46}_{-0.42}$ & $-1.16^{+0.72}_{-0.79}$ & $ 0.26^{+0.01}_{-0.01}$\\
M2+CARMA+ACA\!\!\! & $ 0.65^{+0.33}_{-0.32}$ & $-0.69^{+0.56}_{-0.56}$ & $ 0.25^{+0.01}_{-0.00}$\\
\end{tabular}
\end{center}
\end{table}

% Include this line if you are using the \added, \replaced, \deleted

%\begin{thebibliography}{}

%\end{thebibliography}

\end{document}